\documentclass[11pt]{article}

\usepackage{graphicx}
\usepackage[lmargin=2cm,rmargin=2cm,tmargin=2cm,bmargin=2cm]{geometry}
\usepackage[misc]{ifsym}
\usepackage{color}
\usepackage{empheq}
\usepackage{stmaryrd}
\usepackage{comment}
\usepackage{subcaption}
\usepackage{gensymb}
\usepackage{gensymb}
\usepackage{amsfonts} 
\usepackage{amssymb} 
\usepackage[normalem]{ulem}
\usepackage{authblk}
\usepackage{graphics}

\newcommand{\dV}{\textnormal{ d}V}
\newcommand{\ds}{\textnormal{ d}s}

\newcommand{\grad}{\boldsymbol{\nabla}}
\newcommand{\n}{\boldsymbol{n}}
\newcommand{\bs}{\boldsymbol}

\newcommand{\Otips}{\Omega_\textnormal{tips}}
\newcommand{\Oxfem}{\Omega_\textnormal{xfem}}

\long\def\symbolfootnote[#1]#2{\begingroup
	\def\thefootnote{\fnsymbol{footnote}}\footnote[#1]{#2}\endgroup}

\begin{document}
\title{A combined XFEM phase-field computational model for crack growth without remeshing}

\author{Alba Muix\'i,       
		Onofre Marco,
        Antonio Rodr\'iguez-Ferran,
        Sonia Fern\'andez-M\'endez\footnote{Corresponding author. E-mail: sonia.fernandez@upc.edu}  
}

\affil[]{Laboratori de C\`{a}lcul Num\`{e}ric (LaC\`{a}N) \\ Universitat Polit\`{e}cnica de Catalunya \\ BarcelonaTech, Barcelona, Spain}

\date{}

\maketitle

\begin{abstract}
This paper presents an adaptive strategy for phase-field simulations with transition to fracture.
The phase-field equations are solved only in small subdomains around crack tips to determine propagation, while an XFEM discretization is used in the rest of the domain to represent sharp cracks, enabling to use a coarser discretization and therefore reducing the computational cost. 
Crack-tip subdomains move as cracks propagate in a fully automatic process. 
The same computational mesh is used during all the simulation, with an $h$-refined approximation in the elements in the crack-tip subdomains. 
Continuity of the displacement between the refined subdomains and the XFEM region
is imposed in weak form via Nitsche's method. 
The robustness of the strategy is shown for some numerical examples in 2D and 3D, including branching and coalescence tests.

\vspace{2mm}

\end{abstract}

\section{Introduction} \label{sec:intro}

Typical approaches to model brittle or quasi-brittle fracture can be classified in discontinuous and continuous models, depending on the way cracks are described. 

In \textit{discontinuous models}, cracks are represented as discontinuities in the displacement field (\textit{sharp cracks}), embedded in an elastic bulk. Specific mechanical criteria -- typically based on linear elastic fracture mechanics (LEFM) concepts and based on crack-tip information -- are needed to model crack inception, propagation and branching. A source of difficulty is the singularity of the strain and stress fields in classical linear elasticity at crack tips. For this reason, stress intensity factors (SIF) are often used \cite{MoesBelytschko2002}.

From a computational viewpoint, two different strategies may be used to describe the displacement jump associated to a sharp discontinuity: node and edge duplication, in the spirit of the original cohesive zone model (CZM), see \cite{OrtizPandolfi1999}, or the extended finite element method (XFEM) and related techniques, where the crack path is not constrained to follow element edges and therefore remeshing is avoided \cite{BelytschkoBlack1999,MoesDolbowBelytschko1999}. 
Together with a certain mesh-dependence of the crack path (for CZM) or geometrical complexity (for XFEM), discontinuous models offer various advantages, namely: \textit{i)} the explicit representation of the crack facilitates the modeling of crack-surface physics; \textit{ii)} the sharp discontinuity avoids any spurious interaction between the two faces of the crack; \textit{iii)} coarse discretizations may be used in the wake of the crack tip.

\textit{Continuous models}, usually phase-field or gradient-damage models, assume continuous displacement fields and represent cracks as damaged regions that have lost their load-carrying capacity (\textit{diffuse cracks}). Information on the evolution of cracks is incorporated into the field equations; initiation, propagation, branching and coalescence of cracks are automatically handled by the model. 

The main drawback of these models is their high computational cost. Very fine meshes are needed to capture narrow damage bands that resemble sharp cracks. For cases in which the crack path is not known a priori, one may use an adaptive strategy to automatically refine the discretization  \cite{Nagarajaetal2019,MuixiHDG,MuixiNitsche,GeelenPlewsTupekDolbow2020}. Continuous models do not explicitly identify a crack surface within damage bands and spurious interaction remains between physically separated parts. Several works adressing this issue have been presented which combine continuous and discontinuous approaches \cite{TamayoMasRodriguezFerran2015,GeelenLiuDolbowRodriguezFerran2018}.

In continuous-discontinuous models both sharp and diffuse representations of cracks are overlapped. The discontinuous model is used to represent material separation while the continuous one is used to determine the evolution of cracks.  Critical issues are related to the introduction of discontinuities: when and where (switching and locating criteria, respectively). Still, these models have a high computational cost from solving the continuous model in the whole domain. 

Recent approches consider a phase-field problem only in moving subdomains containing crack tips to improve the efficiency of the simulations. Giovanardi et al \cite{GiovanardiScottiFormaggia2017} propose a two-scale strategy. A global solution for the displacement field is obtained with XFEM. Then, the propagation is determined by solving a phase-field problem in subdomains containing the crack tips with a finer mesh. The strategy is applied to propagation examples in 2D, with no branching nor coalescence. 

Patil et al \cite{PatilMishraSingh2018} present a similar approach: phase-field is used only in small crack-tip regions, and cracks are described by an XFEM discretization in the whole domain. 
In the so-called crack-tip regions, the mesh is adaptively refined with a uniform submesh. The multiscale FEM is used to couple the finer discretization with the coarser one, by means of transition elements and the construction of multi-scale shape functions relating the degrees of freedom from the two levels of refinement. The strategy is successfully applied to branching and coalescence of cracks in 2D. 

With the same philosophy as the above-mentioned proposals, we present a model combining phase-field in small subdomains around crack tips, and a discontinuous model in the rest of the domain. Here, the two descriptions of cracks are not overlapped.
The phase-field model is used to handle the propagation and coalescence of cracks, while the discontinuous model explicitly describes cracks and enables to use a coarser discretization in almost the whole domain.
The presented methodology is stated in 2D and 3D. As a simplification, we use the term crack tips to refer both to crack tips, in 2D, and crack fronts, in 3D.

The small phase-field subdomains consist of a set of elements for which the discretization is uniformly $h$-refined to capture the solution, following the refinement strategy proposed in \cite{MuixiNitsche}, and move with crack tips as cracks propagate in a fully automatic process. The mesh is fixed during all the simulation. As the diffuse cracks evolve, refined elements are nested near crack tips. At the same time, refined elements in the wake of the crack transition into an XFEM coarser approximation. 

On the interface between continuous and discontinuous subdomains, continuity of the displacement field is imposed in weak form by means of Nitsche's method. Imposing continuity in weak form avoids the definition of transition elements and having to constrain the degrees of freedom from the richer part. This leads to a very local refinement, which is handled by the method for any refinement factor.

The paper is structured as follows. First, in Section \ref{sec:models}, we recall the discontinuous and the phase-field models. 
Section \ref{sec:theConcept} describes the idea behind the combined computational model. 
In Section \ref{sec:Nitsche} we
give the FE formulation, together with some implementation details. 
Sections \ref{sec:update}, \ref{sec:sharpCrack} and \ref{sec:refinement} are devoted to the evolution of the crack-tip subdomains and the transition to fracture as cracks propagate.
In Section \ref{sec:numerical-experiments}, we present some representative numerical examples in 2D and a fully 3D test to show the capabilities of the strategy. 
The conclusions in Section \ref{sec:conclusions} close the paper.

\section{XFEM and phase-field computational models} \label{sec:models}

In this paper, we combine a discontinuous model and a phase-field model to simulate crack propagation in the quasi-static regime.
In the discontinuous model, cracks are described by discontinuous displacements fields. In phase-field, cracks are approximated by diffuse bands with a loss of stiffness, and the displacement field is continuous across them.
In this section we give a brief overview of both approaches. 
As a representative phase-field model, we use here the hybrid model proposed by Ambati et al \cite{AmbatiGerasimovDeLorenzis2015}.

\begin{figure}[h]
	\vspace{4mm}
	\centering
	\raisebox{1.3cm}{
		\includegraphics[width=0.29\textwidth]{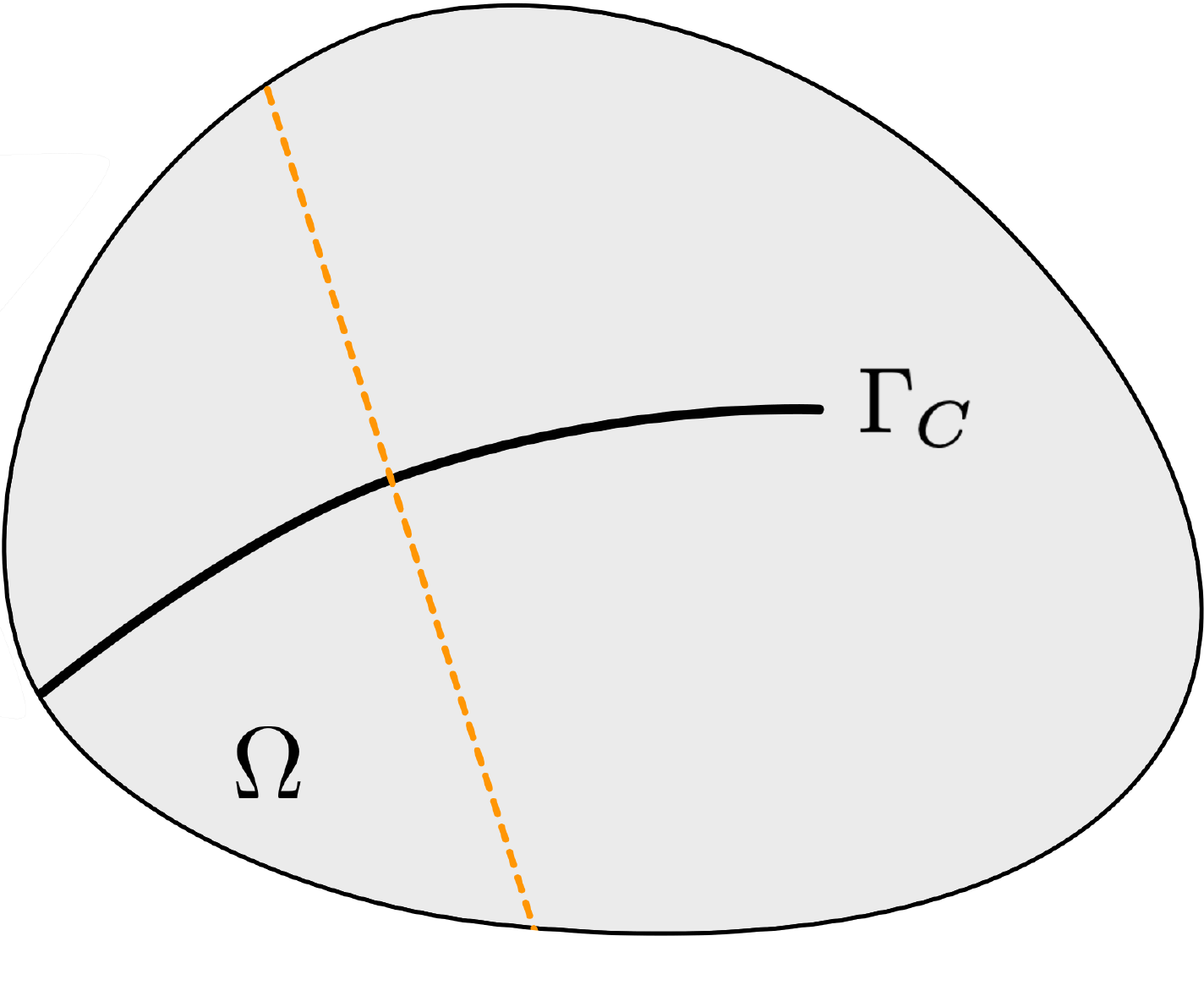}}
	\hspace{-5mm}
	\includegraphics[width=0.18\textwidth]{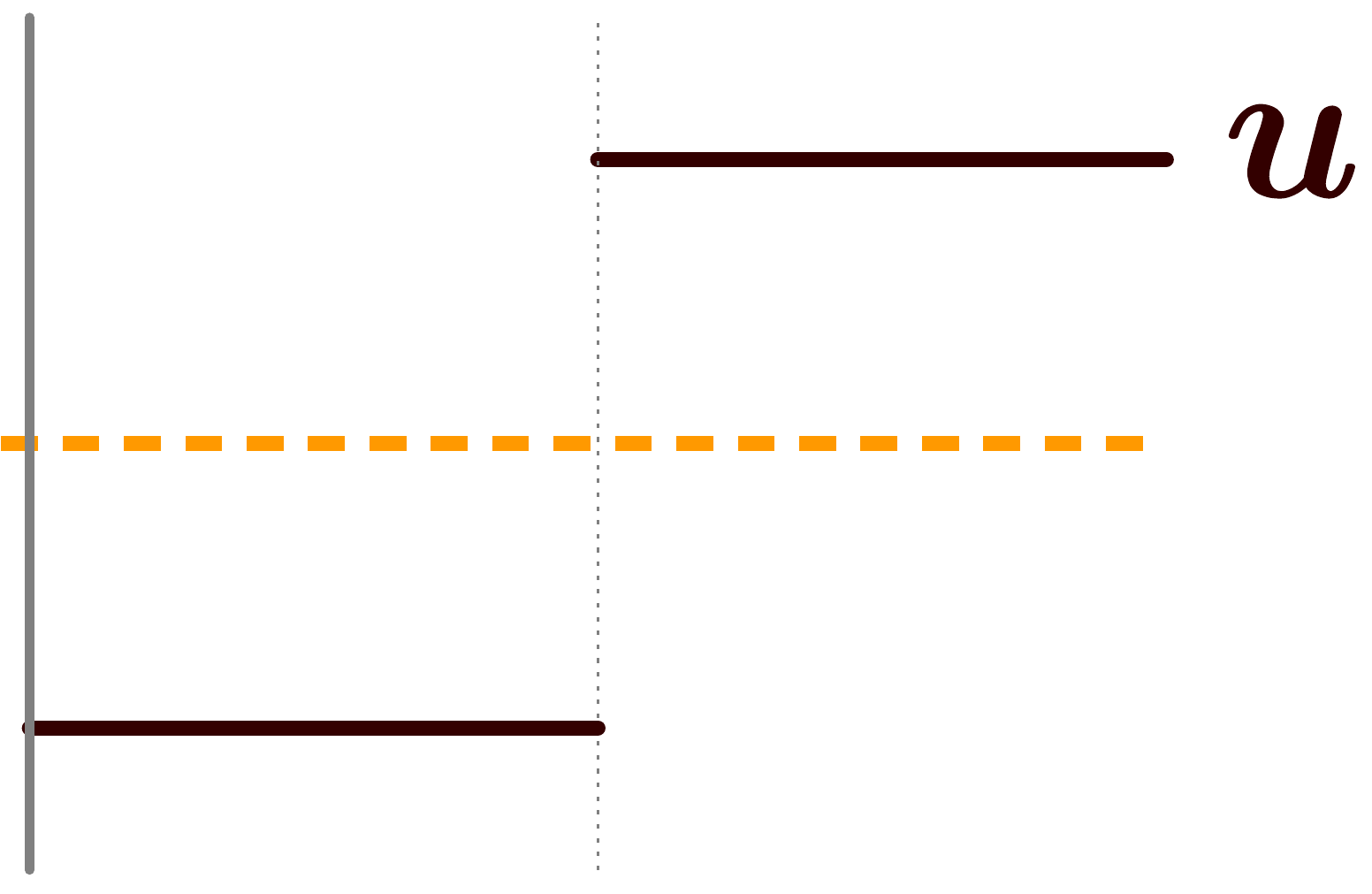}
	\hspace{0mm}
	\raisebox{1.3cm}{
		\includegraphics[width=0.29\textwidth]{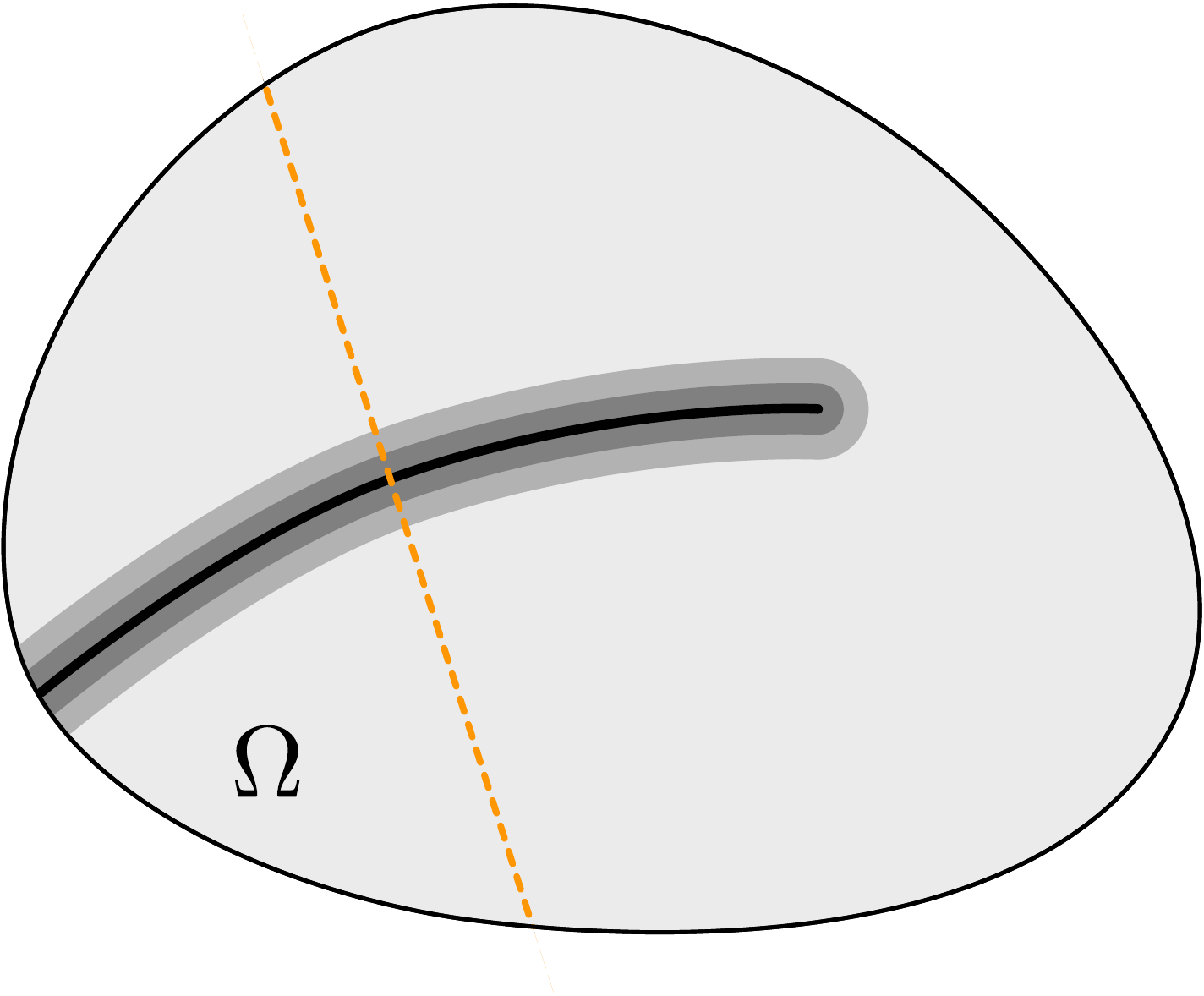}}
	\hspace{-2mm}
	\includegraphics[width=0.2\textwidth]{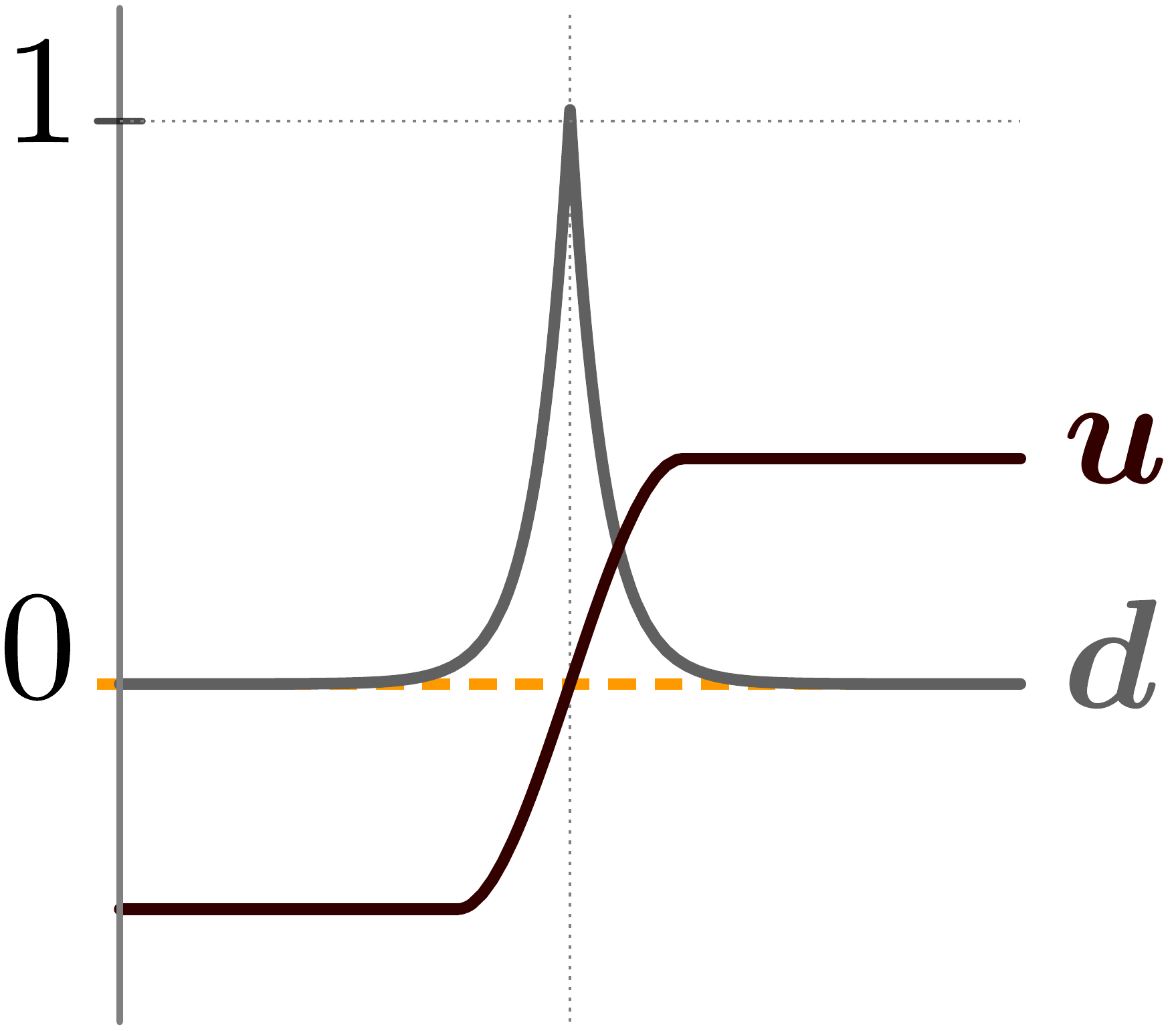}
	\caption{Body with a sharp crack $\Gamma_C$ (left) and a diffuse phase-field representation of the crack with damage variable $d$ (right). The profile of the displacement field along the dotted section shows how the discontinuity from the sharp crack is approximated by a continuous function with a steep variation in phase-field models.}
	\label{fig:disc-cont}       
\end{figure}

Throughout the paper, we consider a linear elastic body with a traction-free crack, occupying a domain $\Omega\subset\mathbb{R}^{n_{sd}}$, with $n_{sd}=2,3$, and under the assumption of small deformations, as illustrated in Figure \ref{fig:disc-cont}. 
We restrict to linear isotropic materials.

\subsection{Discontinuous model}

If no body forces are applied and inertia effects are neglected, the balance of linear momentum 
for the body in $\Omega$
leads to the system 
\begin{subequations}\label{disc-model}
	\begin{empheq}[left=\empheqlbrace]{align}
	&\grad \cdot \bs{\sigma} = \bs{0} && \textnormal{in } \Omega_C, \\
	&\bs{u} = \bs{u}_D && \textnormal{on } \Gamma_D, \\
	&\bs{\sigma} \cdot \n = \bs{t}_N && \textnormal{on } \Gamma_N, \\
	&\bs{\sigma} \cdot \n = \bs{0} && \textnormal{on } \Gamma^+_C \cup \Gamma^-_C, \label{disc-crack} 
	\end{empheq}
\end{subequations}
where $\Omega_C = \Omega\setminus\Gamma_C$ is the cracked domain.
The relation between the displacement field $\bs{u}$ and the stress tensor  $\bs{\sigma}$ is given by the constitutive equation
\begin{equation}\label{discontinuous-constitutive}
\boldsymbol{\sigma} = \dfrac{\partial\Psi_0(\bs{\varepsilon})}{\partial\bs{\varepsilon}}, 
\end{equation}
where $\bs{\varepsilon}$ is the small strain tensor and $\Psi_0$ is the elastic energy density. 
For linear isotropic materials,  
$\Psi_0 = \left( \bs{\varepsilon} : \bs{C} : \bs{\varepsilon} \right) / 2$ and $\bs{\sigma} = \bs{C} : \bs{\varepsilon}$,
with $\boldsymbol{C}$ a fourth order positive definite tensor depending on the Lam\'e parameters.

The Dirichlet and Neumann boundaries are denoted by $\Gamma_D$ and $\Gamma_N$, respectively, and satisfy $\partial\Omega =\Gamma_D \cup  \Gamma_N$ and $\Gamma_D \cap \Gamma_N = \emptyset$. Prescribed displacements and tractions are $\bs{u}_D$ and $\bs{t}_N$, and $\n$ is the exterior unit normal.

Equation \eqref{disc-crack} imposes traction-free conditions on the crack faces, $\Gamma^+_C$ and $\Gamma^-_C$.

Numerical techniques to solve the system in \eqref{disc-model} need to account for the discontinuity of displacements across the crack. The most popular method used to this intent is XFEM \cite{BelytschkoBlack1999,MoesDolbowBelytschko1999}, which enriches the discretization space to properly represent discontinuous displacements with no need to adapt the mesh to the crack geometry. XFEM has demonstrated its applicability and computational efficiency in many applications \cite{FriesBelytschko2010}, but it presents some limitations for crack simulations.

It can be analytically shown that stresses around crack tips vary according to $r^{-1/2}$, where $r$ is the distance to the crack tip \cite{SunJin2012}. This stress singularity is physically unrealistic and, from our experience, it leads to difficulties in the definition of a proper XFEM enrichment around crack tips.  More precisely, to ensure convergence under uniform mesh refinement, the area were the crack-tip enrichment is applied should be kept fixed; but this strategy leads to severe ill-conditioning. 
On the other hand, if the area for the crack-tip enrichment is reduced with the element size, asymptotic convergence cannot be observed \cite{LabordePommierRenardSalau2005}. Thus, from a practical point of view, crack-tip enrichment is not straightforward.
An overview of the advances of the method to solve problems in fracture mechanics can be found in Sukumar et al \cite{SukumarDolbowMoes2015}.

However, the main limitation of the discontinuous approach is that crack inception and propagation are not described by the model.
Different criteria have been proposed to determine the direction and velocity of propagation using local information of the solution at crack tips, see for instance \cite{Erdogan1963,Sih1974,Chang1981}, but a robust theory is not well-established yet.
This is not the case for phase-field computational models, which handle crack evolution in a natural way.

\subsection{The hybrid phase-field model}

In phase-field models, cracks are represented by means of the so-called \textit{phase-field variable} or \textit{damage field}, which is denoted by $d$. The damage takes value $1$ on the fracture path, value $0$ in intact parts of the material and smoothly varies between the two values, as sketched in Figure \ref{fig:disc-cont}. 
The damage field is introduced as an additional unknown into the system of equations. Thus, evolution of cracks is determined by the model itself.

The hybrid phase-field model by Ambati et al \cite{AmbatiGerasimovDeLorenzis2015} is characterized by incorporating a tension-compression splitting, while maintaining a linear equilibrium equation within a staggered scheme to solve the system.
The governing equations are
\begin{subequations}\label{hybrid_system}
	\begin{empheq}[left=\empheqlbrace]{align}
	&	\grad\cdot\bs{\sigma} = \bs{0}\ \textnormal{ with }\ \bs{\sigma} =  g(d)  \dfrac{\partial\Psi_0(\bs{\varepsilon})}{ \partial\bs{\varepsilon}},\label{eqEquation} \\
	&-l^2 \Delta d + d = \dfrac{2l}{G_C}(1-d)\mathcal{H}^+, \label{damageEquation} \\
	& g(d):=  
	\begin{cases}
	(1-d)^2 &\text{ where }\ \Psi_0^+ \ge \Psi_0^-, \label{hybridCondition} \\
	1 &\text{ otherwise,}
	\end{cases} 
	\end{empheq}
\end{subequations}
where $G_C$ is the energy release rate of the material. The length parameter $l$ is  related to the width of the diffuse cracks and it is usually taken very small to mimic sharp cracks. This entails the need of very fine meshes around cracks to properly approximate the solution.

The stress tensor in the equilibrium equation \eqref{eqEquation} is degraded by the function $g(d)$. In fully broken parts of the material, where $d = 1$, this implies a complete loss of stiffness. In  parts of the material with $d = 0$, the linear elastic stress-strain constitutive relation \eqref{discontinuous-constitutive} is recovered.

The model considers an additive decomposition of $\Psi_0$, namely, $\Psi_0 = \Psi_0^+ + \Psi_0^-$ with $\Psi_0^+$ and $\Psi_0^-$ the tensile and compressive parts of $\Psi_0$, respectively. Here, we assume the tension-compression splitting by Miehe et al \cite{Miehe2,Miehe1} which is based on the spectral decomposition of $\bs{\varepsilon}$. 
Denoting the principal strains by $\{ \varepsilon_i \}_{i=1,...,n_{sd}}$ and the principal strain directions by $\{\bs{d}_i\}_{i=1,...,n_{sd}}$, 
\begin{equation}
\Psi_0^{\pm}(\bs{\varepsilon}) = \frac{1}{2} \lambda \langle \textnormal{tr} (\bs{\varepsilon}) \rangle^2_\pm + \mu \textnormal{tr} \left(\bs{\varepsilon}^2_\pm \right),
\end{equation}
where  $\bs{\varepsilon}_\pm = \sum_{i=1}^{n_{sd}} \langle \varepsilon_i \rangle_\pm \bs{d}_i \otimes \bs{d}_i$ and $\langle \odot \rangle_\pm = \left( \odot \pm |\odot| \right)/2$.

The evolution of the damage field $d$ is modeled by equation \eqref{damageEquation}. The history field variable, $\mathcal{H}^+$, is defined as
\begin{equation}\label{splitting_H}
 \mathcal{H}^+(\bs{x},t) = \displaystyle\max_{\tau\in[0,t]} \Psi_0^+\bigl(\bs{\varepsilon}\left(\bs{x},\tau\right)\bigr),
 \end{equation}
and therefore the propagation of $d$ is only caused by the tensile $\Psi_0^+$. The definition of $\mathcal{H}^+$ also enforces irreversibility of fracture \cite{Miehe2,Miehe1}. 

The condition in equation \eqref{hybridCondition} recovers the original stiffness of the material under compression, in order to prevent interpenetration of crack faces. Alternatively, the splitting of $\Psi_0$ may be included into the stress-strain equation in \eqref{eqEquation}, but then the equilibrium equation becomes nonlinear.

The system in \eqref{hybrid_system} is solved with boundary conditions
\begin{equation}\label{BCs_hybrid}
\left\{
\begin{aligned}
&\bs{\sigma}\cdot\n = \bs{t}_N &\textnormal{ on } \Gamma_N, \\
&\bs{u} = \bs{u}_D  &\textnormal{ on } \Gamma_D,\\
&\grad d \cdot \n = 0 &\textnormal{ on } \partial\Omega.
\end{aligned}
\right.
\end{equation}

It is important noting that the value of $l$ determines the element size required around cracks. 
Thus, differently to other applications, adaptive refinement can be directly driven by the crack growth, \emph{with a refinement element size that is known a priori}. This is the idea exploited in \cite{MuixiNitsche}, where elements along the crack are automatically refined considering a refined reference element, and keeping the original background finite element mesh in the whole computation.
The weaknesses of the adaptive phase-field model are the inability of phase-field to explicitly describe crack opening, which is necessary in some applications, and, more importantly,  the computational cost associated to the refinement in all elements along the crack path. 
Thus, the next natural step is derefining the elements in the wake of the crack tips and replacing the diffuse band by a sharp crack, where no variation of the damage is expected, by means of an XFEM enrichment.

In the next section, XFEM and the adaptive phase-field model in \cite{MuixiNitsche} are combined in a new method that has the computational efficiency of XFEM, and also the ability of phase-field models to handle crack evolution.

\section{The concept: sharp cracks with phase-field crack tips} \label{sec:theConcept}

A novel computational model combining a phase-field model and an XFEM discontinuous approach is proposed.
The phase-field equations are solved in small subdomains around crack tips, $\Otips$, and the discontinuous model is used in the rest of the domain, $\Oxfem$. The corresponding diffuse and sharp definitions of cracks are not overlapped and we denote by $\Gamma$ the interface between subdomains, this is, 
\begin{equation*}
\overline{\Omega} = \overline{\Otips} \cup \overline{\Oxfem},\ 
\Otips \cap \Oxfem = \emptyset, \  
\Gamma=\overline{\Otips} \cap \overline{\Oxfem}.
\end{equation*}

The model overcomes the limitations of both approaches. The phase-field model drives the crack propagation and deals with branching and merging of cracks, whereas the crack is explicitly described in almost all the domain by a sharp discontinuity. 
Thus, the high spatial resolution needed around crack paths in standard phase-field models is now only necessary in small neighborhoods of the crack tips, with the consequent important saving in computational cost.

The partition of the domain is based on the computational mesh. 
Crack-tip subdomains $\Otips$ are defined as the set of elements close to crack tips and tips of notches in the domain, where crack growth is expected. 
The rest of elements are $\Oxfem$. The initial definition of $\Otips$ is crucial to detect crack inception, since the damage field is solved only in this part of the domain.

\begin{figure}
	\centering
	\includegraphics[width=0.3\textwidth]{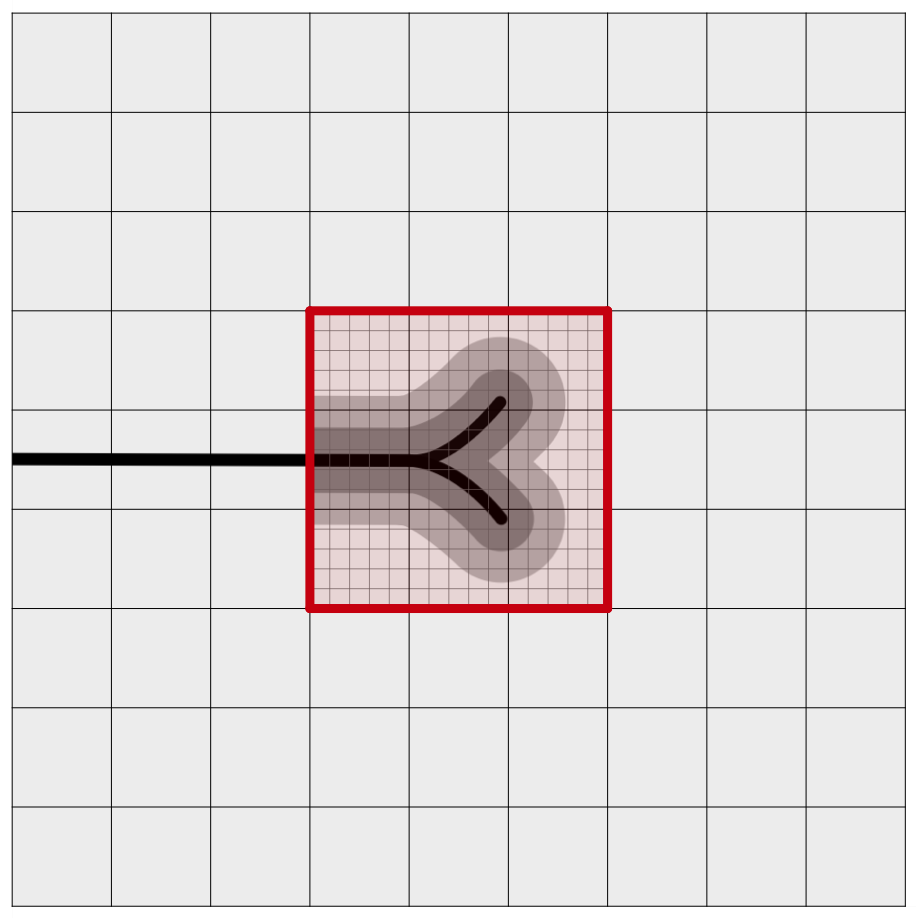}
	\hspace{3mm}
	\includegraphics[width=0.3\textwidth]{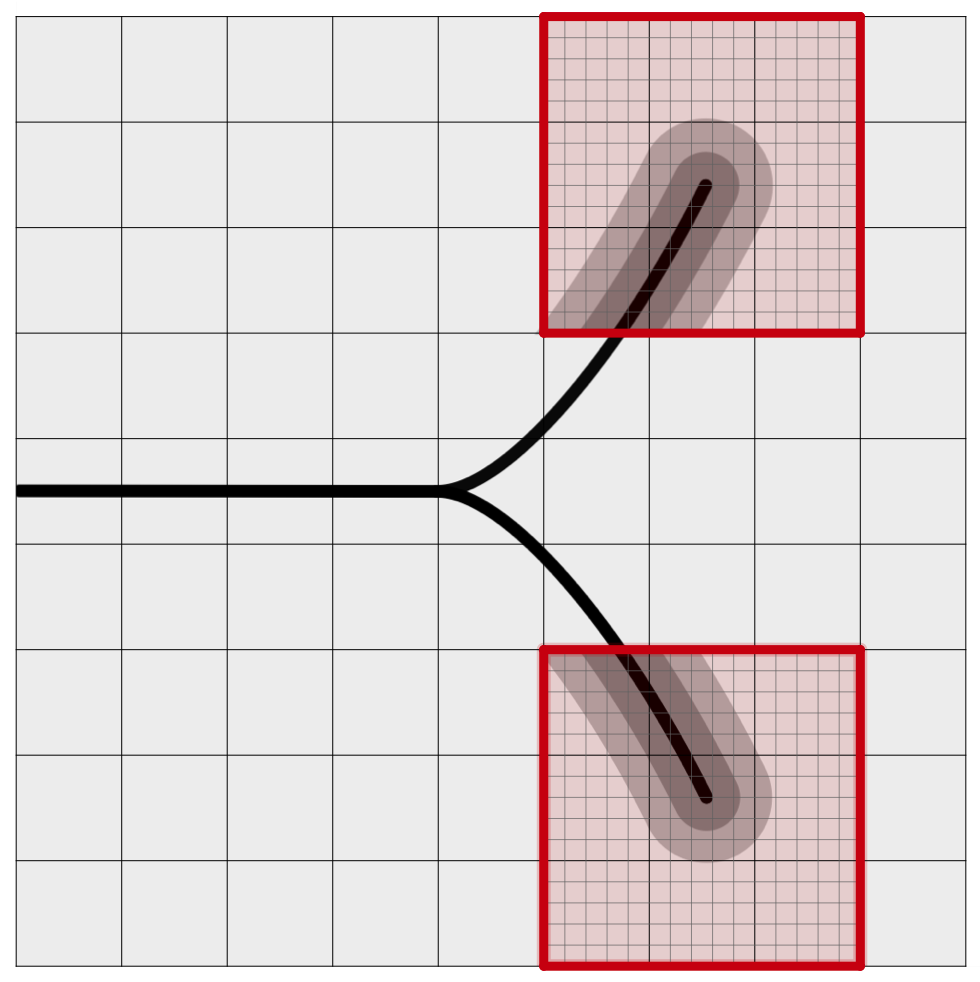}
	\hspace{2mm}
	\raisebox{1.5cm}{
		\begin{subfigure}[b]{0.1\textwidth}
			\includegraphics[width=0.4\textwidth]{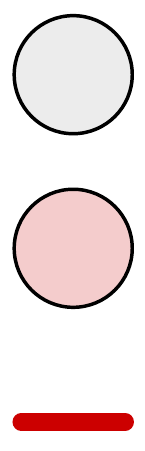}
			\raisebox{0mm}{
				\begin{subfigure}[b]{0.45\textwidth}
					\includegraphics[height=4.5mm]{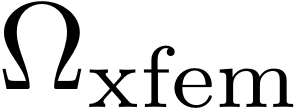}
					
					\vspace{3.6mm}
					\includegraphics[height=4.75mm]{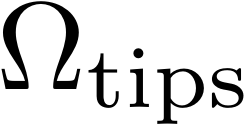}
					
					\vspace{3.2mm}
					\includegraphics[height=3.5mm]{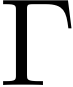}
			\end{subfigure}}
	\end{subfigure}}
	\caption{Scheme of a discretization in two consecutive load steps. In $\Otips$, elements are uniformly $h$-refined and cracks are represented by diffuse bands. The FE discretization in $\Oxfem$ is the one corresponding to the initial mesh, with an XFEM Heaviside enrichment to introduce sharp cracks, represented with black line. The interface $\Gamma$ is depicted in red.}
	\label{fig:strategy-cartoon}       
\end{figure}

As crack tips advance during the computation, the domains $\Otips$ and $\Oxfem$ are repeatedly updated, refining elements in the nose of the cracks and derefining elements in their wake.
When derefining, diffuse cracks are replaced by XFEM sharp traction-free cracks. 

Following the concept in \cite{MuixiNitsche}, elements in $\Otips$ are automatically identified and uniformly refined to properly approximate the phase-field solution, now only around crack tips. The refinement is based on a refined reference element, keeping the original background mesh during the whole computation.

Figure \ref{fig:strategy-cartoon} illustrates the discretizations for two consecutive load steps. 
The elements in $\Otips$ are $h$-refined with a uniform submesh with $m^{n_{sd}}$ subelements. 
The refinement factor $m$ is chosen a priori depending on the length-scale parameter $l$ of the phase-field model.
As the crack evolves, the small subdomains move with the crack tips and the diffuse band becomes a sharp discontinuity in the wake of the crack.

The main challenge is gluing the two models on the interface $\Gamma$, with different discretizations and representations of cracks on each side.

For the equilibrium equation, continuity of the displacement is imposed in weak form by means of Nitsche's method. The Nitsche's formulation, recalled in Section \ref{sec:Nitsche}, is the one considered in \cite{MuixiNitsche}, but now with an XFEM enriched approximation in $\Oxfem$. In addition, when imposing continuity, a small portion of the interface around the intersections with the crack is discarded.

Numerical experiments show that unrealistic displacement fields are obtained if the continuity is enforced on the whole interface $\Gamma$. The reason is that the description of the crack on both sides of the interface is different.
From $\Oxfem$ the crack is sharp and the material retains all the stiffness, whereas on $\Otips$ we have a smooth representation of the crack with a material degradation given by $g(d)$, leading to almost fully damaged material close to the crack path. Thus, for parts of $\Gamma$ which are crossed by cracks, the difference of stiffness between the two subdomains is quite significant.

\begin{figure}[]
	\centering
	\includegraphics[height=3.7cm]{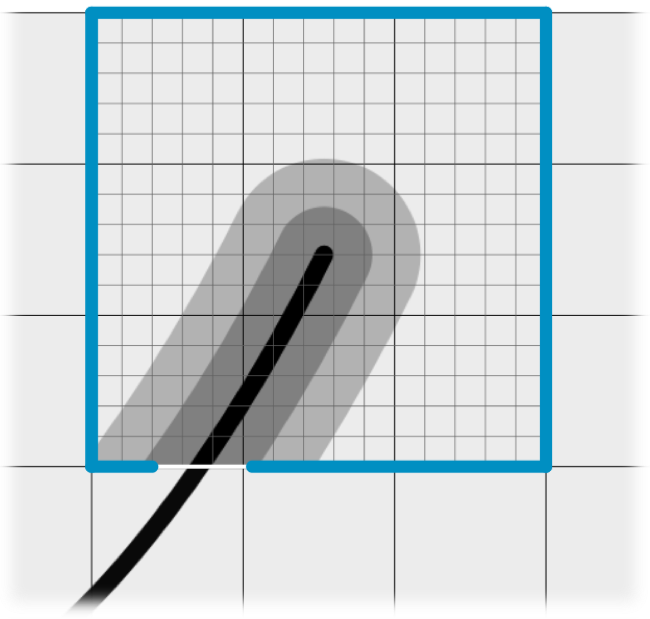}
	\hspace{2mm}
	\includegraphics[height=3.69cm]{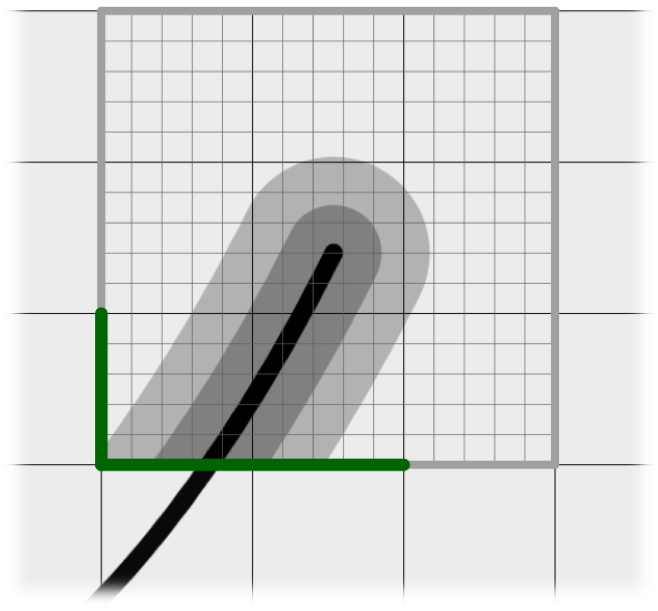}
	\caption{Continuity of displacements is imposed in weak form on the interface $\widehat\Gamma$, in blue (left). Cracks do not necessarily intersect $\Gamma$ perpendicularly. Dirichlet boundary conditions for the damage field $d$ are imposed on $\Gamma_D^d$, i.e. on the side intersected by the crack and on the adjacent sides, in green (right).}
	\label{fig:gluing-u-d}             
\end{figure}

Continuity is then enforced on a slightly cropped interface defined as
\begin{equation}\label{eq:GammaCropped}
\widehat\Gamma = \{\bs{x}\in\Gamma : d(\bs{x})<0.9 \}.
\end{equation}
Figure \ref{fig:gluing-u-d} illustrates the concept. A  simple test comparing the results obtained imposing continuity on $\Gamma$ and on the cropped interface, $\widehat\Gamma$, is presented in Section \ref{test-continuity}. 

It is worth noting that imposing continuity in weak form, by means of Nitsche's method, handles the non-conformal approximations in a natural way. That is, refined discretizations are directly attached to non-refined discretizations, with a standard finite element approximation or with XFEM enrichment. No transition elements are required, avoiding the spreading of the refinement or the enrichment, and without having to deal with hanging nodes.

The other important issue for gluing the computational models is setting proper boundary conditions for the damage variable $d$, since we solve for it only in $\Otips$. 
In standard phase-field models, the boundary conditions for the damage equation are usually taken as homogeneous Neumann on $\partial\Omega$. 
However, we cannot assume that cracks intersect the interface  $\Gamma$ with perpendicularity. 
With this in mind, Dirichlet boundary conditions are imposed on the faces that have been crossed by a crack. More precisely, the Dirichlet boundary, $\Gamma_D^d$, consists of the element sides (in 2D, faces in 3D) on $\Gamma$ that are cut by the crack, i.e. with  $d > 0.95$ at some node, together with their neighboring faces on $\Gamma$. See an example in Figure \ref{fig:gluing-u-d} (right). 
When transitioning elements intersected by the crack from $\Otips$ to $\Oxfem$, the damage field on $\Gamma_D^d$ is saved to be used as Dirichlet value for the damage equation. That is, the damage is assumed to be frozen in the wake of crack tips.
On the rest of the boundary, $\partial\Otips \setminus \Gamma_D^d$, homogeneous Neumann conditions are imposed  not to favor any direction of crack growth.

\section{Finite element formulation} \label{sec:Nitsche}

In this section, we present the formulation, commenting also on some implementation details.
The fracture model is solved in a staggered scheme. Thus, the numerical formulations for the equilibrium and for the damage equation are independent. 

In the equilibrium equation, the discretization has to account for discontinuous displacements across cracks in $\Oxfem$. Also, continuity of displacements between $\Otips$ and $\Oxfem$ has to be handled by the formulation. 
On the other hand, the damage equation is solved with the standard finite element method with refined approximation in $\Otips$.

\subsection{Equilibrium equation: Nitsche's method}

The equilibrium equation is solved in $\Omega$, with different non-conformal approximations for the displacement field in the partition given by $\Otips$ and $\Oxfem$. 
Thus, the displacement is assumed to be in the space $[\mathcal{V}(\Omega)]^{n_{sd}}$ with
\begin{equation*}
	\mathcal{V}(\Omega) = \big\{ v\in L^2(\Omega) :\   
	v|_{\Otips} \in H^1(\Otips) ,
	v|_{\Oxfem} \in H^1(\Oxfem\backslash\Gamma_{c}) \big\},
\end{equation*}
which includes functions that are discontinuous across the interface $\Gamma$, and across the sharp crack $\Gamma_{c}$; the second one, handled by the XFEM enrichment.

Our choice to weakly impose continuity on the interface $\Gamma$ is Nitsche's method. The formulation is based on the Nitsche's formulation for linear elasticity applied to interface problems by Hansbo \cite{Hansbo2005-interfaceCM}, see also \cite{MuixiNitsche}.

In Nitsche's method, the weak form of classical finite element approach is modified in order to impose continuity of the solution on the interface, while keeping a symmetric and coercive bilinear form, and also imposing equilibrium of tractions. 
Differently to Lagrange multipliers, no additional unknowns are introduced to the system.

The weak form reads: 
find $\bs{u}\in[\mathcal{V}(\Omega)]^{n_{sd}}$ such that $\bs{u} = \bs{u}_D$ on $\Gamma_D$ and
\begin{equation}\label{full_weak_form_eq}
\begin{aligned}
&\int_{\Omega} \grad \bs{v} : \bs{\sigma}(\bs{u}) \dV 
- \int_{\widehat\Gamma} \llbracket \bs{v} \otimes \n \rrbracket : \{ \bs{\sigma}(\bs{u})  \} \ds 
 & -\int_{\widehat\Gamma} \{ \bs{\sigma} (\bs{v} ) \} : \llbracket \bs{u}\otimes\bs{n} \rrbracket \ds
+ \beta_{\text{E}} \int_{\widehat\Gamma} \llbracket \bs{u}\otimes \bs{n} \rrbracket : \llbracket \bs{v}\otimes\bs{n} \rrbracket \ds
\\ & -\int_{\Gamma_N} \bs{v}\cdot \bs{t}_N \ds = 0,
\end{aligned}
\end{equation}
for all $\bs{v}\in[\mathcal{V}(\Omega)]^{n_{sd}}$ such that $\bs{v} = \bs{0}$ on $\Gamma_D$, where $\beta_{\text{E}}$ is the Nitsche's stabilization parameter, and  $\widehat\Gamma$ is almost the whole interface $\Gamma$, discarding small portions around the intersections with the cracks, as defined in \eqref{eq:GammaCropped} and illustrated in Figure \ref{fig:gluing-u-d} (left).

The mean and jump operators are defined as $\{ \odot \} = \frac{1}{2} \left( \odot_t + \odot_x \right)$ and $\llbracket \odot \n\rrbracket = \odot_t \n_t + \odot_x \n_x = (\odot_t - \odot_x) \n_t$, with the lower indices, $t$ and $x$, indicating the values on $\widehat\Gamma$ from $\Otips$ and $\Oxfem$, respectively, and $\n_t$, $\n_x$ the corresponding unit exterior normals.
A derivation of this weak form can be found in \cite{MuixiNitsche}. The differences here are just the aproximation spaces to be glued, now including sharp cracks via XFEM, and the small cropping of the interface, $\widehat\Gamma$.

Even though it has been omitted for simplicity, note that the stress tensor $\bs{\sigma}$ in \eqref{full_weak_form_eq} depends also on the damage field $d$ in $\Otips$.

The stability of the formulation depends on the value of $\beta_{\text{E}}$. This parameter has to be large enough to ensure coercivity of the bilinear form, thus leading to a discrete system with a positive-definite matrix.  
To obtain optimal orders of convergence, this parameter can be taken of the form 
\begin{equation}\label{betaE}
\beta_{\text{E}} =  \alpha_{\text{E}} E (h/m)^{-1},
\end{equation}
with $E$ the Young's modulus, $h$ the element size in the background mesh, $m$ the refinement factor for refined elements, and $\alpha_{\text{E}}$ a large enough positive constant independent of other numerical and material parameters.
A lower bound for its value can be found by solving an eigenvalue problem \cite{GriebelSchweitzer2003}.
However, the formulation is very robust on the parameter and the value of $\alpha_{\text{E}}$ can be easily tuned by numerical experimentation \cite{MuixiNitsche}.

\subsection{Spaces of approximation: XFEM and refined}

The computation is based on a fixed background mesh during the whole simulation, with elements $\{K_i\}_{i=1}^{n_{el}}$. That is,
\[
\overline\Omega= \bigcup_{i=1}^{n_{el}} \overline K_i,\quad K_i\cap K_j=\emptyset \text{ if } i\neq j.
\]
As cracks evolve, in every load step and iteration of the staggered scheme, the elements in the background mesh are identified as refined, i.e. in $\Otips$, if they are close to crack tips or tips of initial notches, or in $\Oxfem$, otherwise.

In $\Oxfem$, the XFEM approximation is based on a standard finite element space, that is
\begin{equation*}
\mathcal{V}^h(\Oxfem) = \{ v\in\mathcal{H}^1(\Oxfem)  :  v|_{K_i} \in\mathcal{P}^p \\ \text{ for } K_i \subseteq \Oxfem \},
\end{equation*}
with nodal basis $\{N_i\}_{i\in I_{\text{xfem}}}$, where $I_{\text{xfem}}$ is the set of the indices of the nodes in $\Oxfem$ and $\mathcal{P}^p$ is the space of polynomials up to degree $p$.
The XFEM approximation space in $\Oxfem$ in the case of a unique crack is then
\[
 \mathcal{V}^h_{\text{xfem}} = < N_i >_{i\in I_{\text{x}}}\otimes  < HN_i >_{i\in I_{\text{enr}}},
\]
where $I_{\text{enr}}$ is the set of indices of enriched nodes, i.e. the nodes of the elements which are cut by the crack, and $H$ is the Heaviside function taking, for instance, values $1$ and $-1$ on each side of the crack. In the presence of branching or several cracks, the approximation is enriched with additional Heaviside functions 
\begin{equation*}
\mathcal{V}^h_{\text{xfem}} = < N_i >_{i\in I_{\text{x}}}\otimes  < H^1N_i >_{i\in I_{\text{enr}}^1} \otimes \dots \\ \otimes < H^{n_{\text{cracks}}}N_i >_{i\in I_{\text{enr}}^{n_{\text{cracks}}}},
\end{equation*}
where ${n_{\text{cracks}}}$ is the total number of crack branches \cite{MoesDolbowBelytschko1999}.
Here, no asymptotic tip enrichment is necessary since $\Oxfem$ does not contain crack tips.

A proper numerical quadrature must be defined in the elements intersected by the crack to account for the discontinuity of the XFEM approximation. That is, a numerical quadrature must be defined in each portion of cut elements, see for instance \cite{Onofre2015,GurkanSalaKronbichlerFernandez2017}.

\begin{figure}[]
	\centering
	\includegraphics[width=0.35\textwidth]{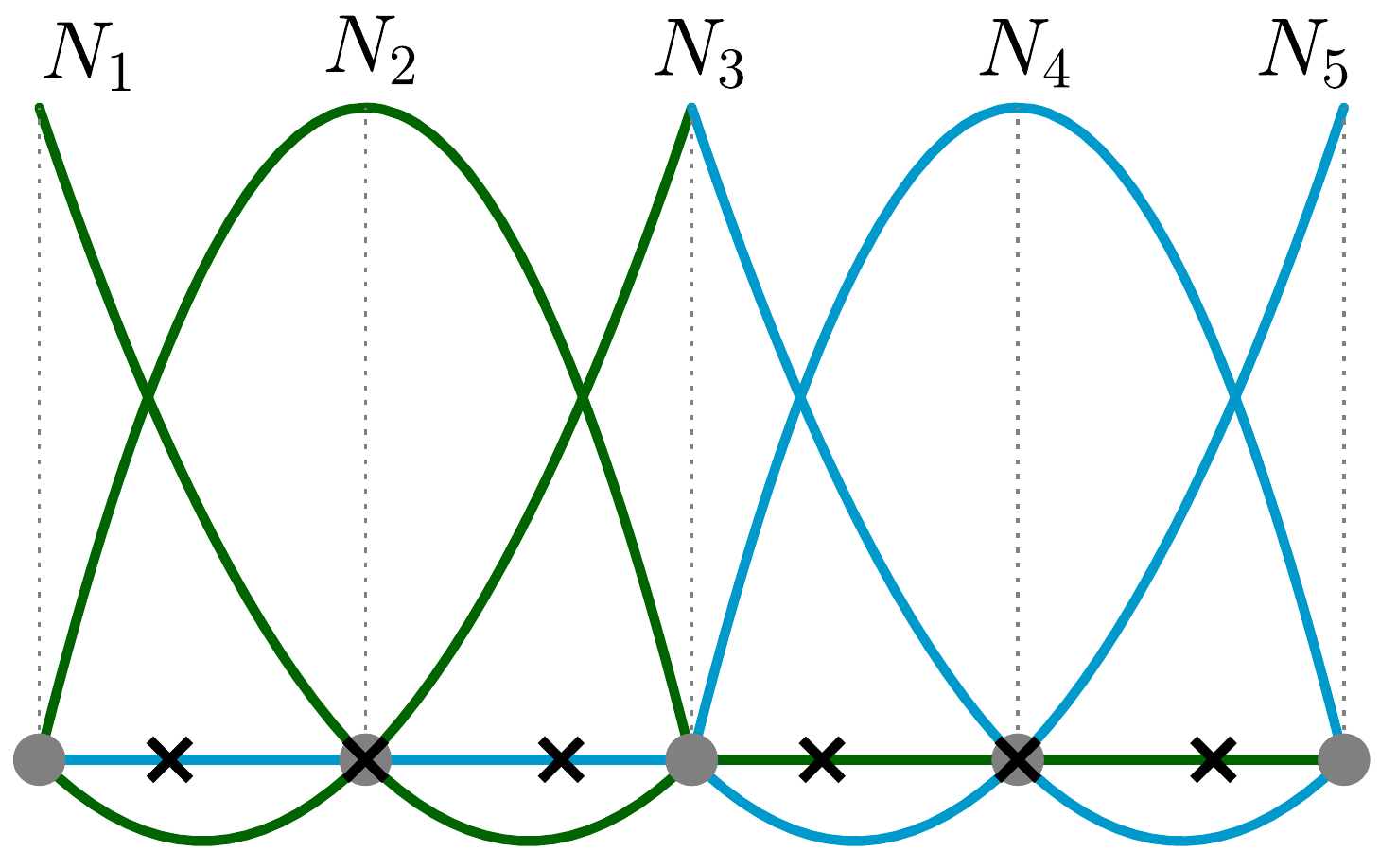}
	\caption{Refined reference element in 1D, for degree $p=2$ and refinement factor $m=2$. Nodes are represented by gray dots and integration points, by black crosses. The refined element has $5$ basis functions and 6 integration points.}
	\label{fig:refinedShapeFunctions}       
\end{figure}

On other hand, in $\Otips$, a uniformly $h$-refined discretization is considered to capture the sharp variation of the phase-field solution. These elements are mapped into a uniformly refined reference element. The resulting approximation space is equivalent to a standard finite element approximation on a finer mesh, non-conformal on the interface $\Gamma$.
The chosen implementation, based on a refined reference element, is simpler since no mesh generation is needed, the assembly and the integration can be done as usual, and the definition of the refined reference element is done as a simple preprocess. This is a viable option in this application since all refined elements have the same refinement factor, which is known from the beginning. The refinement factor is computed from the phase-field length, $l$, the characteristic element size of the background mesh, $h$, and the degree of approximation $p$.  It can be taken as
\begin{equation*}
	m = \left \lceil{\dfrac{ah}{lp}}\right \rceil.
\end{equation*}
In our experience, a reasonable value for the factor $a$ is between $3$ and $5$.
The approximation space in $\Otips$ is then defined as
\begin{equation*}
\mathcal{V}^h_\textnormal{tips}=\{ v\in H^1(\Otips) \;: \; v|_{K_i} \in \mathcal{P}^p_\textnormal{ref}(K_i) \textnormal{ for } K_i \subseteq \Otips \},
\end{equation*}
where $\mathcal{P}^p_\textnormal{ref}(K_i)$ is the refined space in each element. The element is split into $m^{n_{sd}}$ uniform subelements, and the refined approximation space in the element is
\begin{equation*}
\mathcal{P}^p_{\textnormal{ref}}(K_i) = \{ v \in \mathcal{H}^1(K_i) : v|_{K_{ij}} \in \mathcal{P}^p(K_{ij}),\  j = 1...m^{n_{sd}} \},
\end{equation*}
with $\{K_{ij}\}_{j=1}^{m^{n_{sd}}}$ denoting the subelements in the element $K_i$. To illustrate the idea, a 1D refined element is shown in Figure \ref{fig:refinedShapeFunctions}.

Finally, the approximation space to discretize the Nitsche's weak form \eqref{full_weak_form_eq} is the combined space
\begin{equation*}
	\mathcal{V}^h(\Omega) = \{ v\in H^1(\Omega) : v|_{\Oxfem} \in \mathcal{V}^h_{\text{xfem}} \text{, } v|_{\Otips} \in \mathcal{V}^h_{\text{tips}} \}.
\end{equation*}

\subsection{Damage equation}

The damage equation is solved only in $\Otips$, this is, in the small subdomains containing crack tips and tips of notches, where damage evolution is expected. 
As commented in Section \ref{sec:theConcept}, and illustrated in Figure \ref{fig:gluing-u-d} (right), Dirichlet boundary conditions are imposed on $\Gamma_D^d$, defined as a portion of the boundary around the intersection with cracks. 
Homogeneous Neumann boundary conditions are imposed on the rest of the boundary.

The weak form is then: find $d\in \mathcal{H}^1(\Otips)$ such that $d = d_D$ on $\Gamma_D^d$ and
\begin{equation}\label{eq:weakFormDamage}
\begin{aligned}
&\int_{{\Otips}} \left(  \frac{G_C}{l} + 2\mathcal{H}^+ \right) vd \dV + \int_{{\Otips}} G_C l \grad v \cdot \grad d \dV 
 & = \int_{{\Otips}} v2\mathcal{H}^+ \dV,
\end{aligned}
\end{equation}
for all $v\in\mathcal{H}^1(\Otips)$ such that $v = 0$ on $\Gamma_D^d$.

The value set on the Dirichlet boundary, $d_D$, is the value saved from the damage solution in previous iterations, assuming that the damage is frozen on the wake of the crack.

The weak form is discretized with the refined approximation space $\mathcal{V}^h_\textnormal{tips}$.

\subsection{Staggered scheme}\label{sec:staggered}

Numerical simulations take an incremental process in load steps. 
At every load step, the system is solved with a staggered scheme, for which 
the equations are solved alternately until convergence. 
For load step $n$, 
we iterate over
\begin{itemize}
\item[\textit{i.}] \emph{Computing the displacement field} $\bs{u}$ in $\Omega$ by solving the weak form \eqref{full_weak_form_eq}, with the current damage field $d$ in $\Otips$,
and with boundary data $\bs{u}_D=\bs{u}_D^{n}$ on  $\Gamma_D$ and $\bs{t}_N=\bs{t}_N^{n}$ on  $\Gamma_N$. Recall that traction-free conditions on the crack are imposed by means of the XFEM enrichment in $\Oxfem$, and a refined discretization is considered in $\Otips$. 
\item[\textit{ii.}] Updating the history field $\mathcal{H}^+$ in $\Otips$.

\item[\textit{iii.}] \emph{Computing the damage field} $d$ in $\Otips$ by solving the weak form \eqref{eq:weakFormDamage}.

\item[\textit{iv.}] \emph{Updating the partition}, $\Otips$ and $\Oxfem$, and the geometrical description of sharp cracks, as detailed in Sections \ref{sec:partitionUpdate}, \ref{sec:criteria} and \ref{sec:sharpCrack}.
\end{itemize}

As a convergence criterion we check if the error of the damage field in the Euclidean norm is lower than a certain tolerance.

Note that crack-tip subdomains are updated at every staggered iteration. Since we are modeling brittle fracture, cracks can significantly grow at a single load step and the phase-field subdomain has to be accordingly modified to allow the propagation.

\subsection{Geometrical information} \label{sec:partitionUpdate}

The geometry of the domain is defined by the background mesh $(X,T)$, with $X$ the nodal coordinates and $T$ the connectivity matrix. This mesh is kept fixed during all the computation. 

In the preprocess, information of the faces is computed from the connectivity matrix. 
Faces are numbered and, for each one of them, we store the number of the elements sharing the face and the local number of the face in each element. In 3D, the rotation is also stored, that is, which node in the face from the second element corresponds to the first node of the face from the first element. This information on faces is used to compute the integrals over $\hat{\Gamma}$ in the equilibrium weak form \eqref{full_weak_form_eq}.

Information of the partition $\{\Otips,\Oxfem\}$ is given by three vectors storing, respectively, the numbers of the elements in each subdomain and the numbers of the faces composing the interface $\Gamma$. 
Also, we need to save the list of nodes on $\Gamma_D^d$ and the value of the damage field on these nodes from the previous iteration, to be able to impose the boundary conditions in \eqref{eq:weakFormDamage}. 
During the computation, information on the partition is  updated at every staggered iteration according to the propagation of cracks by applying the criteria in Section \ref{sec:criteria}. 

A geometrical description of sharp cracks in $\Oxfem$ is needed to introduce the discontinuities into the XFEM discretization. Here, we store the crack path together with a nodal vector indicating whether each enriched node is in the left or right side of the crack. Some details on the construction of the sharp crack can be found in Section \ref{sec:sharpCrack}.

In addition, to ensure continuity of the approximation in the refined region $\Otips$, we define an auxiliary mesh $(X_\textnormal{ref},T_\textnormal{ref})$. 
This new mesh is constructed by mappings of the refined reference element into the elements in $\Otips$, and it is modified as the list of elements in $\Otips$ is updated. When an element is added to $\Otips$, the nodes in the refined element are added to $X_\textnormal{ref}$, avoiding repetitions, and the corresponding row is added to $T_\textnormal{ref}$. When an element is supressed, nodes not belonging to other refined elements are removed from $X_\textnormal{ref}$. $T_\textnormal{ref}$ is modified according to the new nodal numbering and the row corresponding to the element is removed.
The use of a refined mesh is optional, since the computations are directly done with the refined reference element, but it clearly facilitates the assembly for the elements in $\Otips$.

\section{Update of the partition} \label{sec:update}

The criteria to decide if an element should be in $\Otips$ or in $\Oxfem$
are based on \textit{i)} the maximum value of the damage reached in the element, and \textit{ii)} the distance to elements containing crack tips (or tips of notches).
These criteria are applied at every iteration of the staggered algorithm.

In this section, we describe the criteria and also the algorithm to identify the elements that contain crack tips in 2D. 
The conversions from a phase-field to a sharp representation of cracks and vice versa are discussed in Sections \ref{sec:sharpCrack} and \ref{sec:refinement}, respectively.

\subsection{Refining and switching criteria} \label{sec:criteria}

The refining criterion is applied to elements in $\Oxfem$ which are adjacent to the interface $\Gamma$, in order to decide if they should be transferred to $\Otips$.
Following the idea in  \cite{MuixiNitsche}, an element is refined if the damage field reaches a threshold value $d^*$ at one of its nodes. 
Here, since the damage is computed only in $\Otips$, an element in $\Oxfem$ is refined 
if any of its nodes on the interface reaches the threshold value $d^*$ and it is close to some element containing a crack tip, with a threshold distance $\delta^*$.
This last consideration ensures that elements in the wake of cracks stay in $\Oxfem$.

Then, to decide if some of the elements in $\Otips$ have to switch to the discontinuous approximation, the distance criterion with threshold value $\delta^*$ is also applied.
More precisely, if the minimum distance of an element to all the elements containing a crack tip (or notch tip) is larger than $\delta^*$, the element is transferred to $\Oxfem$.

Distances between elements are computed center-to-center.
In practice, values of $d^*$ between $0.1$ and $0.2$ give accurate phase-field approximations with narrow bands of refinement along cracks. A reasonable value for the distance threshold $\delta^*$ is around $10l$.

\subsection{Identification of elements containing crack tips}

Here, we focus on quadrilateral elements in 2D. 
The logic of the algorithm is easily extendable to triangular elements. As a simplification, we assume a smeared crack can only intersect the same element edge once. 

Elements containing crack tips are identified using the number of sides which are intersected by the smeared band and the area of the band inside the element. We assume a side is intersected if it reaches $d >0.95$ at some of its integration points. The area of the band is approximated as the sum of the weights of integration points in the element with $d > 0.95$.

\begin{figure}[]
	\begin{subfigure}{0.5\textwidth}
		\includegraphics[width=0.835\textwidth]{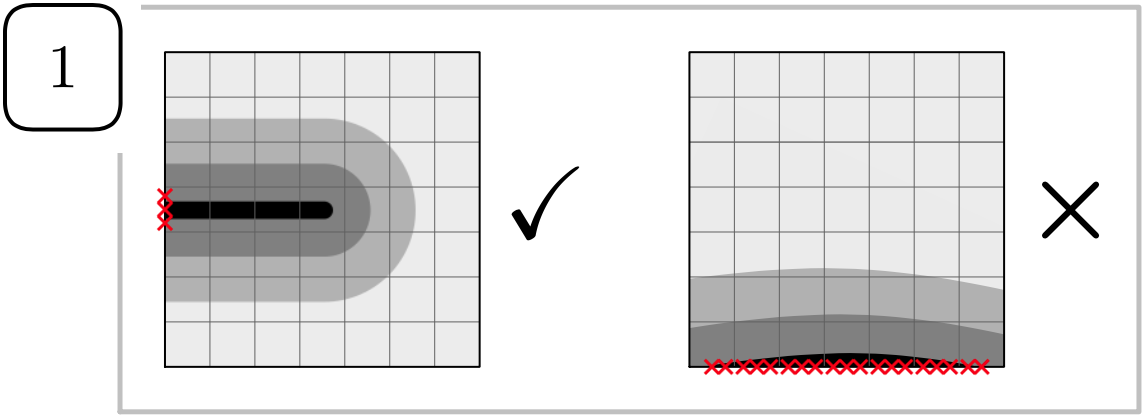}
		
		\vspace{1mm}
		
		\includegraphics[width=\textwidth]{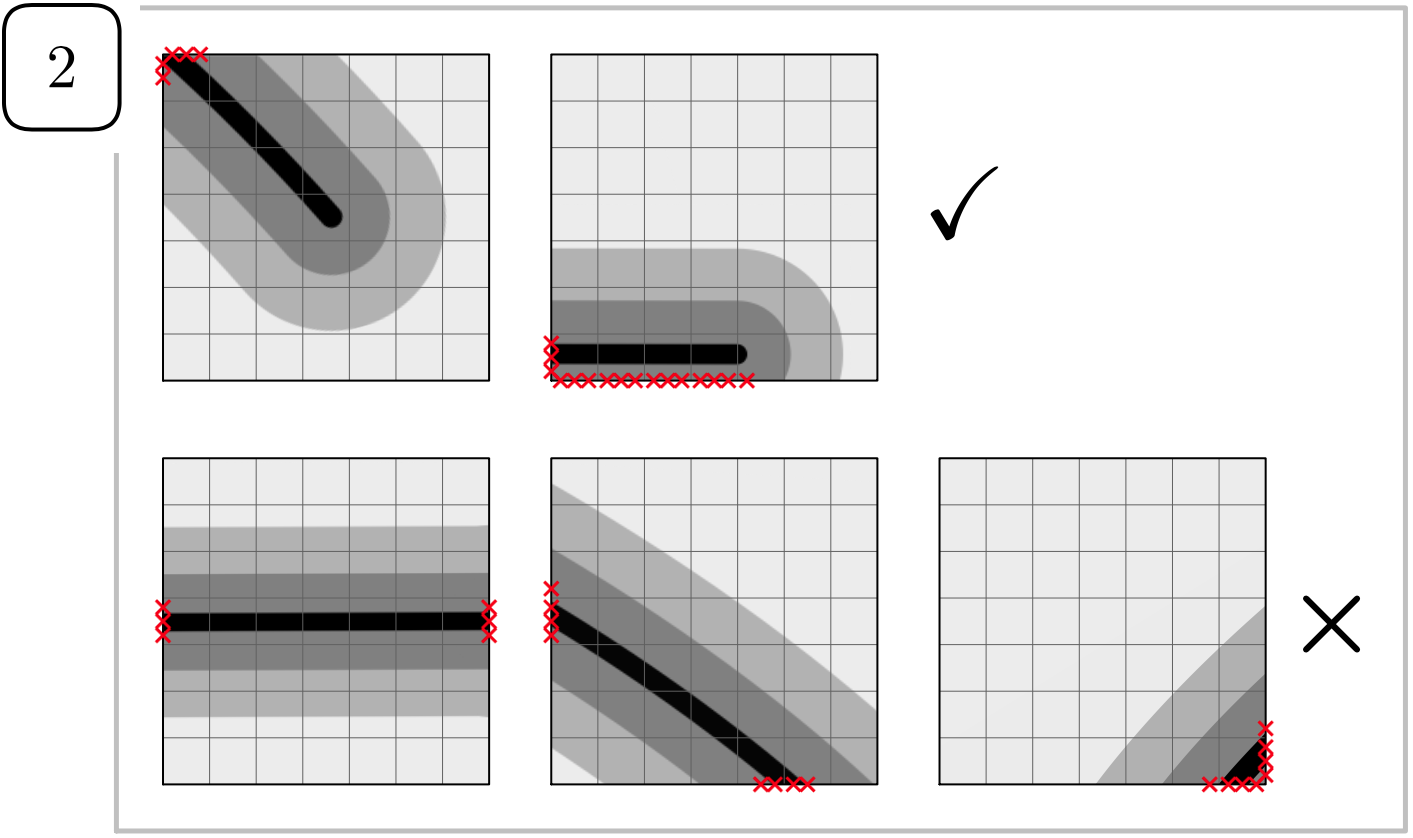}
	\end{subfigure}
	\begin{subfigure}{0.5\textwidth}
		\hspace{7mm}
		\includegraphics[width=0.73\textwidth]{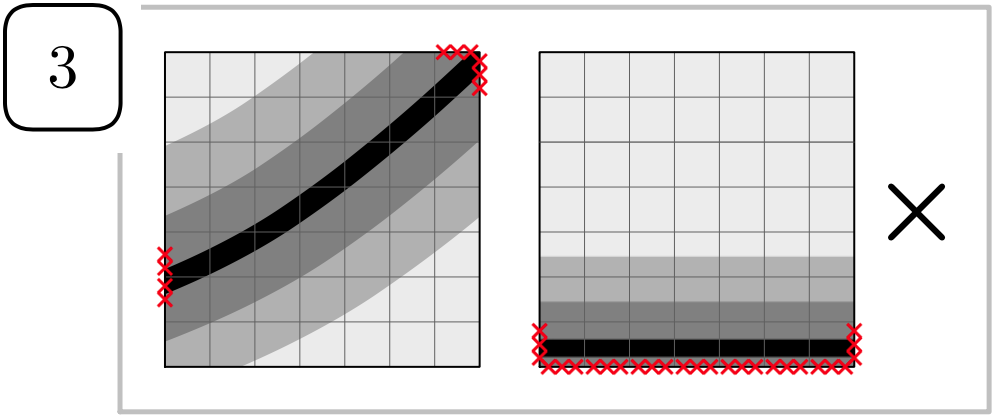}
		
		\vspace{1mm}
		
		\hspace{7mm}
		\includegraphics[width=0.46\textwidth]{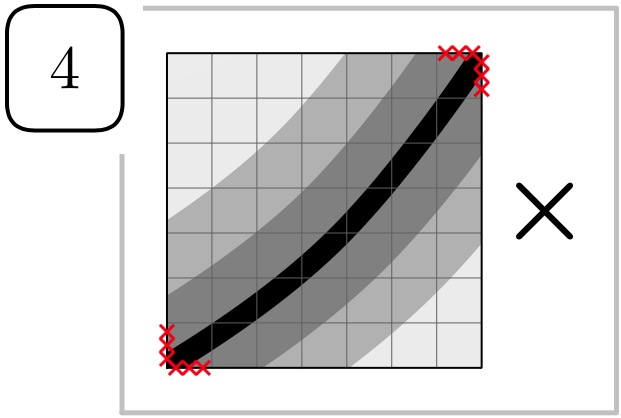}
	\end{subfigure}

	\caption{Detail of a refined element with possible crack paths. Cases are classified depending on the number of intersected sides. Red crosses indicate integration points on the sides with $d > 0.95$.}
	\label{fig:crack-tips}
\end{figure}

First, for all elements in $\Otips$, we count how many sides are intersected by the crack. The considered configurations are summarized in Figure \ref{fig:crack-tips}.

In elements with only 1 intersected side, 
we mark the element as containing a crack tip if the area of the band is larger than a threshold $A^*$; on the other case, the crack is assumed to be tangent to the side.

For 2 intersected sides, the element contains a crack tip if the two intersected sides are adjacent, the shared node satisfies $d > 0.95$ and the area of the band is larger than $A^*$. All other cases are discarded. 

Elements with 3 o 4 intersected sides do not contain a crack tip.

The algorithm does not cover all possibilities.
For instance, crack tips which are almost tangent to an element side may not be correctly identified. 
Robustness is ensured by saving the elements containing crack tips from the previous iteration. If a crack tip from the previous iteration disappears, i.e. it is not found in the same element and has not moved to a neighboring element, we assume the crack tip has advanced to an ambiguous position and consider its previous localization. This is not a problem since the threshold distance in the switching criterion is always taken coarse enough, and propagating cracks leave the ambiguous positions as they evolve in the following iterations. 

This simple method is enough for all the 2D examples presented in this paper, taking $A^* = hl/5$. 
The robust extension of the algorithm to 3D is nontrivial due to the numerous possible cases. Elements containing crack fronts are easily identified for the particular example in Section \ref{sec:twisting} since the direction of crack propagation is known. 

Other strategies may be explored, for instance, applying the medial-axis algorithm in \cite{TamayoMasRodriguezFerran2015}.

\section{Transition to discontinuous fracture and definition of the sharp crack geometry} \label{sec:sharpCrack}

In elements in the wake of crack tips, no variation of crack paths is expected and the phase-field damage band is replaced by a sharp representation.
This implies a significant reduction in the computational cost of simulations.
The transition is done for fully degraded material, so no additional energetic considerations are needed.

Sharp cracks are defined by the union of elemental contributions. Once a cut element transitions from $\Otips$ to $\Oxfem$, the crack path is identified within the phase-field diffuse band and is then added to the existing sharp crack. The process is illustrated in Figure \ref{fig:SharpCrackDefinition}.

\begin{figure}[b]
	\vspace{3mm}
	\centering 
	\includegraphics[width=0.29\textwidth]{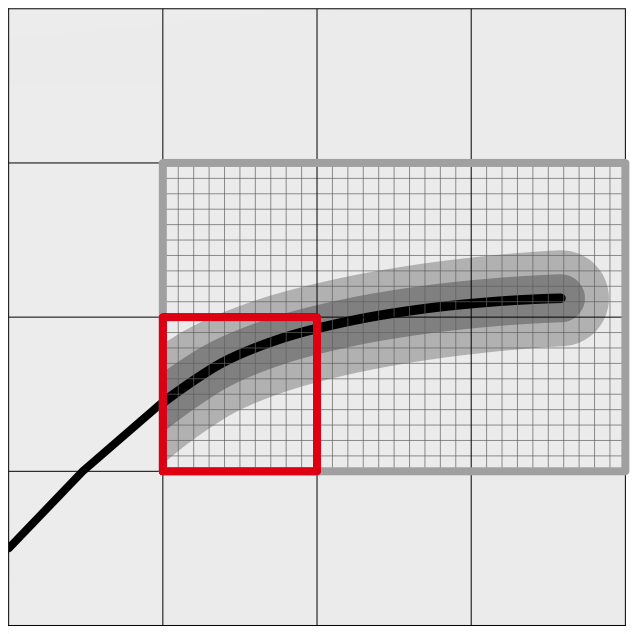}
	\hspace{5mm}
	\includegraphics[width=0.29\textwidth]{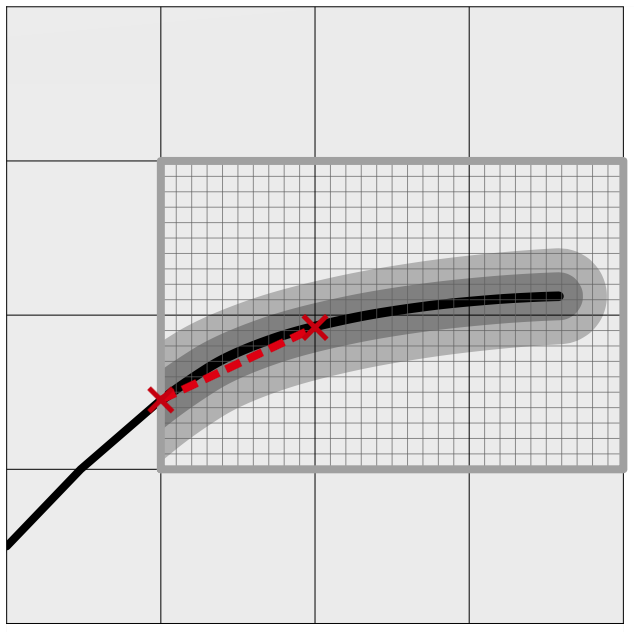}
	\hspace{5mm}
	\includegraphics[width=0.29\textwidth]{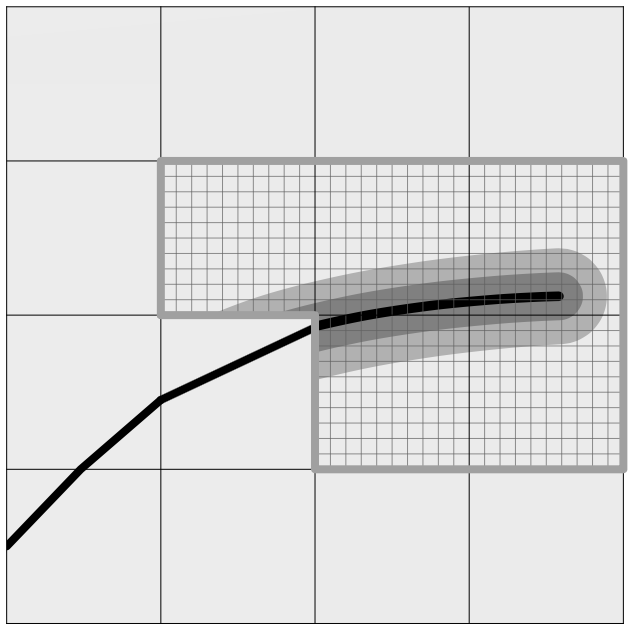}
	\caption{Sketch of transition to sharp crack. The diffuse crack is replaced by a sharp crack in the element selected to be derefined, marked with red square. In 2D the crack is composed by linear segments, defined by the intersection of the diffuse crack with the elements sides.}
\end{figure} \label{fig:SharpCrackDefinition}

For both 2D and 3D problems, we search for intersections of the diffuse cracks along the edges of the mesh of transitioning elements. 
In the $1$-dimensional searchs, intersections are taken as the middle points of the nodes (in the refined discretization) with $d > 0.98$.
In 2D, this leads to piecewise linear cracks, with a segment in each cut element. In the event of crack branching, a proper piecewise representation is considered in the element containing the branching point, as shown in Section \ref{sec:example-branching}.
In 3D, crack surfaces are constructed as the tringular facets defined by the intersection points on edges. Note that these facets can have very different sizes. This is not a problem since the surface is used only to define the integration subdomains in the element.

This algorithm is enough to show the capabilities of the strategy. In our numerical examples, it leads to similar results to the ones obtained by a plain phase-field model. 
More sophisticated techniques can be introduced at this step if more accuracy is needed or crack paths are more  challenging, such as a higher-degree curved representation of the crack path in each cut element, given by more than 2 points in 2D \cite{GurkanSalaKronbichlerFernandez2017}, the medial-axis algorithm  \cite{TamayoMasRodriguezFerran2015} or 
the optimization-based approach in  \cite{GeelenLiuDolbowRodriguezFerran2018}.

It is important for the different representations along a crack (sharp and diffuse) to match on the interface $\Gamma$ with enough accuracy, to avoid the creation of unphysical corners. 
In our experience, level-set representations of the crack in the coarse background mesh are not accurate enough to fulfill this requirement.

Regarding the displacement field, since it is computed in the staggered scheme in Section \ref{sec:staggered}, there is no need to compute its projection into the new approximation space.

\section{Refinement of elements and cracks coalescence}\label{sec:refinement}

When an element in $\Oxfem$ is selected to be refined, nodal values for the damage have to be defined.
If the element is refined for the first time, nodal values of the damage can be set to zero except for the nodes on the interface, where the damage is known. 
If the element is intersected by a sharp crack, the damage field in the element has to approximate the crack. This is the case in cracks coalescence.

In experiments with merging of cracks, we may have elements cut by a sharp crack which are approached by a phase-field crack tip. In this case, the criteria in Section \ref{sec:criteria} lead to refining again the elements, transitioning back to the continuous representation. The coalescence is then handled by the phase-field model. Once the cracks have merged, the elements transition to the discontinuous again with the corresponding update in the involved sharp cracks. 

In the current implementation we store the damage field for elements that transition to discontinuous and recover its value in case these elements need to be refined again.
Another option would be to construct the damage band from scratch using the sharp representation, for instance, defining the corresponding history field variable, $\mathcal{H}^+$, following Borden et al \cite{Bordenetal2012}.

\section{Numerical experiments}\label{sec:numerical-experiments}

In this section, we aim to demonstrate the robustness of the proposed approach to efficiently simulate fracture processes. 
During the computations the background mesh is kept fixed and the discretization is automatically modified to account for the different representations of cracks.

First, a simple test is presented to numerically validate the definition of $\hat{\Gamma}$ to impose continuity of displacements between the subdomains $\Otips$ and $\Oxfem$.
Then, several benchmark tests are visited, covering a wide range of cases in fracture simulations. Examples in 2D prove the applicability of the strategy to branching and coalescence of cracks. The approach is also applied to a fully 3D setting.

Plane strain conditions are assumed in 2D. 
In all examples, we solve for degree of approximation $p = 1$. 
The tolerance for convergence in the staggered scheme is fixed to $10^{-2}$.
The Nitsche's parameter for the equilibrium equation is $\alpha_E = 100$ and 
the refinement of elements is triggered by the threshold value $d^* = 0.2$. A numerical study on the influence of $\alpha_E$ and $d^*$ in adaptive phase-field simulations can be found in \cite{MuixiNitsche}.

Preexisting cracks in the domain are first introduced as diffuse phase-field bands by solving the damage equation with an initial history field, $\mathcal{H}_0^+$, as described in Borden et al \cite{Bordenetal2012}. Then, before initializing the simulation, the procedure of transition to discontinuous is applied to replace the diffuse phase-field band by a sharp crack where needed, according to the respective switching criterion. 

\subsection{Test on the continuity of displacements}
\label{test-continuity}

In this test, we analyse the effect of imposing continuity of displacements
on the whole interface $\Gamma$ and on the cropped interface $\hat{\Gamma}$ (defined in \eqref{eq:GammaCropped} as the part of $\Gamma$ where the material is not significantly degraded) on a simple tension test with a fixed crack.

Consider a square plate in $[-0.4,0.4]\times[-0.5,0.5] \textnormal{ mm}^2$  with a horizontal crack at midheight crossing the whole piece. 
The piece is clamped on its bottom face and has imposed displacements $(0,u_D)$ on its top face, with $u_D = 10^{-4}$ mm. 
The material parameters are $E = 20$ GPa, $\nu = 0.3$ and $G_C = 10^{-4}$ kN/mm.

\begin{figure}[b]
	\centering
	\includegraphics[height=5.5cm]{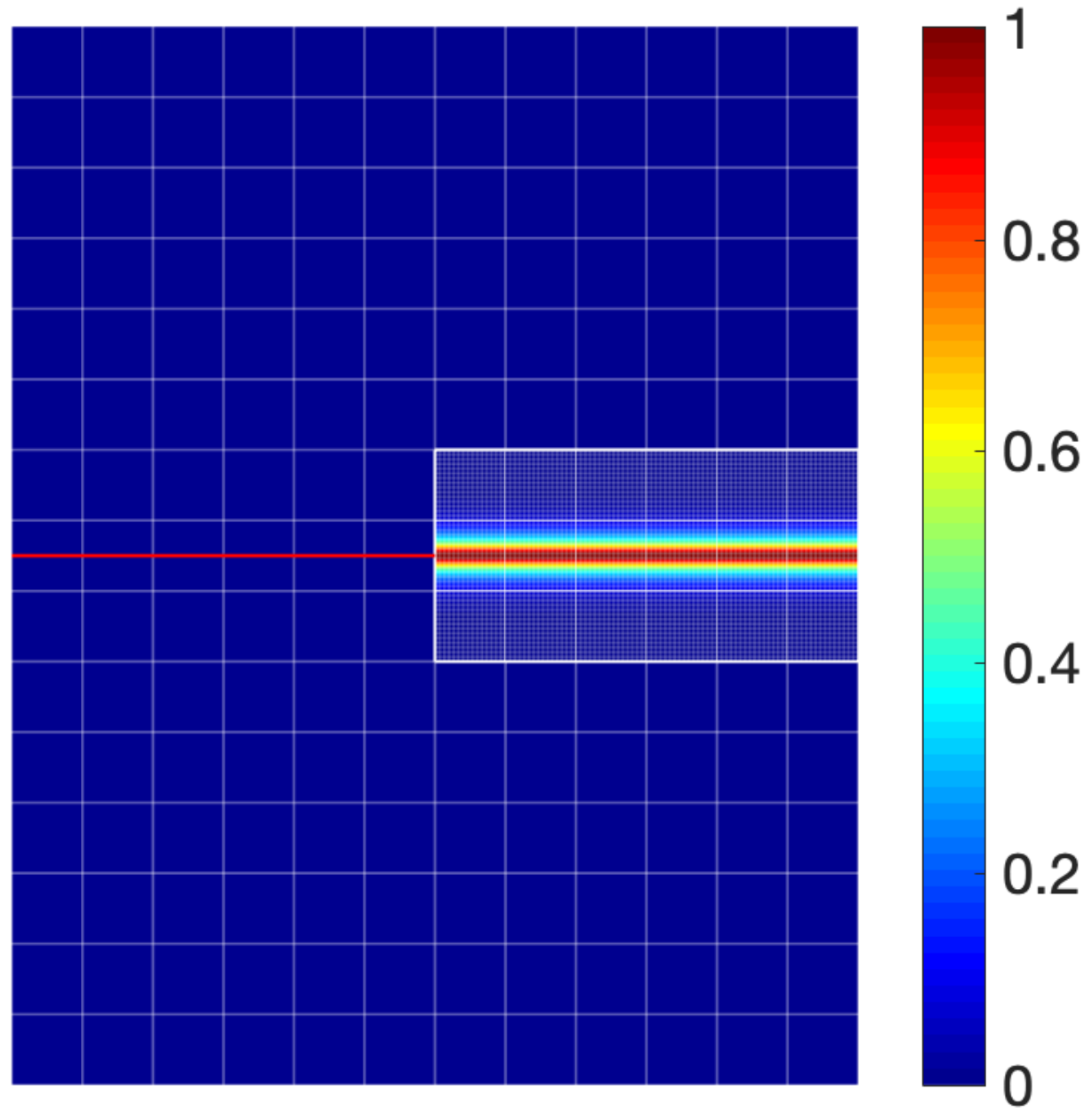}
	\caption{\textit{Test on continuity.} Initial discretization and approximation of the crack. }
	\label{fig:test-continuity-setting}       
\end{figure}

The crack is represented by a sharp crack for $x < 0$ and by a smeared damage band for $x>0$, with a length-scale parameter $l = 0.012$ mm. 
The domain is covered by a uniform quadrilateral mesh with $12\times15$ elements and $\Otips$ is taken as $\{x\ge 0, |y|\le 0.075 \textnormal{ mm} \}$. The rest of the domain is $\Oxfem$. Elements are refined with a uniform submesh of refinement factor $m = 15$.  
The discretization is shown in Figure \ref{fig:test-continuity-setting}.

\begin{figure}[]
	\centering
	\includegraphics[height=6cm]{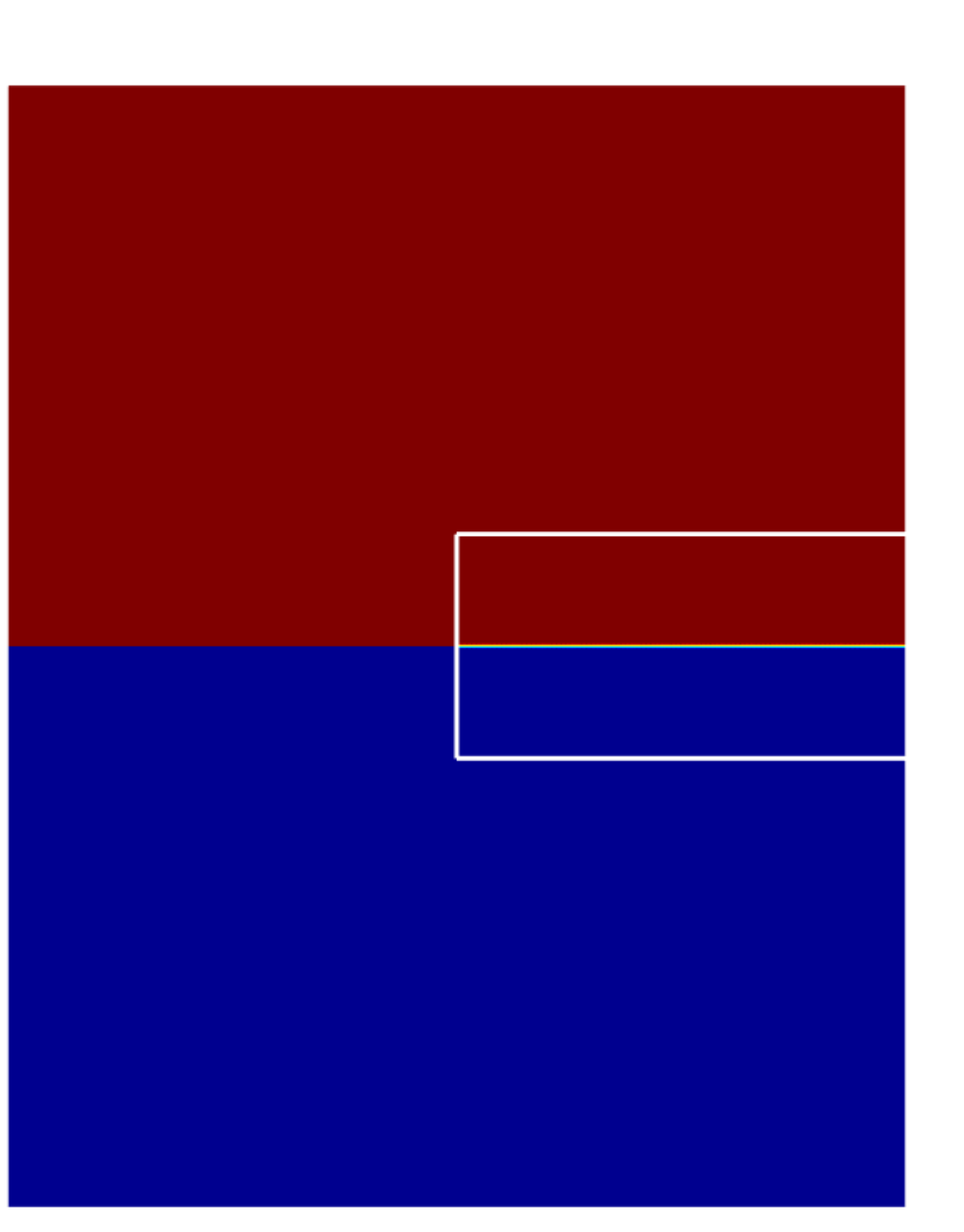}
	\hspace{5mm}
	\includegraphics[height=6cm]{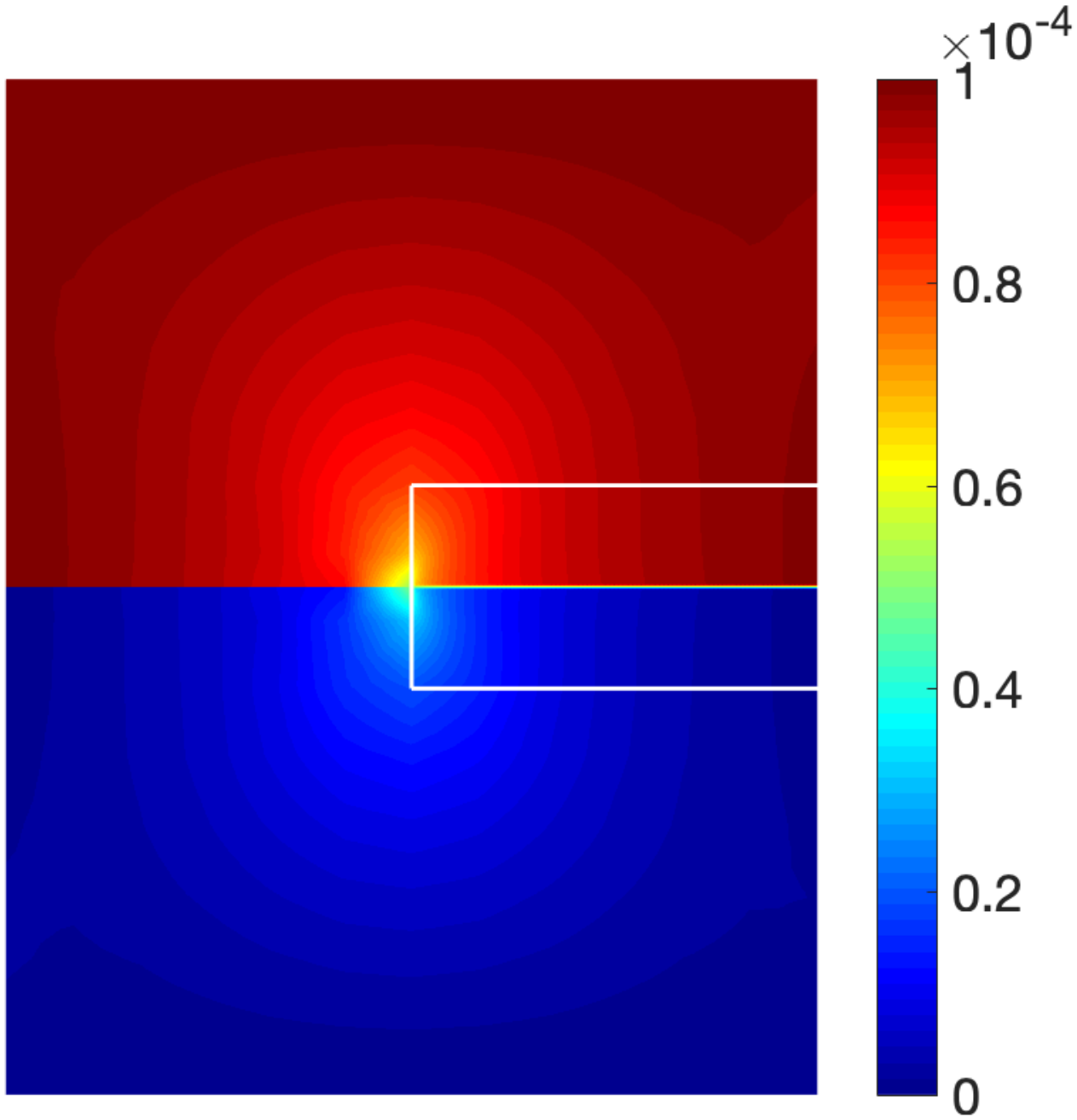}
	
	\raisebox{0cm}{\rotatebox[origin=t]{0}{Continuity on $\hat{\Gamma}$}}
	\hspace{27mm}
	\raisebox{0cm}{\rotatebox[origin=t]{0}{Continuity on $\Gamma$}}
	\hspace{10mm}
	
	\caption{\textit{Test on continuity.} Vertical displacement field $u_y$ imposing continuity on $\hat\Gamma$ and on $\Gamma$.}
	\label{fig:test-continuity-uy}       
\end{figure}

Figure \ref{fig:test-continuity-uy} shows the vertical displacement field  $u_y$ when the equilibrium equation is solved imposing continuity of displacements on $\hat\Gamma$. 
As expected, the upper half moves rigidly, with a \textit{constant} vertical displacement equal to $u_D$, whereas the lower half does not move and has zero vertical displacement. In $\Oxfem$ we obtain the expected discontinuity and in $\Otips$ the displacement is continuous and abruptly varies between these two values.

We repeat the experiment imposing continuity on the whole interface $\Gamma$. The vertical displacements exhibit an unrealistic pattern near the gluing of the two representations of the crack, due to the different material stifnesses, see Figure \ref{fig:test-continuity-uy}. 
Consequently, continuity will be imposed on $\hat\Gamma$ in all examples.

\subsection{L-shaped panel test}\label{sec:exampleL}

\begin{figure}[]
	\centering
	\includegraphics[width=0.38\columnwidth]{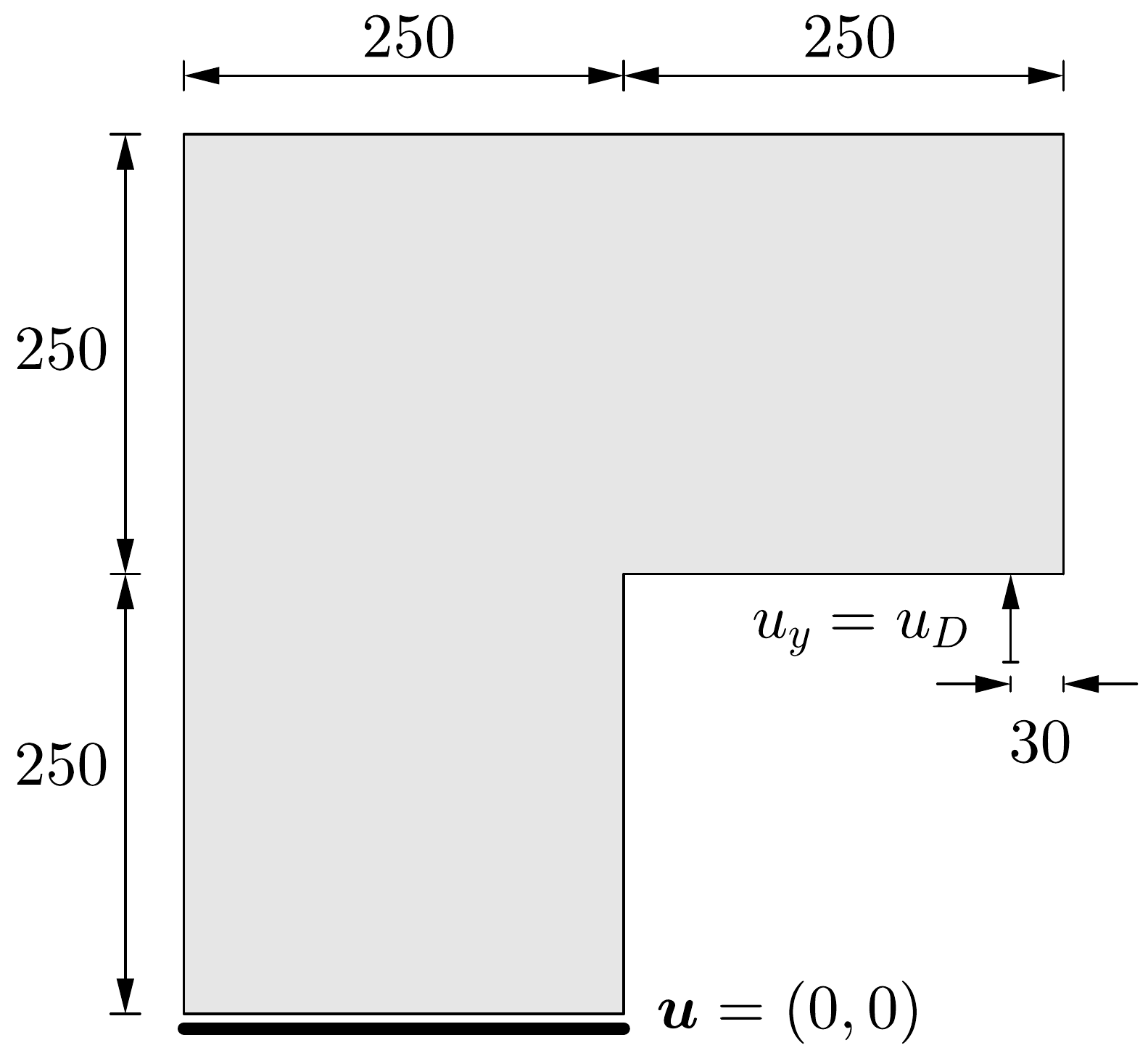}
	\caption{\textit{L-shaped test.} Geometry and boundary conditions. Dimensions in mm. }
	\label{fig:Lshaped-setting}       
\end{figure}

Consider an L-shaped panel with boundary conditions as illustrated in Figure \ref{fig:Lshaped-setting}. Following Ambati et al \cite{AmbatiGerasimovDeLorenzis2015},
the material parameters are
$E = 25.84$ GPa, $\nu = 0.18$ and $G_C = 8.9 \cdot 10^{-5}$ kN/mm. The phase-field length-scale parameter is $l = 2.5$ mm. The load process takes displacement increments of $\Delta u_D = 10^{-3}$ mm. 

The background mesh is a uniform quadrilateral mesh with element size $h = 10$ mm, and elements in $\Otips$ are refined with refinement factor $m = 20$. The subdomain $\Otips$ initially consists of the 3 elements in the corner of the piece, where crack inception is expected. 
The distance of derefinement in the switching criterion is taken as $\delta^* = 2h$.
Recall that a correct initial definition of $\Otips$, including crack tips and notches of the domain, is essential since the damage field is  computed only in this part of the domain.

\begin{figure}
	\centering
	\hspace{3.3cm}
	\raisebox{0cm}{\rotatebox[origin=t]{0}{PF}}
	\hfill
	\raisebox{0cm}{\rotatebox[origin=t]{0}{PF - XFEM}}
	\hspace{5.2cm}
	\vspace{2mm}
	
	\raisebox{1.8cm}{	
		\begin{subfigure}[b]{0.14\textwidth}
			\centering
			\includegraphics[width=\textwidth]{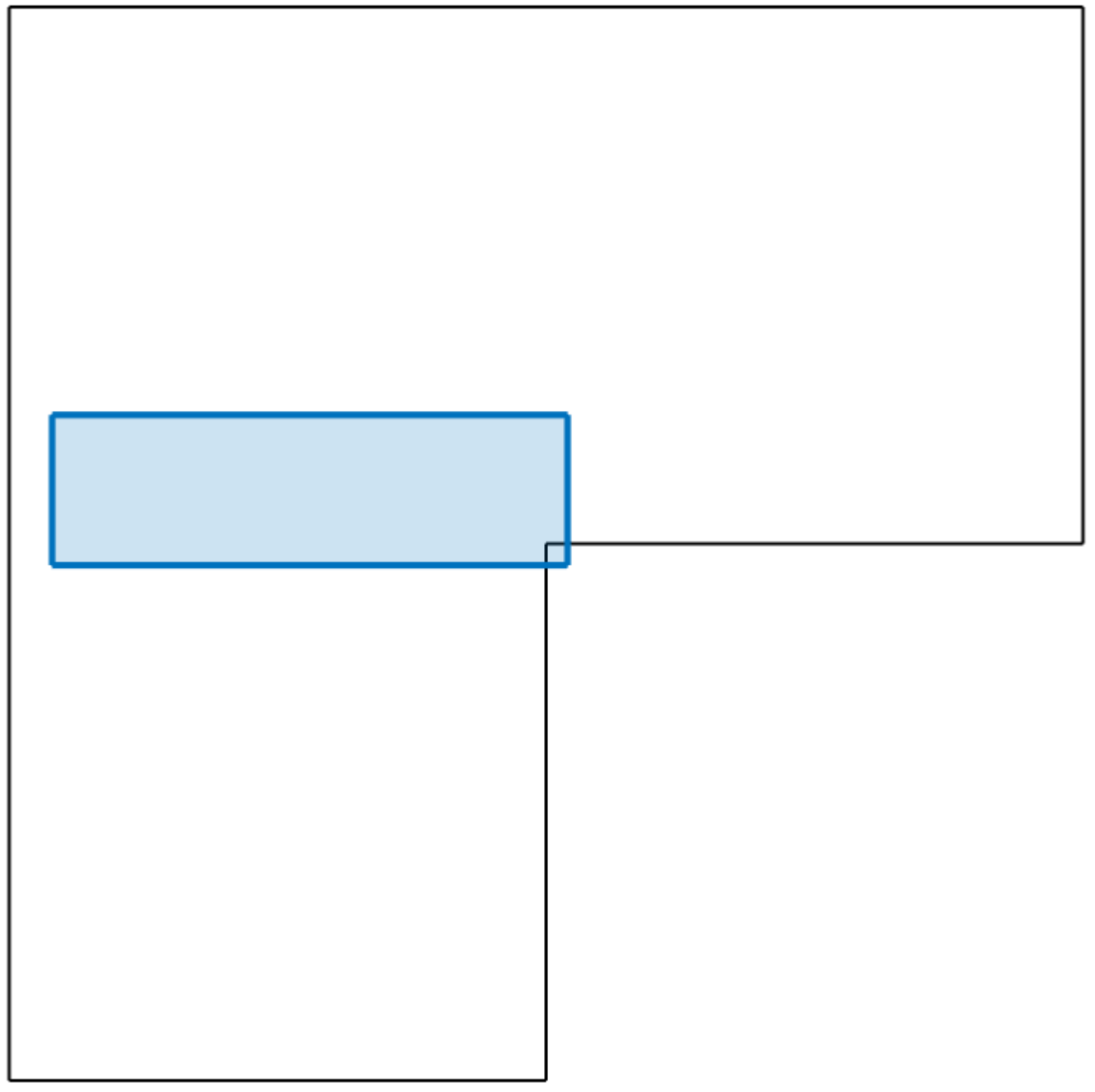}
	\end{subfigure}}
	\hfill
	\begin{subfigure}[b]{0.78\textwidth}
		\centering
		\raisebox{8mm}{\rotatebox[origin=c]{90}{$0.26$ mm}}
		\includegraphics[width=0.48\textwidth]{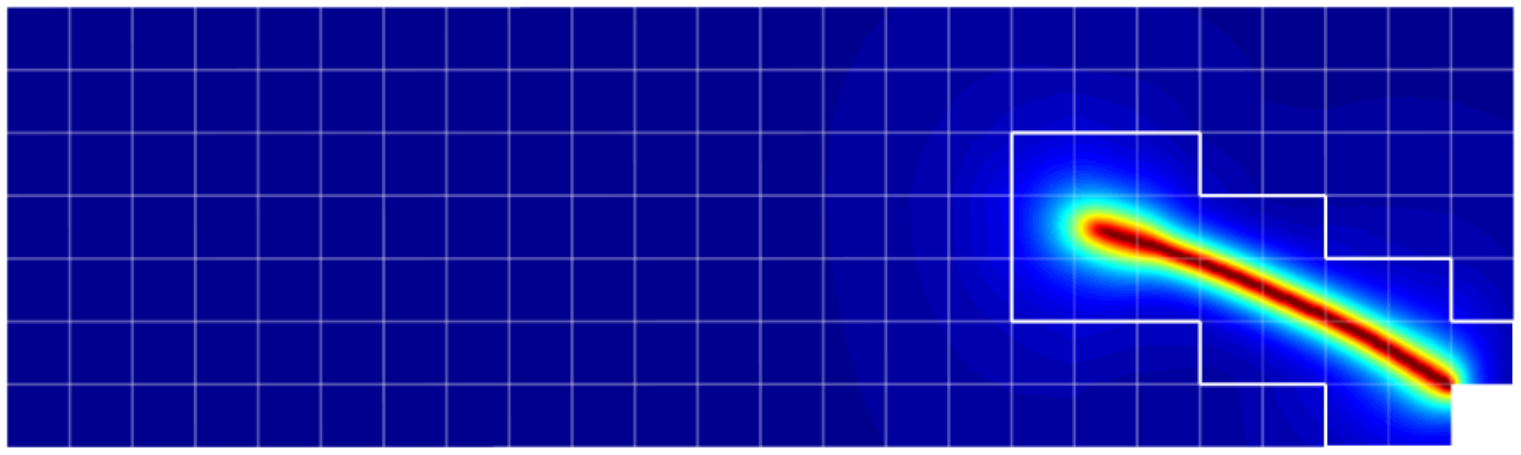}
		\includegraphics[width=0.48\textwidth]{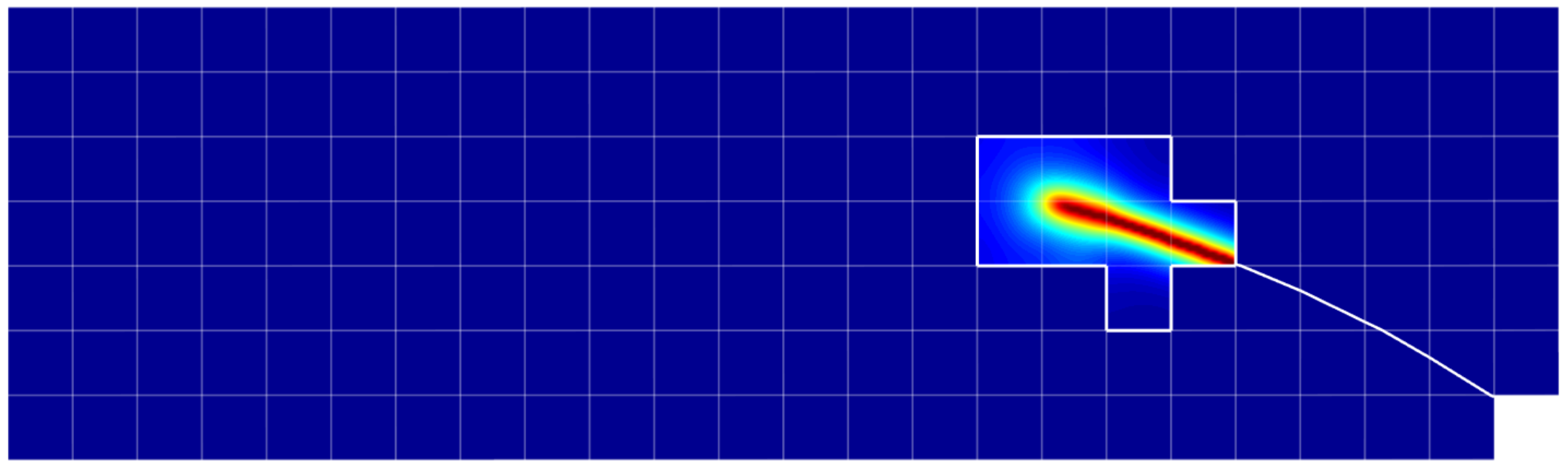}
		
		\raisebox{8mm}{\rotatebox[origin=c]{90}{$0.40$ mm}}
		\includegraphics[width=0.48\textwidth]{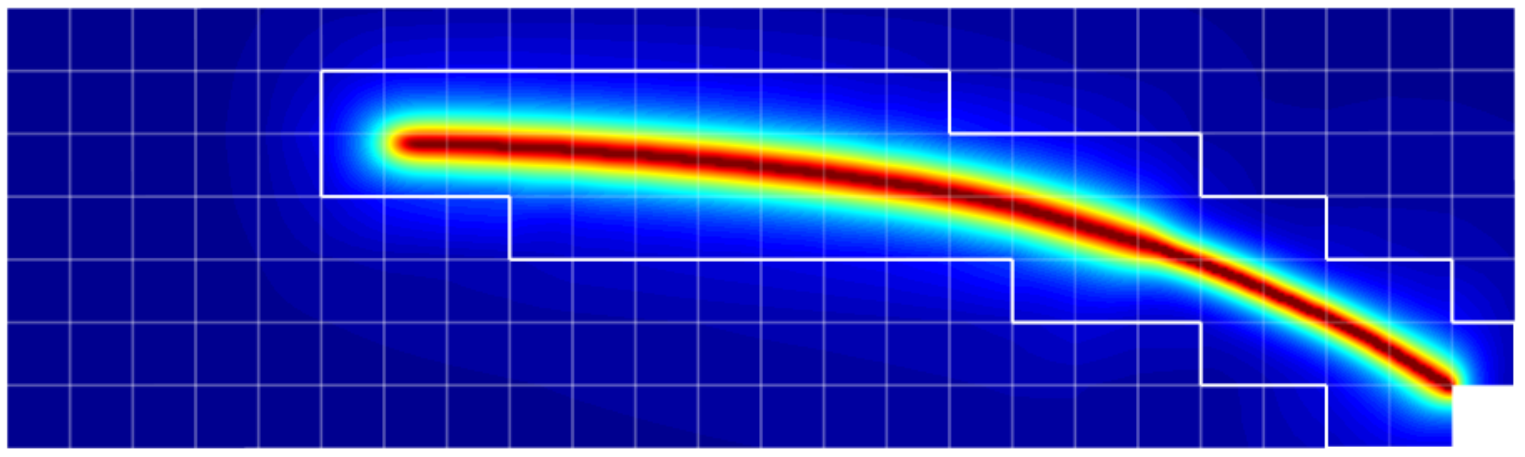}
		\includegraphics[width=0.48\textwidth]{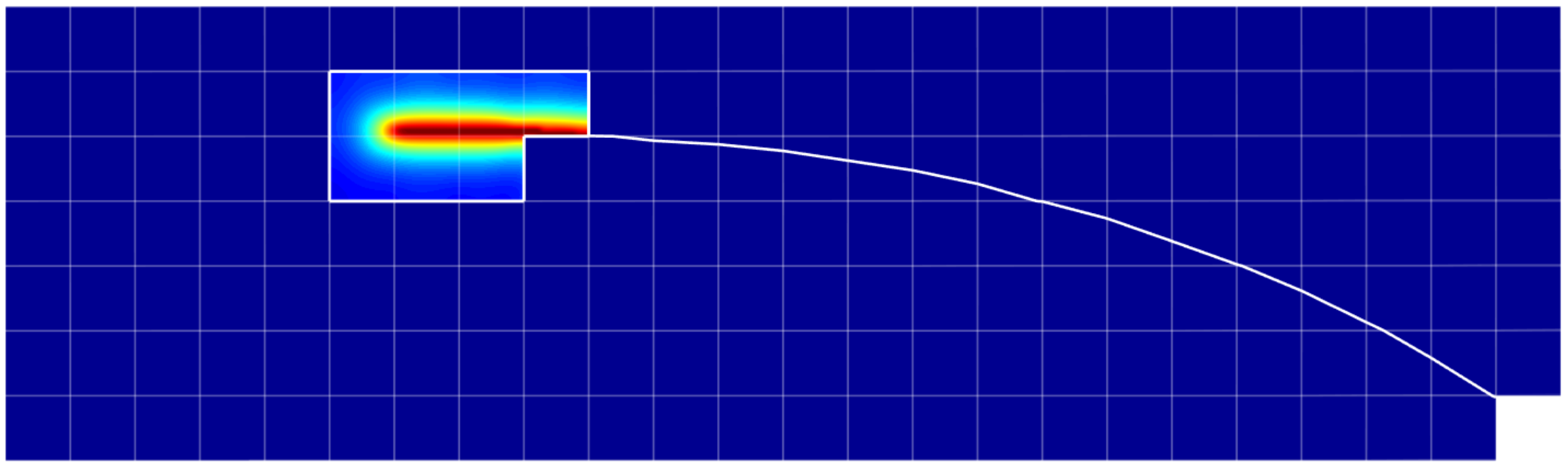}
		
		\raisebox{8mm}{\rotatebox[origin=c]{90}{$0.60$ mm}}
		\includegraphics[width=0.48\textwidth]{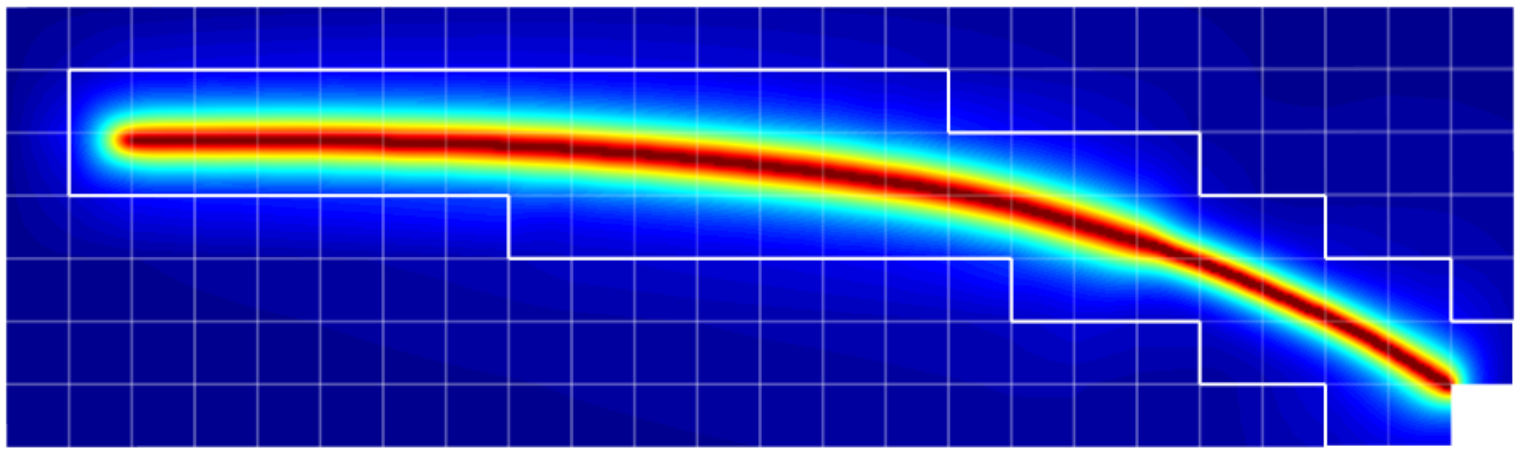}
		\includegraphics[width=0.48\textwidth]{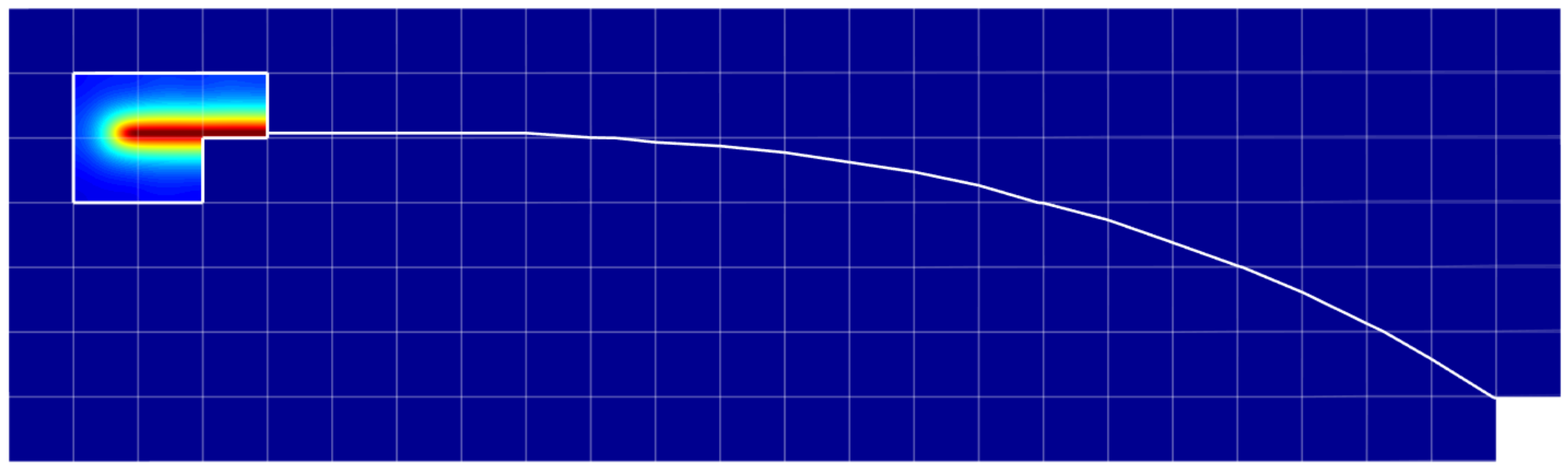}
	\end{subfigure}
	\begin{subfigure}[b]{0.05\textwidth}
		\raisebox{7mm}{
			\includegraphics[width=\textwidth]{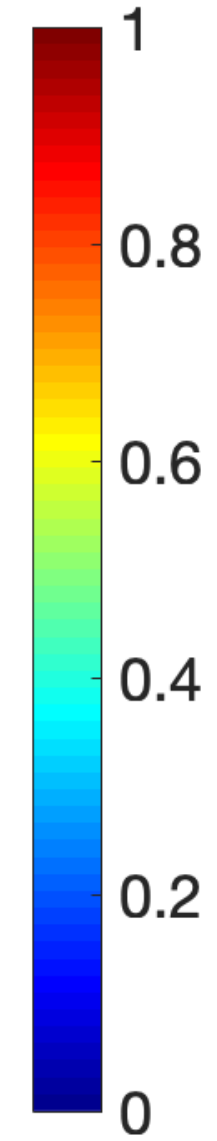}}
	\end{subfigure}
	\caption{\textit{L-shaped test.} Damage field at different load steps for PF and PF-XFEM. The sharp crack in PF-XFEM is plotted in white. Zoom into $[-230,10]\times[-10,60]$ $\textnormal{mm}^2$. }
	\label{fig:Lshaped-damage-zoom}  
\end{figure}

\begin{figure}[]
	\centering
	\includegraphics[width=0.5\columnwidth]{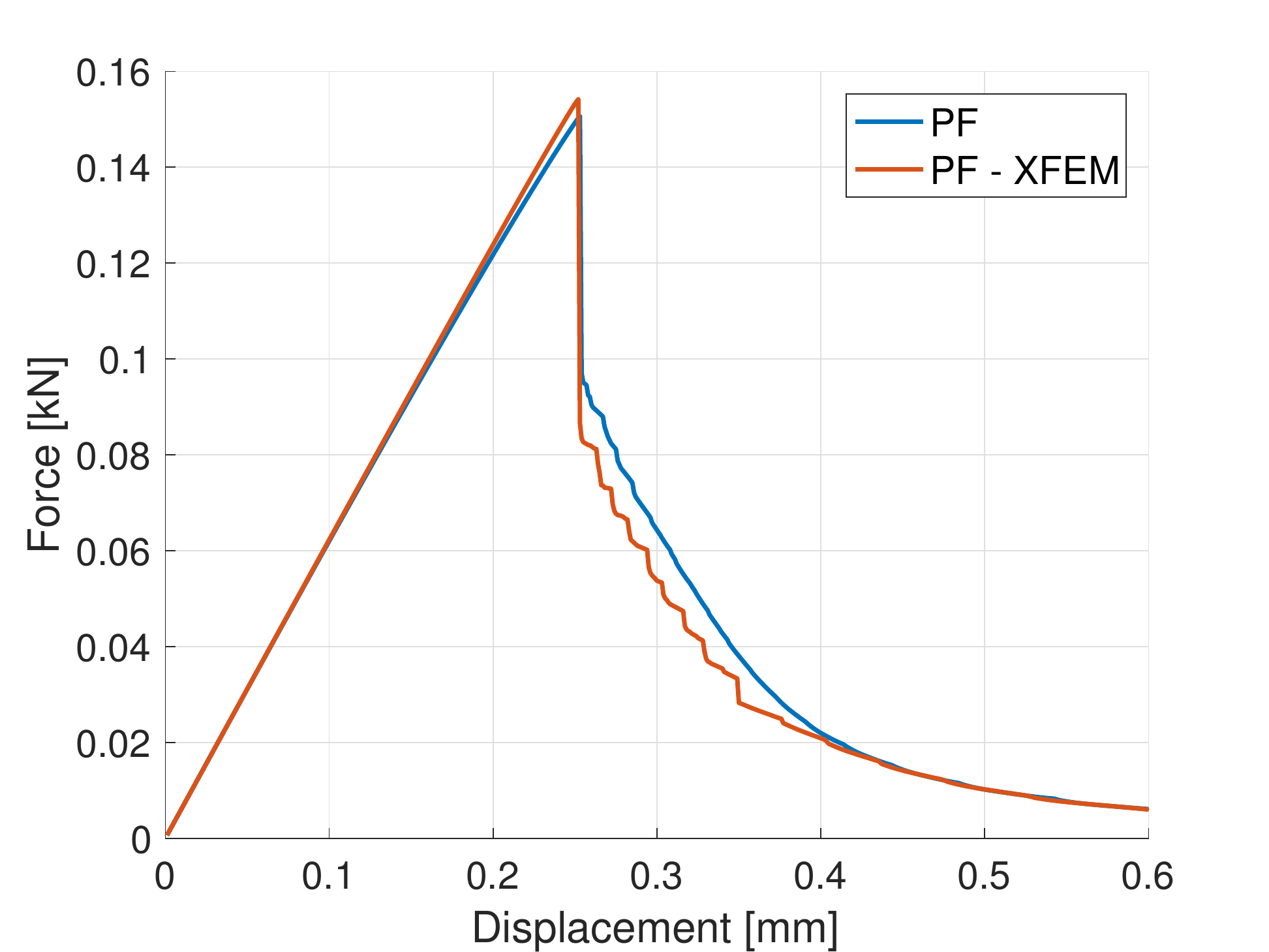}
	\caption{\textit{L-shaped test.} Load-displacement curve for PF and PF-XFEM.}
	\label{fig:Lshaped-loaddisp}       
	
	\centering
	\includegraphics[width=0.5\columnwidth]{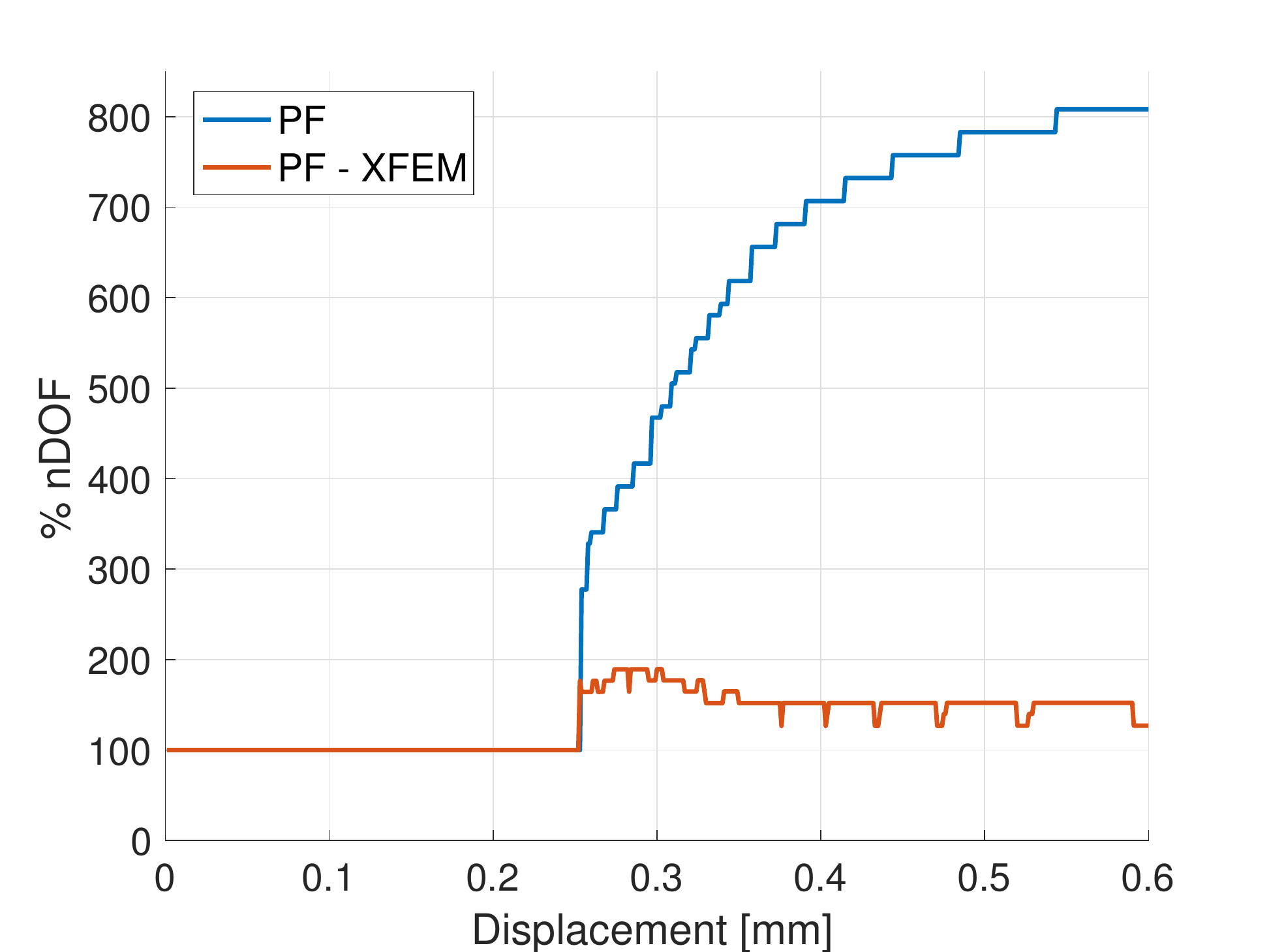}
	\caption{\textit{L-shaped test.} Evolution of nDOF in the equilibrium equation for PF and PF-XFEM. The percentage is computed with respect to the nDOF of the initial discretization.}
	\label{fig:Lshaped-ndofs}       
\end{figure}

As a reference solution, we run the simulation with a plain phase-field model with adaptivity, following the refinement strategy proposed in \cite{MuixiNitsche}. 
The equations of the hybrid phase-field model are solved in the whole domain. Elements along cracks are dynamically refined, but no transition to a sharp representation is considered. 
Thus, more elements become refined as the crack propagates. 
On the contrary, in the newly proposed approach, only elements near the crack tip are refined. Thus, the size of $\Otips$ does not increase constantly. Elements which are far enough from the crack tip transition to an XFEM coarser representation, and the damage band is replaced by a sharp crack. 
Remeshing is avoided in both cases. 
In what follows, we abbreviate these approaches by PF and PF-XFEM, respectively. 

The crack path obtained with the two strategies at different load steps is depicted in Figure \ref{fig:Lshaped-damage-zoom}. 
We obtain similar results, with a slightly faster propagation for PF-XFEM at load step $u_D = 0.26$ mm. This can be explained by the spurious transmission of forces across phase-field cracks, while sharp cracks are completely traction-free.
In both cases, the discretization is updated according to crack growth.

The slightly faster propagation when introducing sharp cracks is also observed in the load-displacement curves in Figure \ref{fig:Lshaped-loaddisp}. After the peak, when the crack starts propagating, we observe a steeper descent of the curve, meaning the crack is slightly longer. 
As the simulations evolve, the difference between approaches diminishes. 
In the PF curve, we observe the characteristic loss of stiffness prior to the peak of these models; solving the damage equation in the whole domain  with a quadratic degradation function $g(d)$ propagates the damage, softening the piece in the simulation. 
In PF-XFEM the behavior of the material before cracking is closer to linear elasticity because the damage is computed only in small subdomains. 

For PF-XFEM, $\Otips$ takes a very small part of the domain.
This leads to a reduction of the number of degrees of freedom (nDOF) for the equilibrium equation. 
The reduction is substantial even though in PF the mesh is  refined only in a narrow band containing the crack. 
Figure \ref{fig:Lshaped-ndofs} shows the evolution of nDOF with respect to its initial value. For PF, the number increases up to $800\%$, while for PF-XFEM nDOF the value increases only up to $150\%$.

\subsection{Branching test}\label{sec:example-branching}

This example was first proposed in Muix\'i et al \cite{MuixiHDG} and is revisited here to test the capability of the strategy to reproduce bifurcations, as well as to handle multiple disconnected subdomains in $\Otips$. 

\begin{figure}[]
	\centering
	\includegraphics[width=0.4\columnwidth]{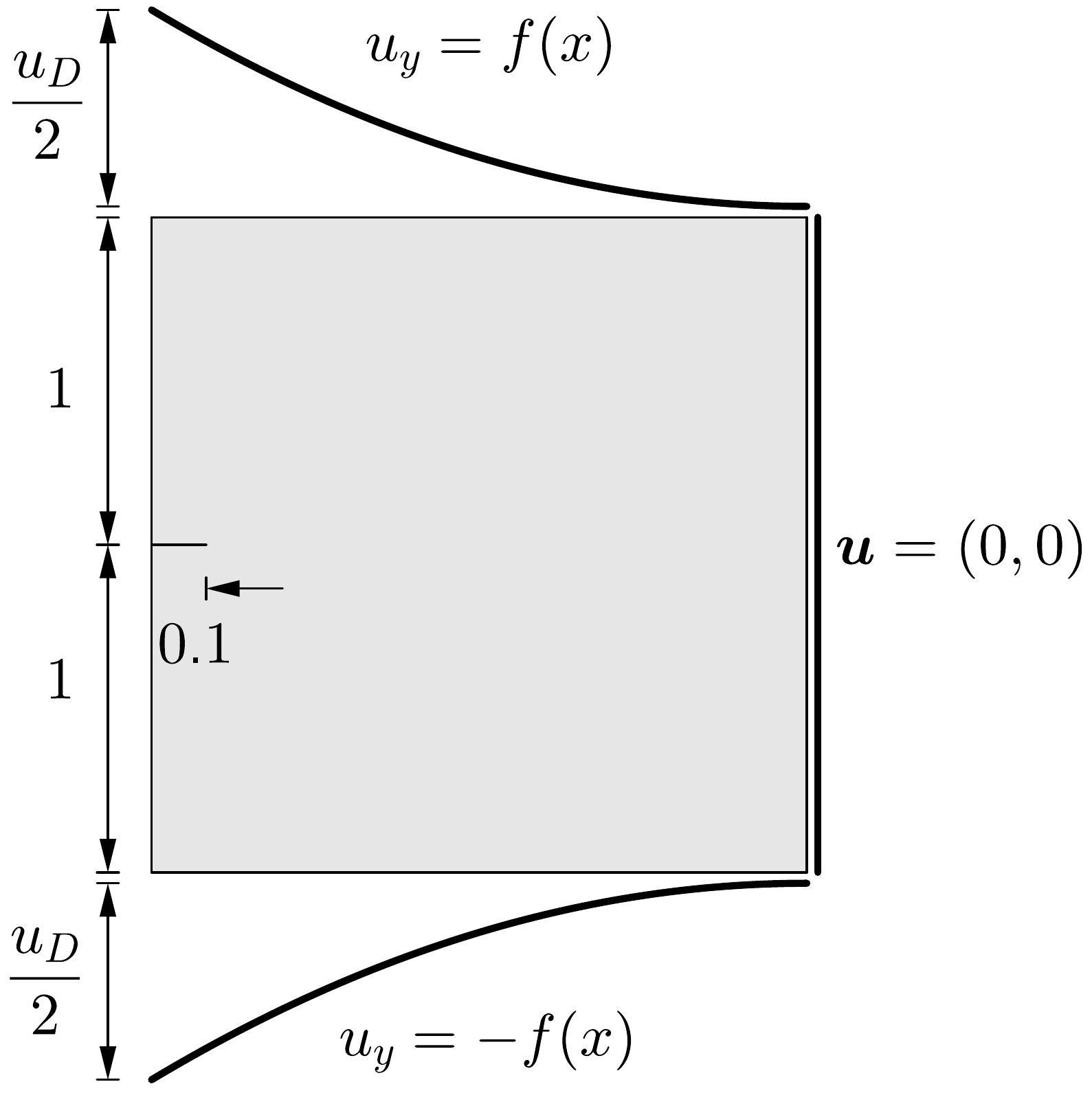}
	\caption{\textit{Branching test.} Geometry and boundary conditions. Dimensions in mm.}
	\label{fig:branching-setting}       
\end{figure}

We consider a square plate in $[-1,1]^2$ $\textnormal{mm}^2$, precracked at midheight and with boundary conditions as illustrated in Figure \ref{fig:branching-setting}. The piece is fixed on its right edge and has imposed vertical displacements on its top and bottom edges, with
$f(x) = u_D (x-1)^2 / 8$.
The material parameters are $E = 20$ GPa, $\nu = 0.3$ and $G_C = 8.9 \cdot 10^{-5}$ kN/mm. We take $l = 0.01$ mm and load increments of $\Delta u_D = 5 \cdot 10^{-5}$ mm.
The domain is discretized into a uniform quadrilateral mesh of $45\times45$ elements. Elements are refined with refinement factor $m = 15$. The threshold distance in the switching criterion is taken as $\delta^* = 3h$. The initial $\Otips$ consists of the elements containing the preexisting crack and the rest of the domain is part of $\Oxfem$. 

\begin{figure}[]
	\centering
	\begin{subfigure}[b]{0.3\textwidth}
		\centering
		\includegraphics[width=\textwidth]{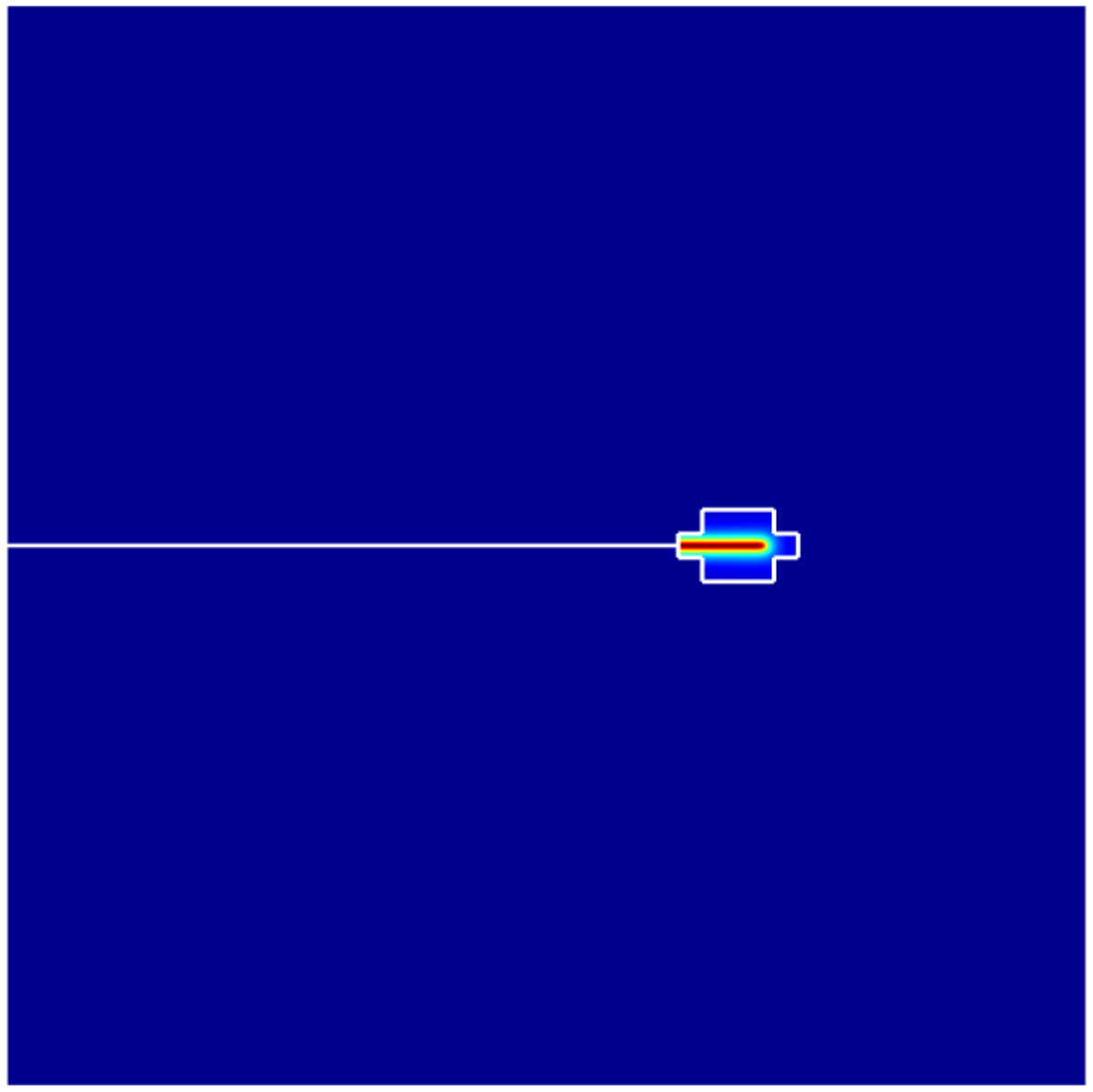}
	\end{subfigure}
	\hspace{0.1mm}
	\begin{subfigure}[b]{0.3\textwidth}
		\centering
		\includegraphics[width=\textwidth]{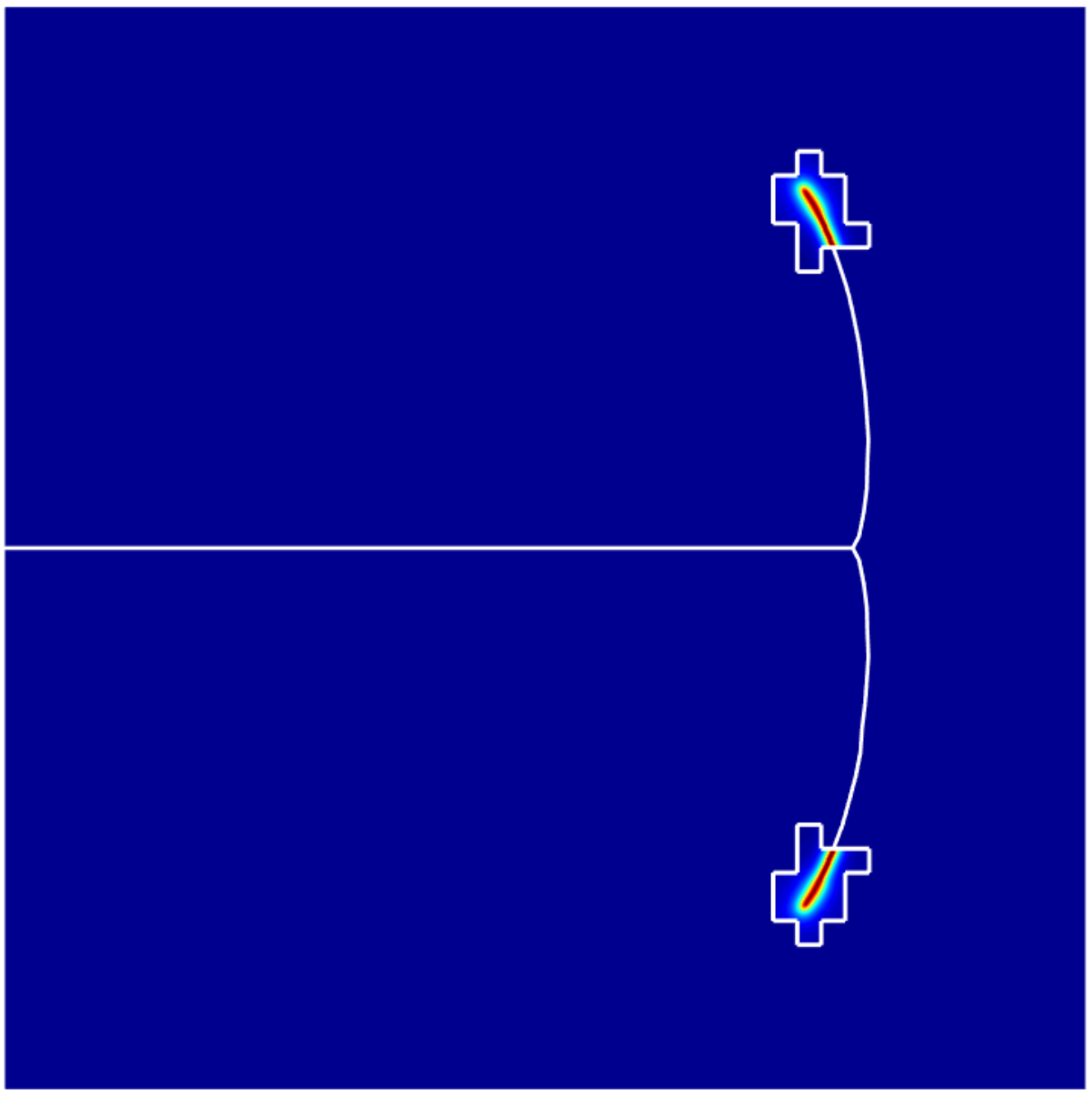}
	\end{subfigure}
	\hspace{0.1mm}
	\begin{subfigure}[b]{0.3\textwidth}
		\centering
		\includegraphics[width=\textwidth]{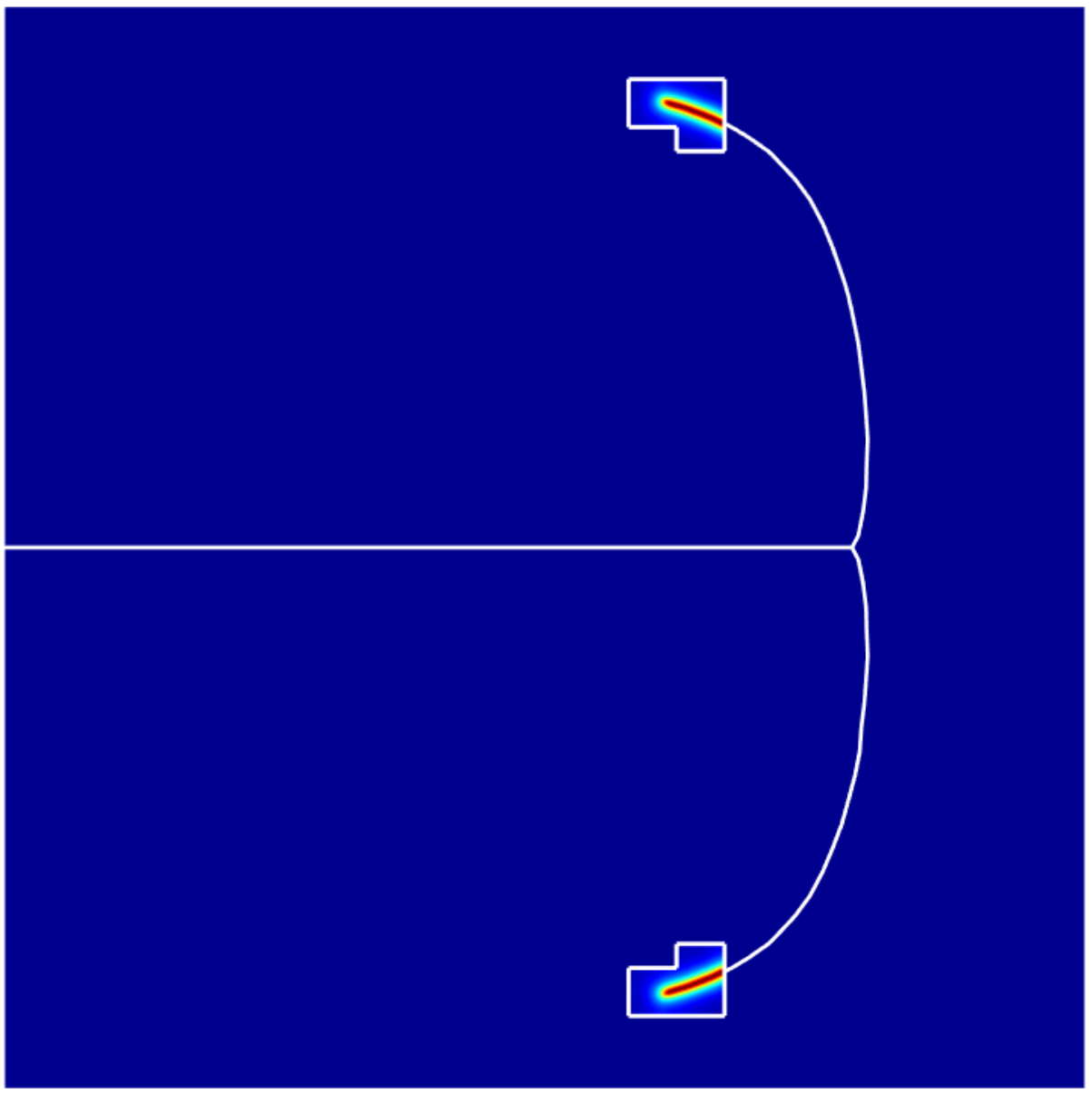}
	\end{subfigure}
	\hspace{0.1mm}
	\begin{subfigure}[b]{0.057\textwidth}
		\centering
		\includegraphics[width=\textwidth]{figures/colorbar.pdf}
	\end{subfigure}

	\hspace{1.4cm}
	\raisebox{0cm}{$u_D = 0.030$ mm}
	\hfill
	\raisebox{0cm}{$u_D = 0.055$ mm}
	\hfill
	\raisebox{0cm}{$u_D = 0.080$ mm}
	\hspace{2.5cm}
	\vspace{1mm}
	
	\caption{\textit{Branching test.} Evolution of the crack at different load steps. The sharp crack is plotted in white.}
	\label{fig:branching-damage}       
\end{figure}

The obtained crack at different stages is shown in Figure \ref{fig:branching-damage}. 
The crack propagates horizontally and then branches abruptly, maintaining the symmetry of the solution in the whole simulation. 
When the branching occurs, we start having two crack tips and $\Otips$ contains the elements near both of them. In $\Oxfem$ we have two separate sharp cracks, each one of them contributing to the XFEM discretization with an independent Heaviside function.

\begin{figure}[]
	\centering
	\begin{subfigure}[b]{0.3\textwidth}
		\centering
		\includegraphics[width=\textwidth]{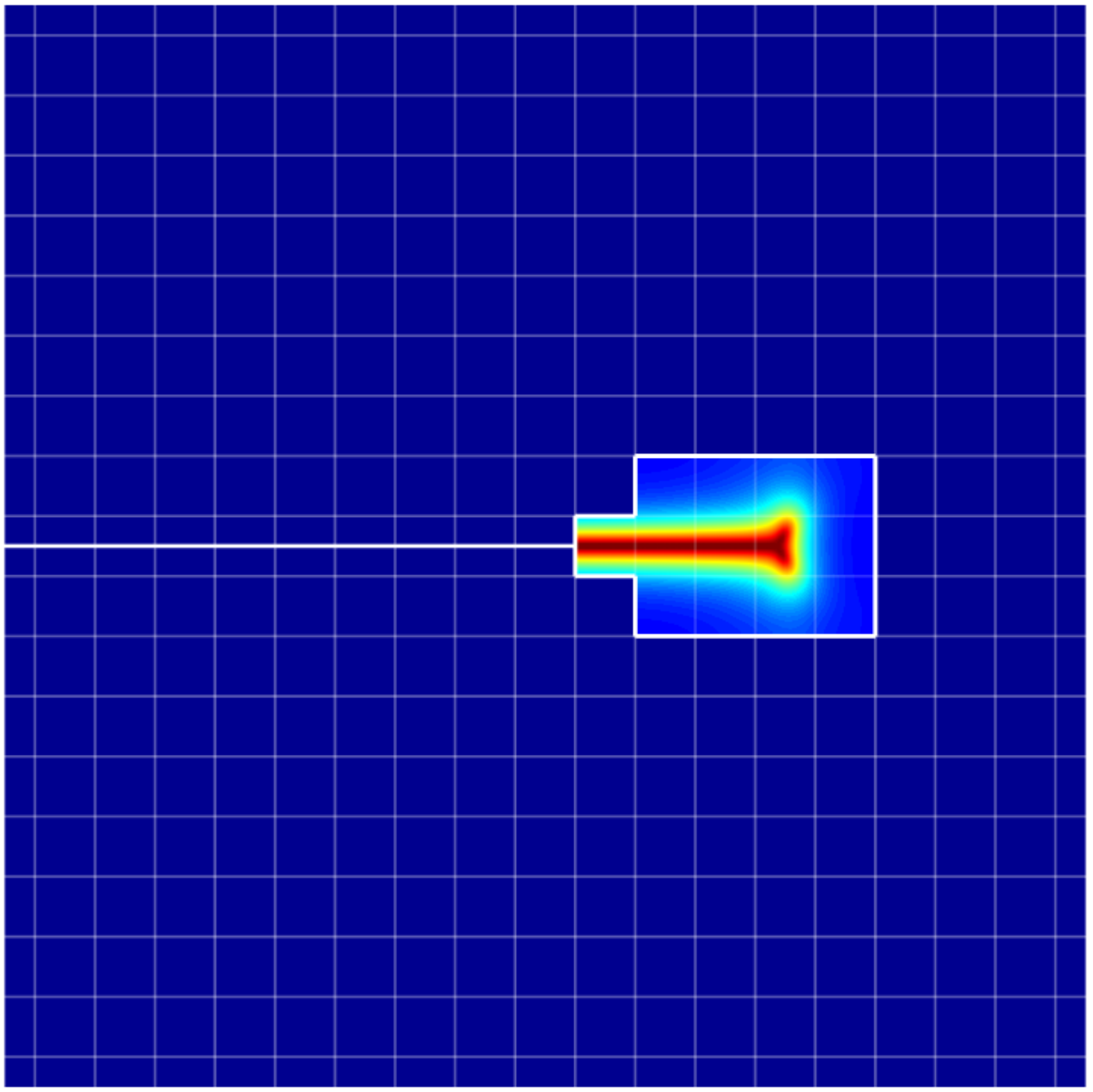}
	\end{subfigure}
	\hspace{0.1mm}
	\begin{subfigure}[b]{0.3\textwidth}
		\centering
		\includegraphics[width=\textwidth]{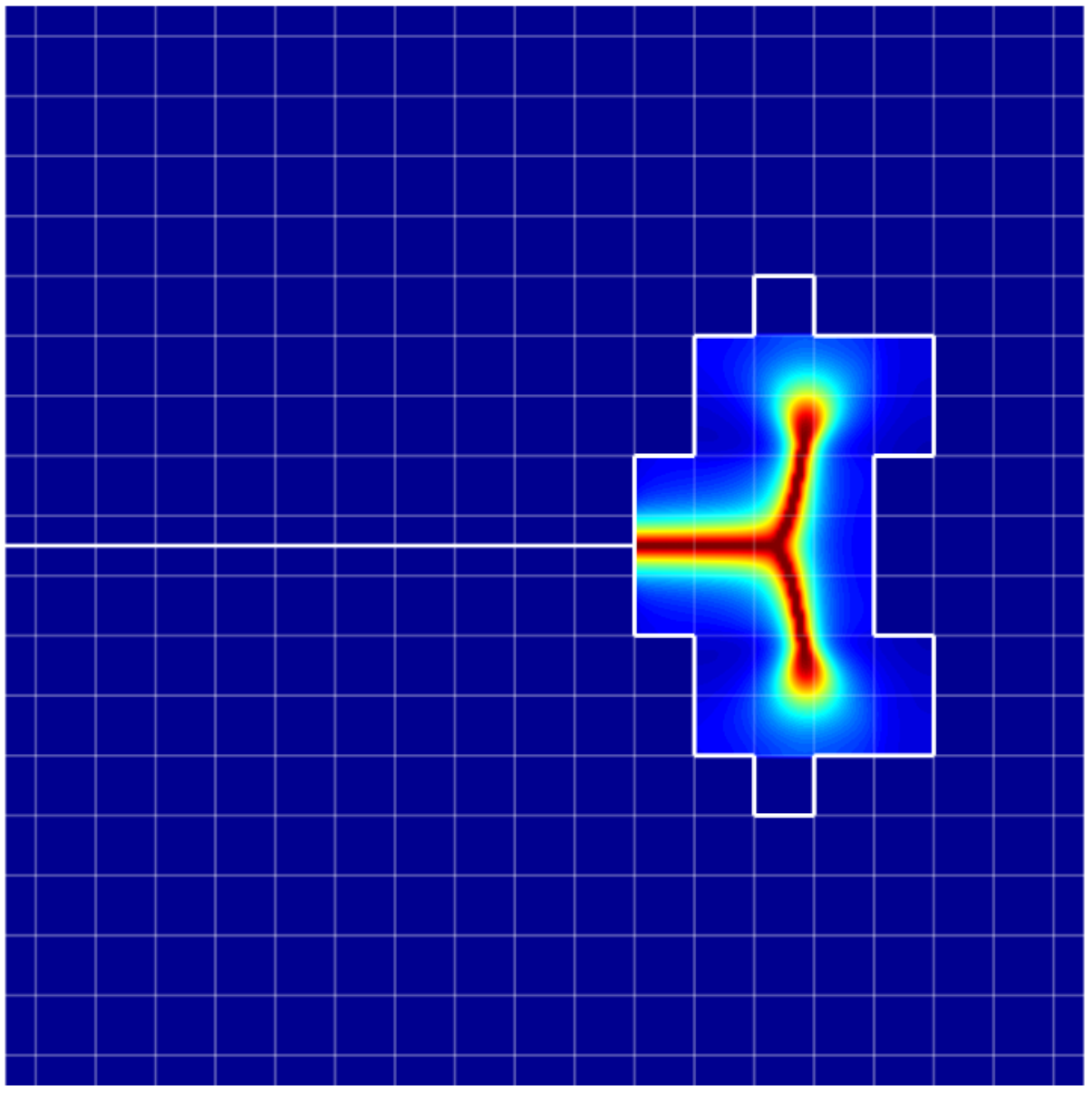}
	\end{subfigure}
	\hspace{0.1mm}
	\begin{subfigure}[b]{0.3\textwidth}
		\centering
		\includegraphics[width=\textwidth]{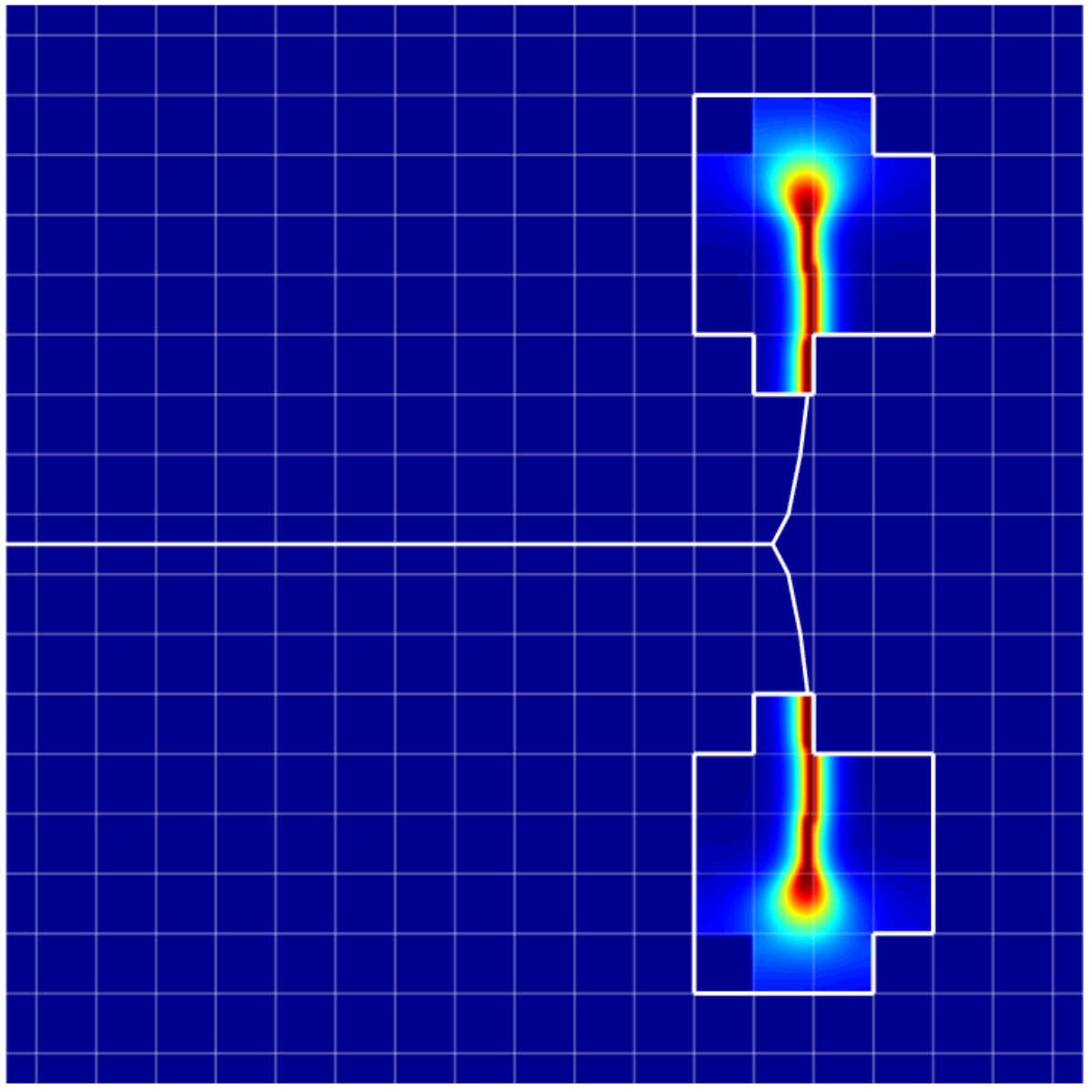}
	\end{subfigure}
	\hspace{0.1mm}
	\begin{subfigure}[b]{0.057\textwidth}
		\centering
		\includegraphics[width=\textwidth]{figures/colorbar.pdf}
	\end{subfigure}
	
	\hspace{1.7cm}
	\raisebox{0cm}{$s = 0$}
	\hfill
	\raisebox{0cm}{$s = 20$}
	\hfill
	\raisebox{0cm}{$s = 40$}
	\hspace{2.7cm}
	\vspace{1mm}
	
	\caption{\textit{Branching test.} Detail of the crack path at some staggered iterations for imposed displacement  $u_D = 0.05195$ mm. Zoom into $[0,0.8] \times [-0.4,0.4]$ $\textnormal{mm}^2$.}
	\label{fig:branching-damage-zoom}       
\end{figure}

The evolution of the crack at the load step when it branches, $u_D = 0.05195$ mm, is plotted in Figure \ref{fig:branching-damage-zoom} at some illustrative staggered iterations. 
The existing sharp crack is updated with the contribution of one of the branches and the other branch defines the new sharp crack. 
If the branching point is interior to the element, the representation of the crack is piecewise linear and contains the point, i.e. it is defined by a Y-shaped approximation, as can be seen in the third plot of Figure \ref{fig:branching-damage-zoom}.

\begin{figure}[]
	\centering
	\includegraphics[height=6.0cm]{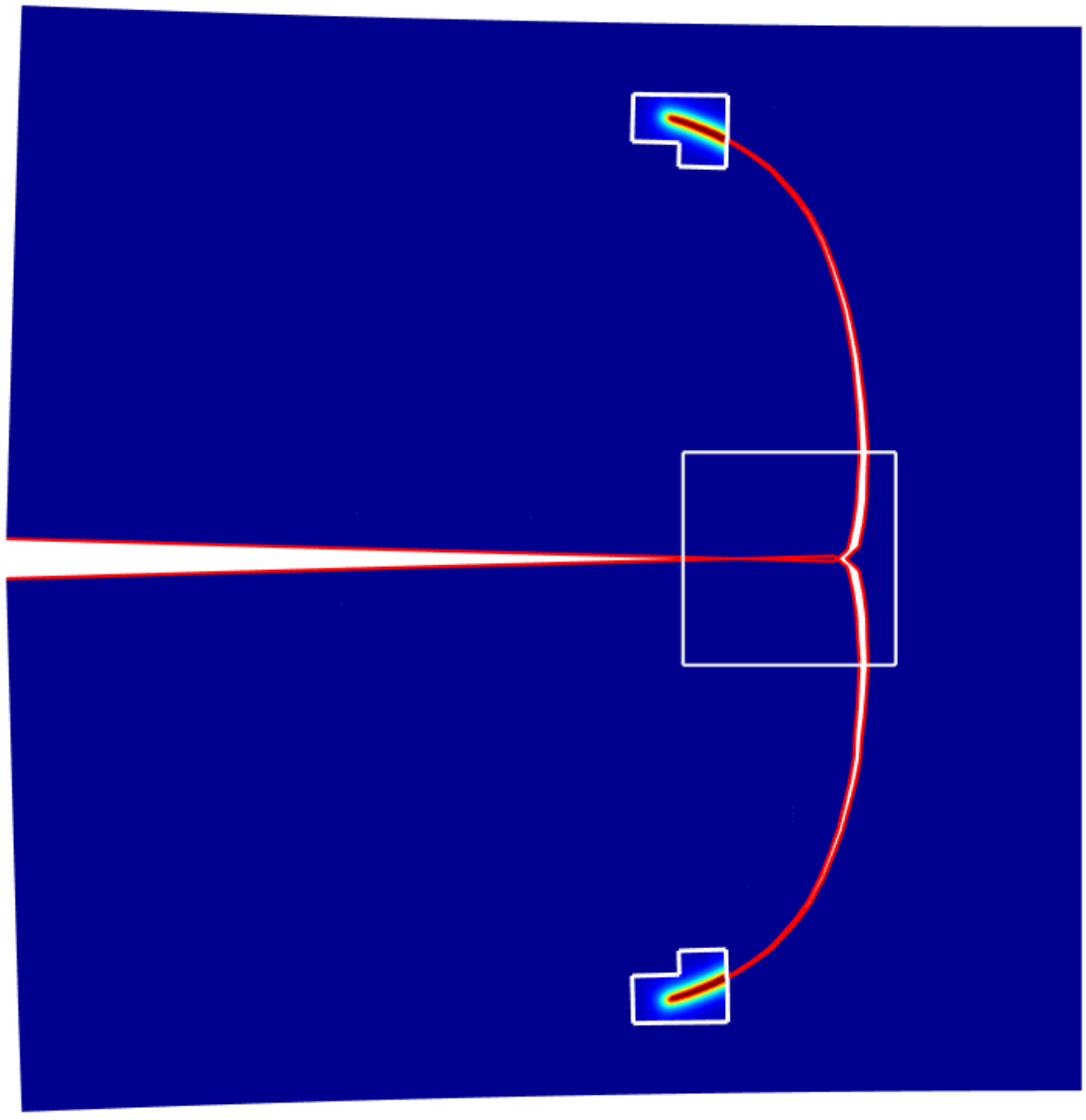}
	\hspace{2mm}
	\raisebox{0.45cm}{
		\includegraphics[height=5.1cm]{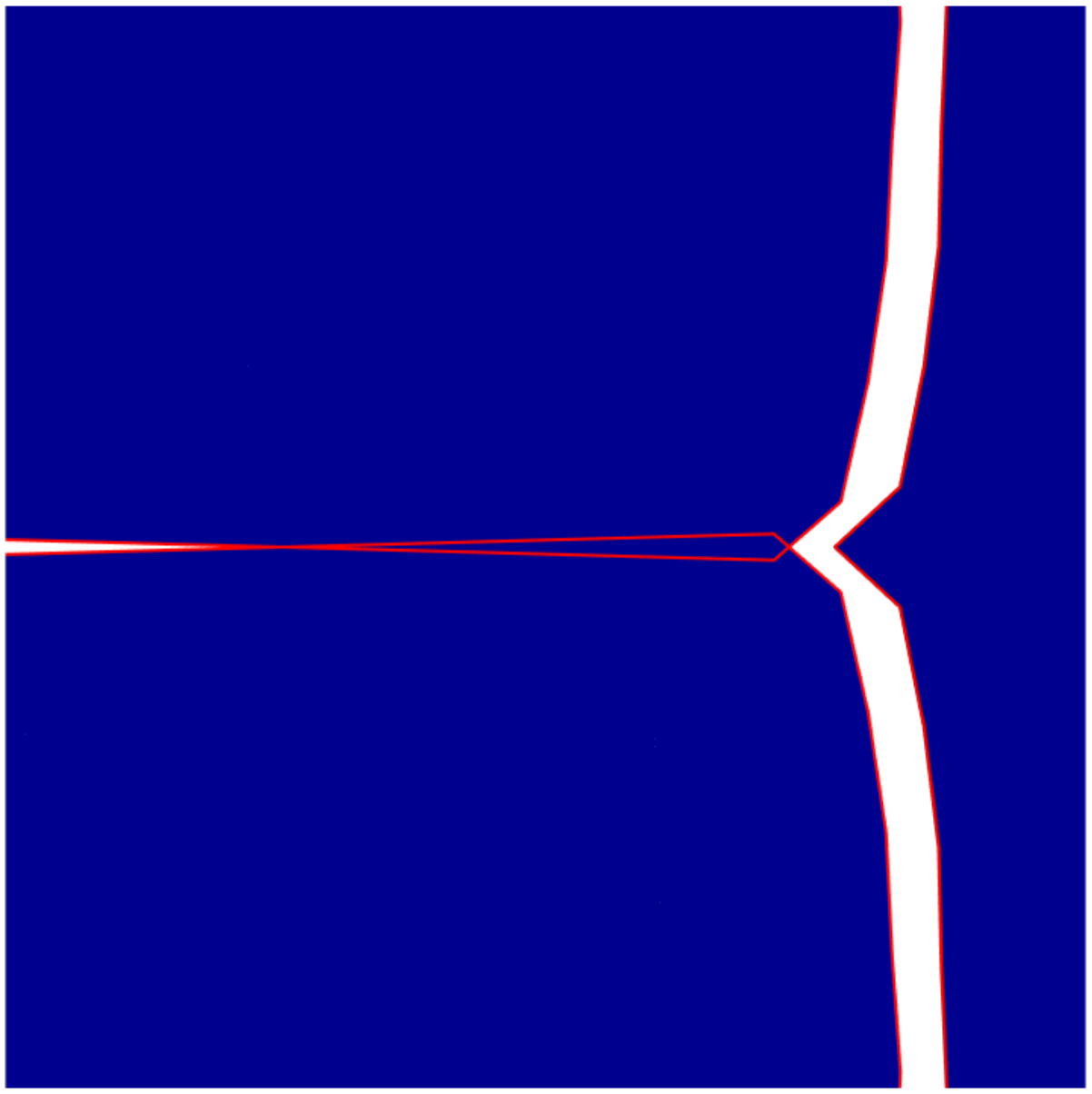}
		\hspace{2mm}
		\includegraphics[height=5.1cm]{figures/colorbar.pdf}
	}	
	\caption{\textit{Branching test.} Deformed piece at imposed displacement $u_D = 0.08$ mm. Crack faces, highlighted in red, show a slight interpenetration. Zoom into $[0.25,0.65]\times[-0.2,0.2]$ $\textnormal{mm}^2$.}
	\label{fig:branching-interpenetration}       
\end{figure}

As reported in \cite{MuixiHDG}, we observe a slight interpenetration of the crack faces near the branching point. In the phase-field model, interpenetration is prevented by the definition of $g(d)$ in \eqref{hybridCondition}, but when transitioning to the XFEM representation the crack faces intersect as shown in Figure \ref{fig:branching-interpenetration}. This issue can be tackled by implementing contact conditions into the XFEM discretization, see for instance \cite{KimDolbowLaursen2007,GinerTurTaranconFuenmayor2009}. 

\subsection{Multiple cracks test}

This example, proposed in \cite{MuixiNitsche}, exhibits the robustness of the proposed strategy to simulate coalescence of cracks.

\begin{figure}[]
	\centering
	\includegraphics[width=0.5\columnwidth]{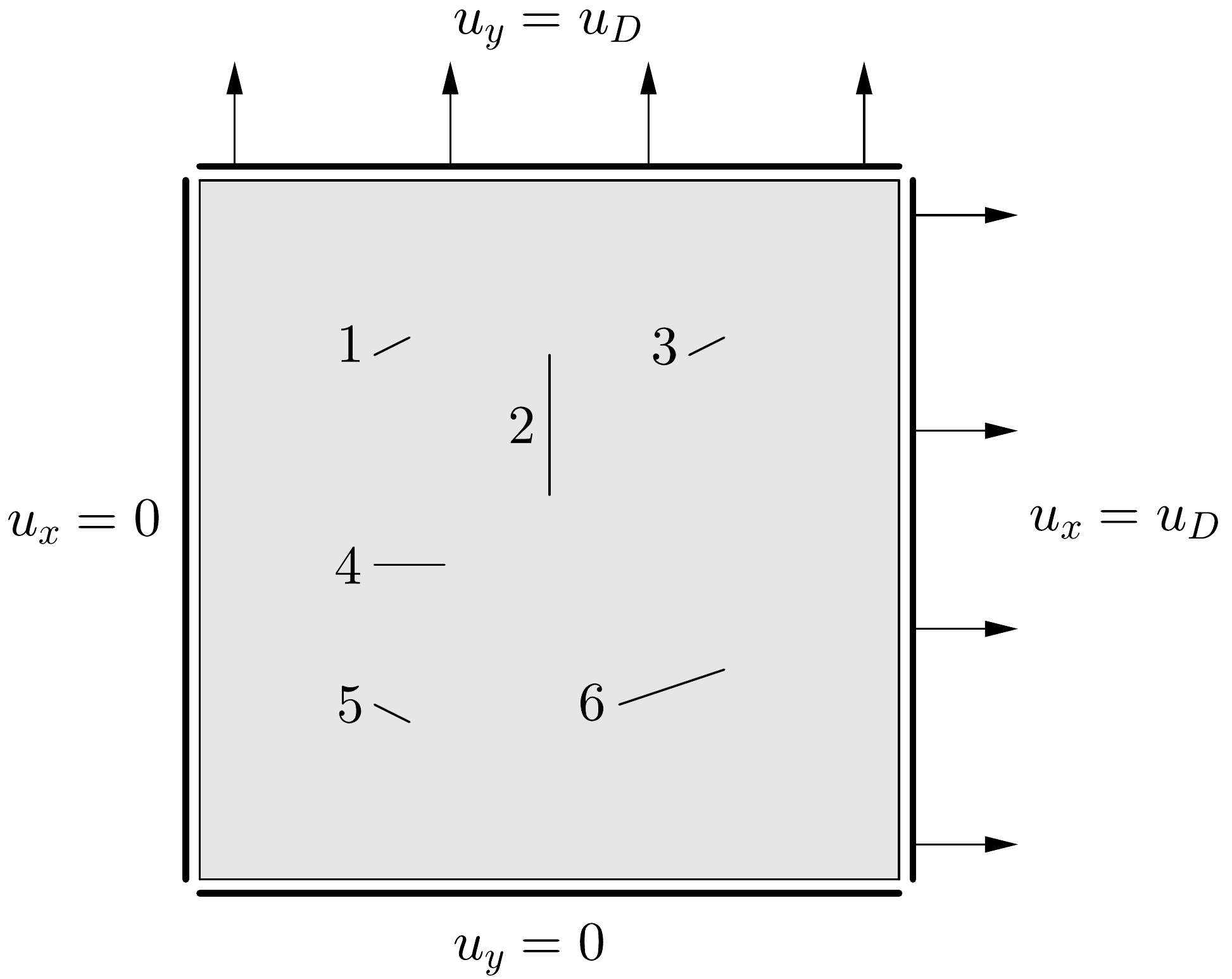}
	\caption{\textit{Multiple cracks test.} Geometry and boundary conditions. }
	\label{fig:multi-setting}       
\end{figure}

Consider a square plate in $[0,2]^2$ $\textnormal{mm}^2$, with 6 initial cracks and subjected to biaxial tension as depicted in Figure \ref{fig:multi-setting}.
The tips of the initial cracks are in Table \ref{tab:multi}.
Material parameters are
$E = 20$ GPa, $\nu = 0.3$ and $G_C = 10^{-3}$ kN/mm. We use $l = 0.012$ mm and take increments of $\Delta u_D = 5 \cdot 10^{-5} $ mm. The background mesh is a uniform quadrilateral mesh with $47\times47$ elements and the refinement factor is $m = 17$. The distance to switch to discontinuous is $\delta^* = 3h$.

\begin{figure}[h!]
	\centering
	\vspace{6mm}
	
	\includegraphics[height=0.27\textwidth]{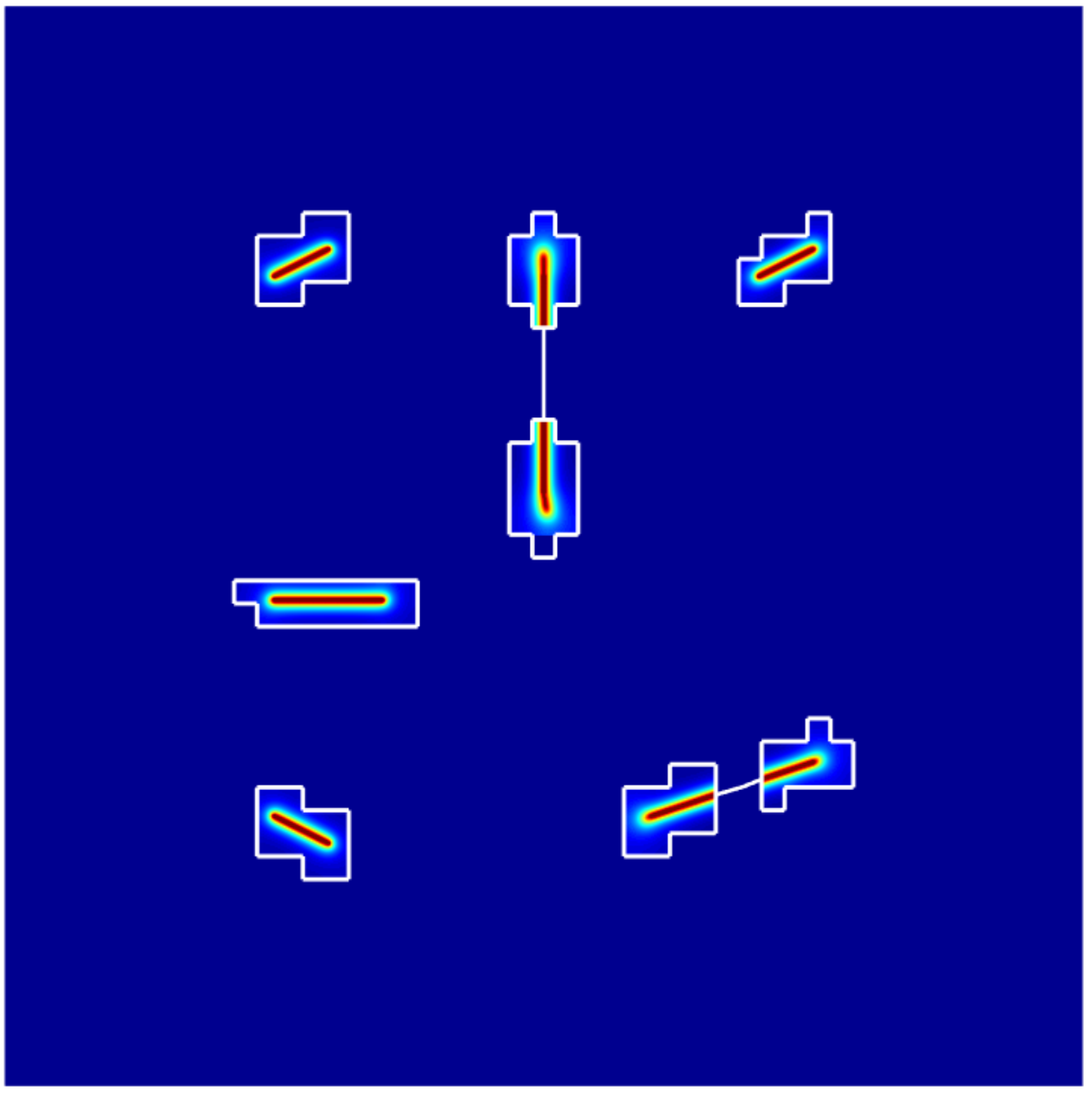}
	\hspace{1mm}
	\includegraphics[height=0.27\textwidth]{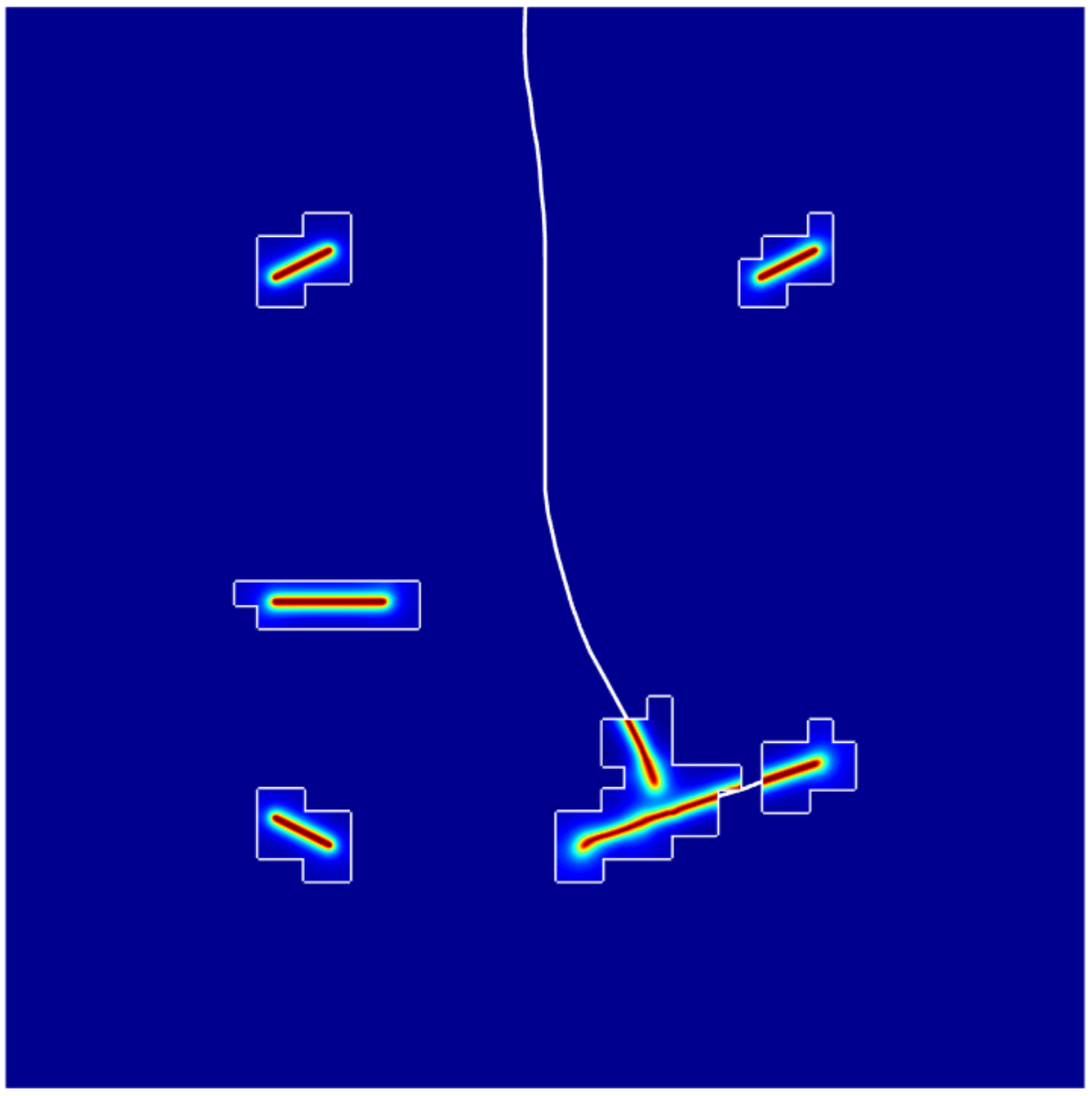}
	\hspace{1mm}
	\includegraphics[height=0.27\textwidth]{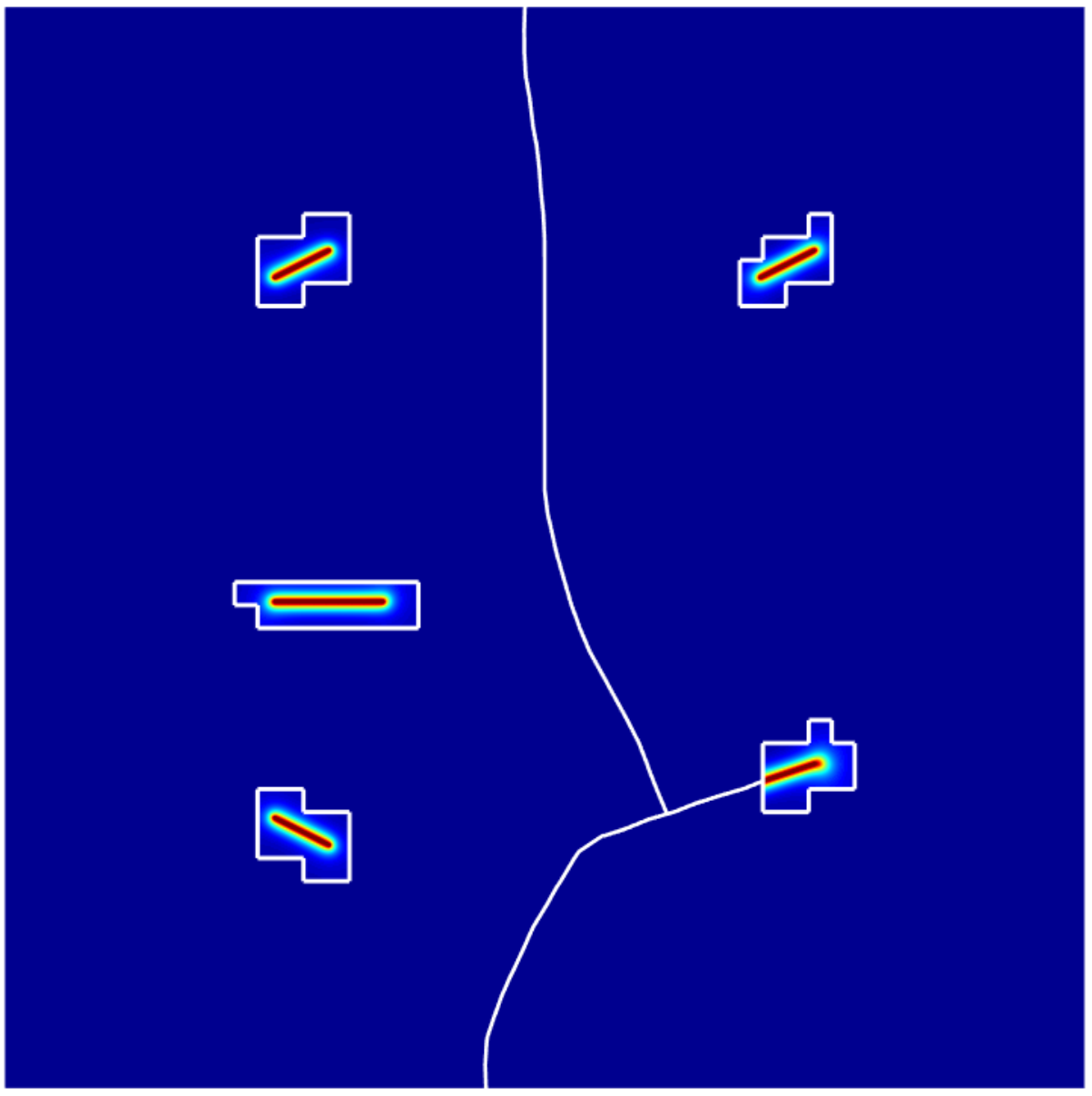}
	\hspace{1mm}
	\includegraphics[height=0.27\textwidth]{figures/colorbar.pdf}
	
	\hspace{1.8cm}
	\raisebox{0cm}{$u_D = 0.0120$ mm}
	\hfill
	\raisebox{0cm}{$u_D = 0.0125$ mm}
	\hfill
	\raisebox{0cm}{$u_D = 0.0130$ mm}
	\hspace{2.9cm}
	\vspace{3mm}

	\includegraphics[height=0.27\textwidth]{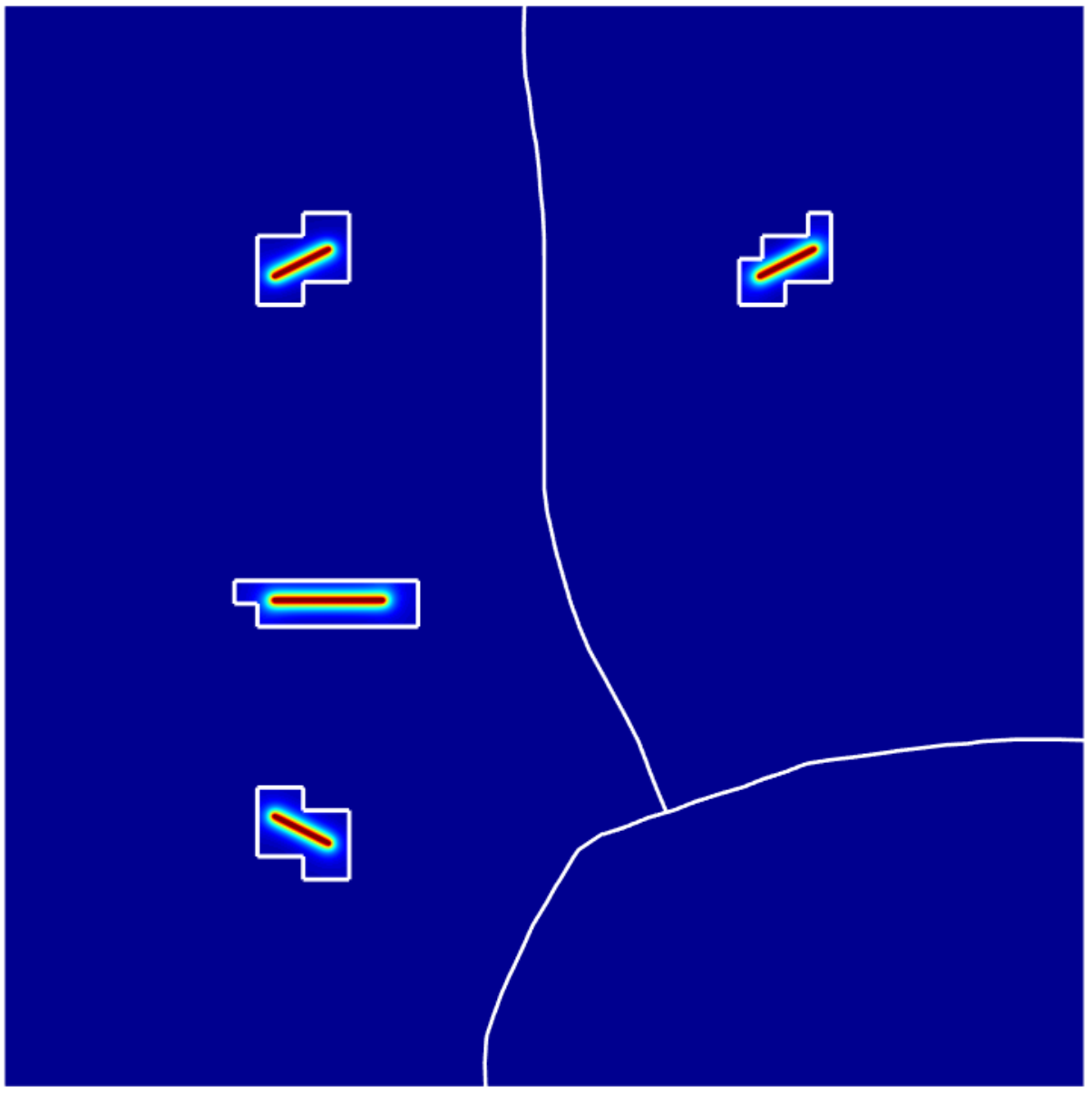}
	\hspace{1mm}
	\includegraphics[height=0.27\textwidth]{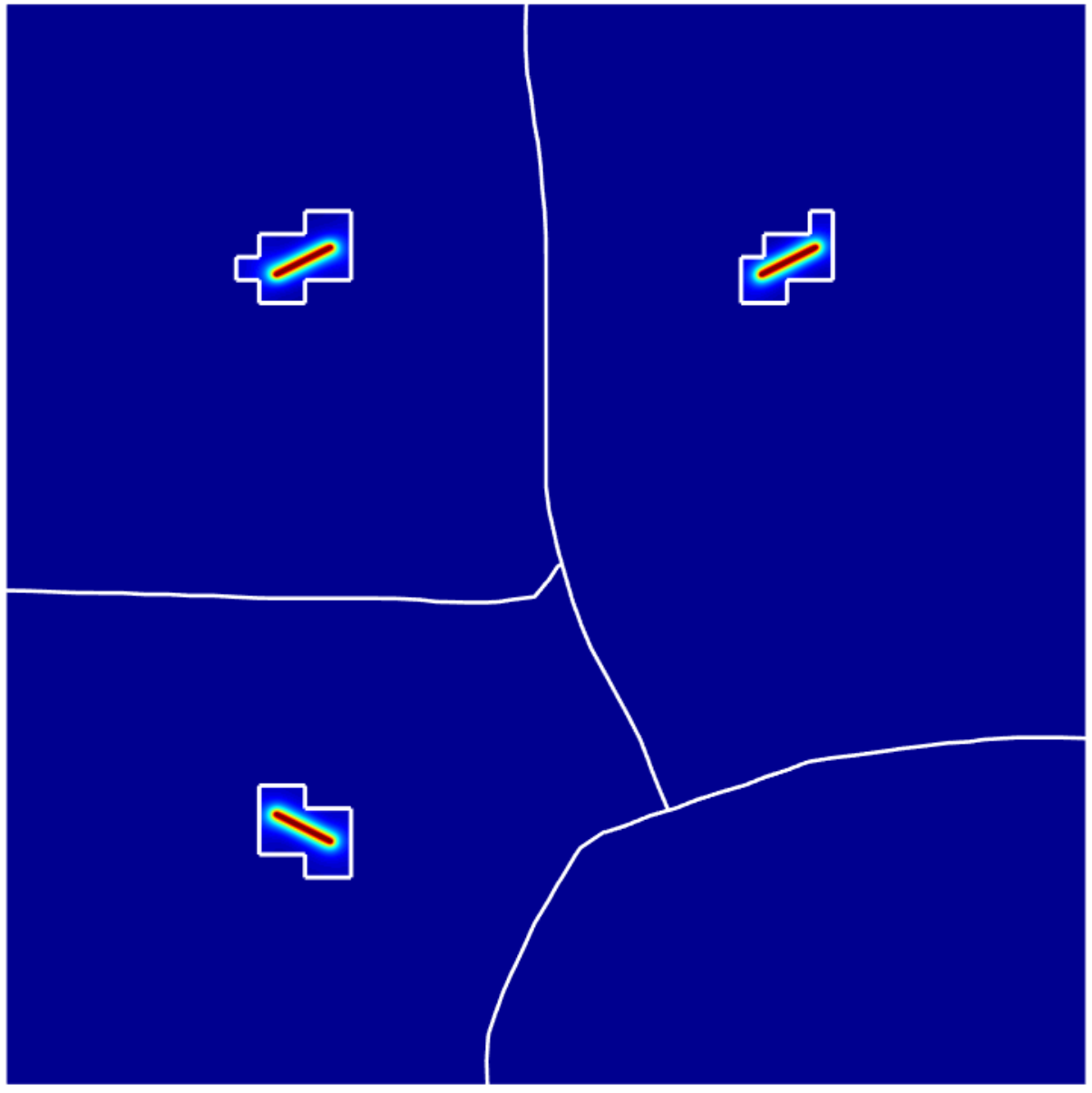}
	\hspace{1mm}
	\includegraphics[height=0.27\textwidth]{figures/colorbar.pdf}
	
	\hspace{4.35cm}
	\raisebox{0cm}{$u_D = 0.019$ mm}
	\hfill
	\raisebox{0cm}{$u_D = 0.027$ mm}
	\hspace{5.6cm}
	
	\caption{\textit{Multiple cracks test.} Damage field at several imposed displacements.}
	\label{fig:multi-damage}       
\end{figure}

\begin{figure}[]
	\centering
	\raisebox{1.8cm}{
		\begin{subfigure}[b]{0.14\textwidth}
			\includegraphics[width=\columnwidth]{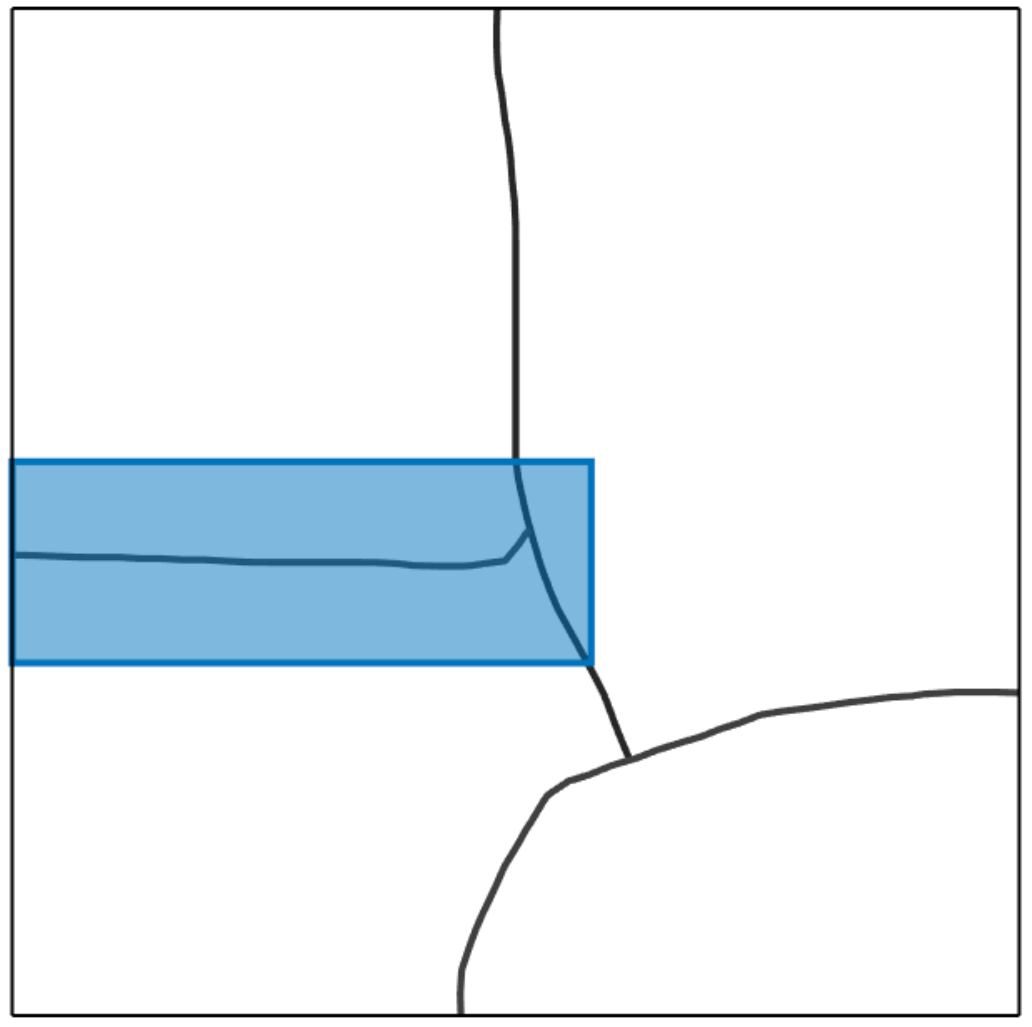}
	\end{subfigure}}
	\hspace{0.1mm}
	\begin{subfigure}[b]{0.78\textwidth}	
		\includegraphics[width=0.5\textwidth]{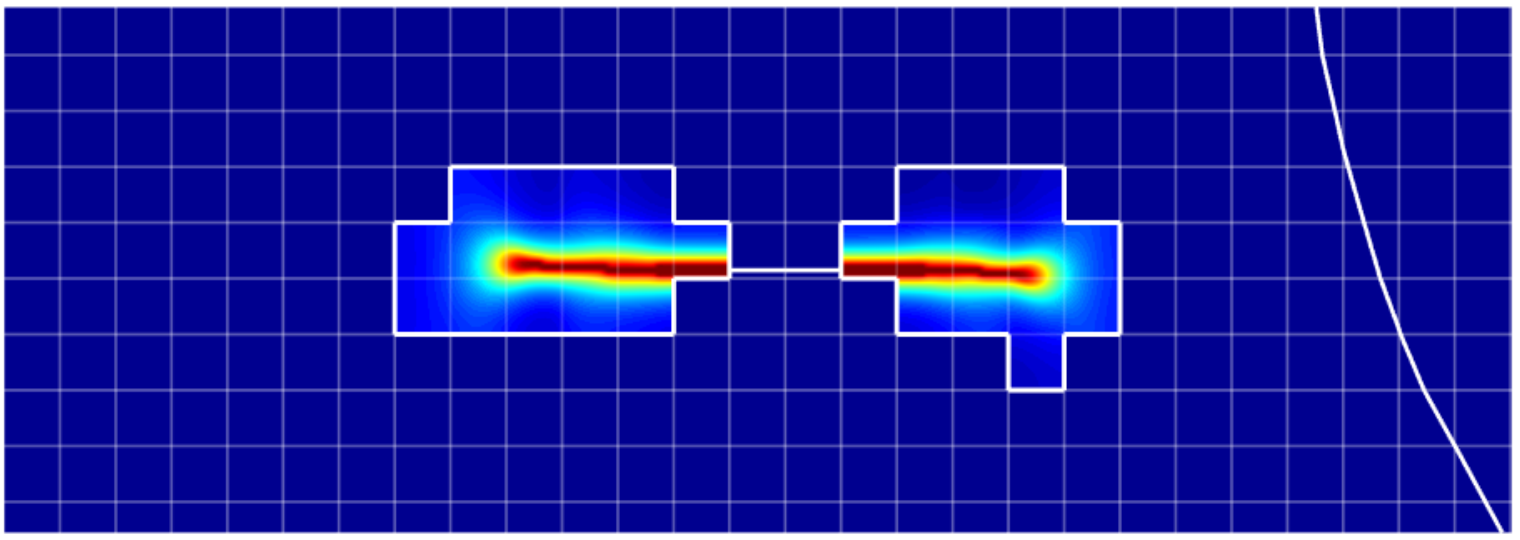}
		\includegraphics[width=0.5\textwidth]{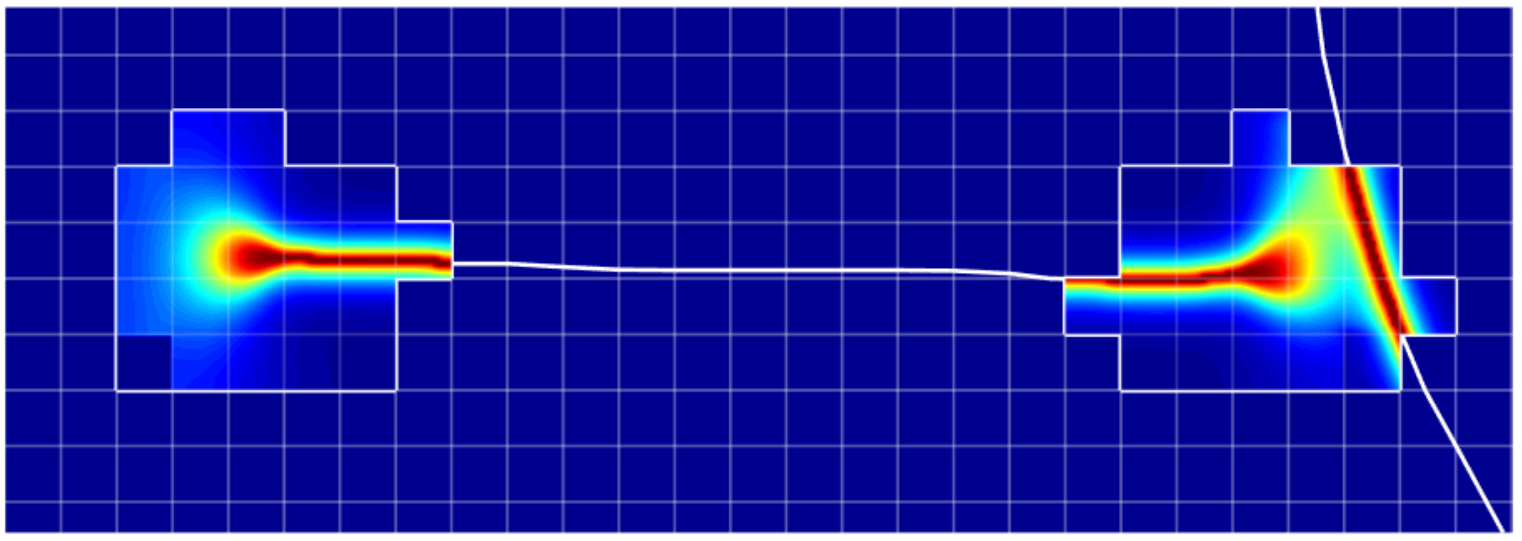}
		
		\hspace{0.22\textwidth}
		\raisebox{0cm}{$s = 30$}
		\hspace{0.42\textwidth}
		\raisebox{0cm}{$s = 60$}
		\vspace{1mm}
		
		\includegraphics[width=0.5\textwidth]{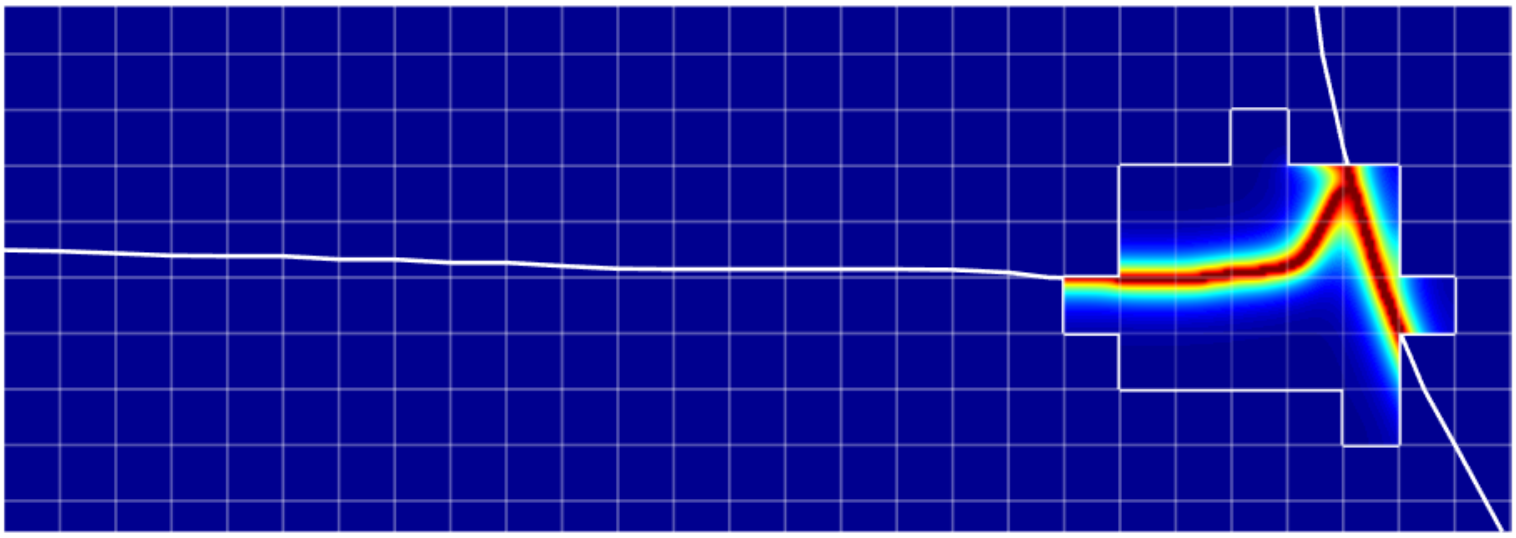}
		\includegraphics[width=0.5\textwidth]{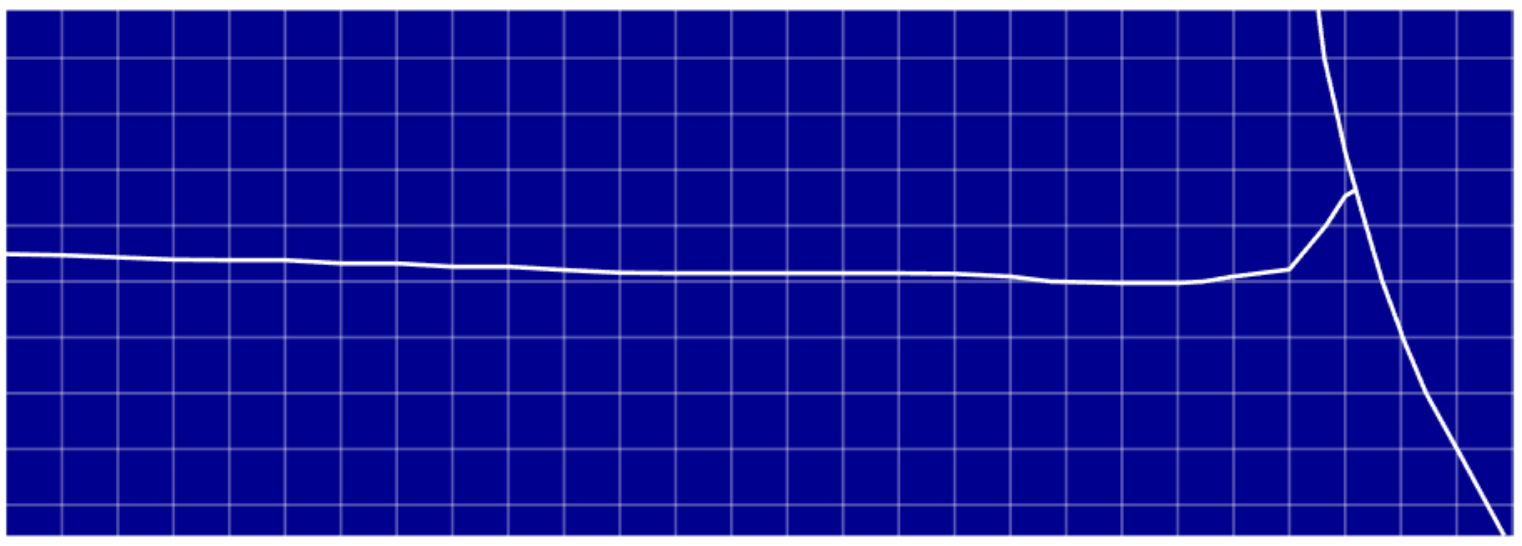}
		
		\hspace{0.22\textwidth}
		\raisebox{0cm}{$s = 67$}
		\hspace{0.42\textwidth}
		\raisebox{0cm}{$s = 68$}
	\end{subfigure}
	\begin{subfigure}[b]{0.05\textwidth}
		\raisebox{0.5cm}{
			\includegraphics[width=\textwidth]{figures/colorbar.pdf}}
	\end{subfigure}
	\caption{\textit{Multiple cracks test.} Crack pattern at different staggered iterations for imposed displacement $u_D = 0.262$ mm. Zoom into $[0,1.15]\times[0.7,1.1]$ $\textnormal{mm}^2$. }
	\label{fig:multi-zoom-coalescence}       
\end{figure}

The obtained crack pattern is plotted in Figure \ref{fig:multi-damage}. The initial cracks propagate, coalescing between them, until the piece is broken into 4 independent pieces. There are two mergings of cracks in the process. 
Since cracks grow abruptly, we need to plot the crack at staggered iterations to see the detail of how coalescence is handled by the method. Figure \ref{fig:multi-zoom-coalescence} shows some representative staggered iterations for imposed displacement $u_D = 0.262$ mm. In the figure, we can see how the sharp crack in the right is replaced by its diffuse representation as the crack tip approaches and the cut elements are refined. Then, after the merging has occured, all the elements in this subdomain of $\Otips$ transition back to $\Oxfem$.

\begin{table}[]
	\centering
	\caption{\textit{Multiple cracks test.} Tip coordinates for the initial cracks in the domain $[0,2]^2 \textnormal{ mm}^2$.}
	\label{tab:multi}       
	\begin{tabular}{cll}
		\hline\noalign{\smallskip}
		Crack & $P_1$ (mm) & $P_2$  (mm) \\
		\noalign{\smallskip}\hline\noalign{\smallskip}
		$1$ & $(0.5,1.5)$ & $(0.6,1.55)$ \\
		$2$ & $(1,1.1)$ & $(1,1.5)$ \\	
		$3$ & $(1.4,1.5)$ & $(1.5,1.55)$ \\
		$4$ & $(0.5,0.9)$ & $(0.7,0.9)$ \\
		$5$ & $(0.5,0.5)$ & $(0.6,0.45)$ \\
		$6$ & $(1.2,0.5)$ & $(1.5,0.6)$ \\
		\noalign{\smallskip}\hline
	\end{tabular}
\end{table}

\begin{figure}[]
	\centering
	\includegraphics[height=5.5cm]{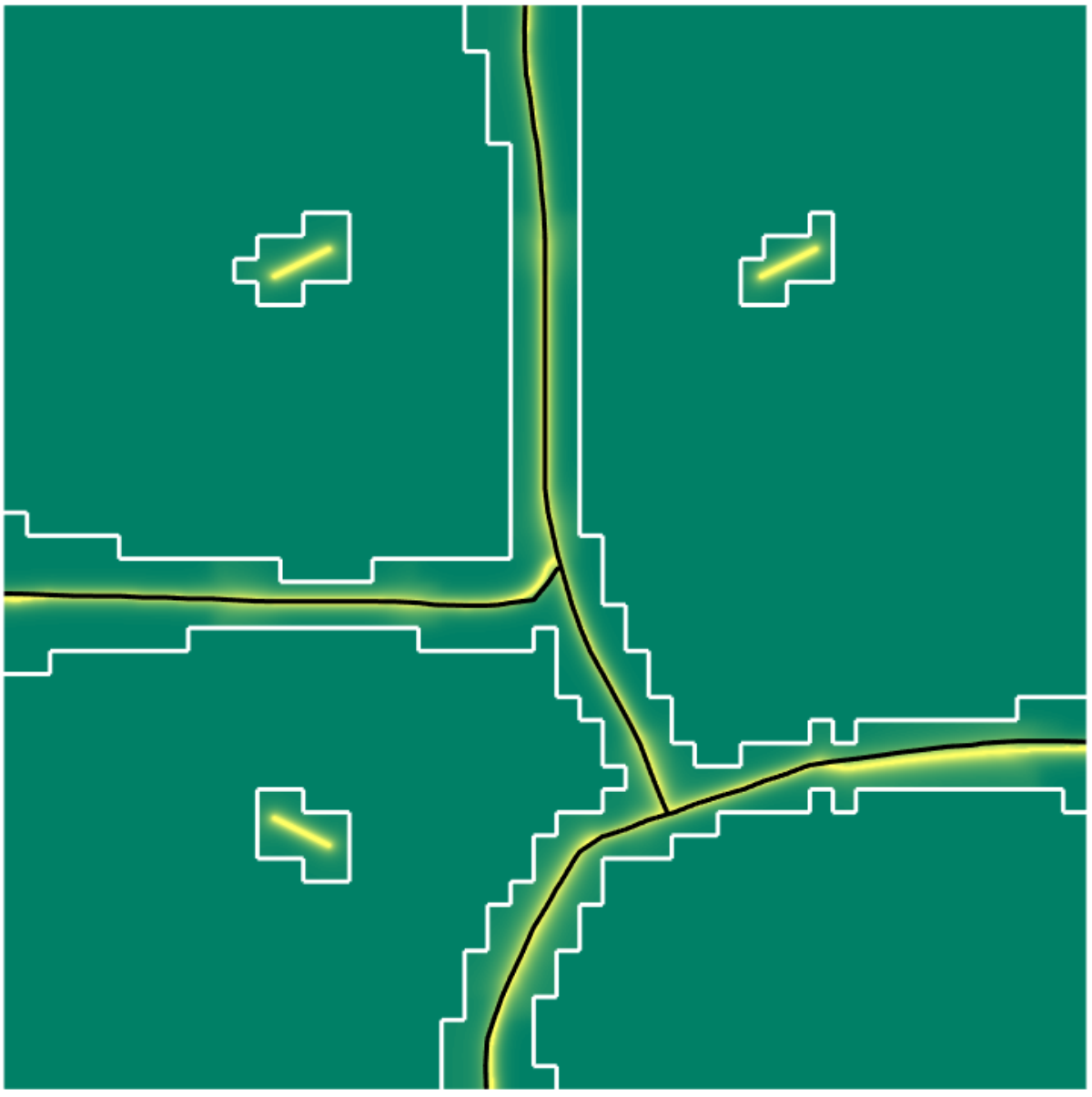}
	\hspace{1mm}
	\includegraphics[height=5.5cm]{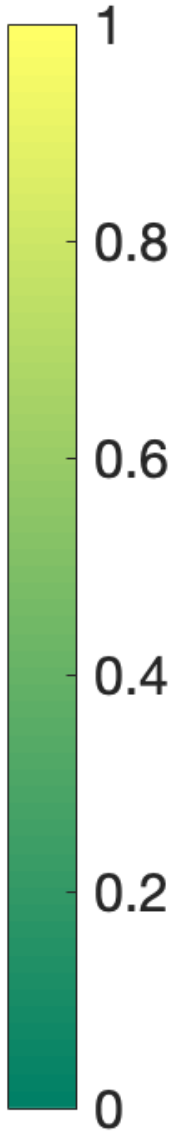}
	\caption{\textit{Multiple cracks test.} Comparison with PF solution.}
	\label{fig:multi-comparison}       
\end{figure}

\begin{figure}[]
	\centering
	\includegraphics[width=0.55\columnwidth]{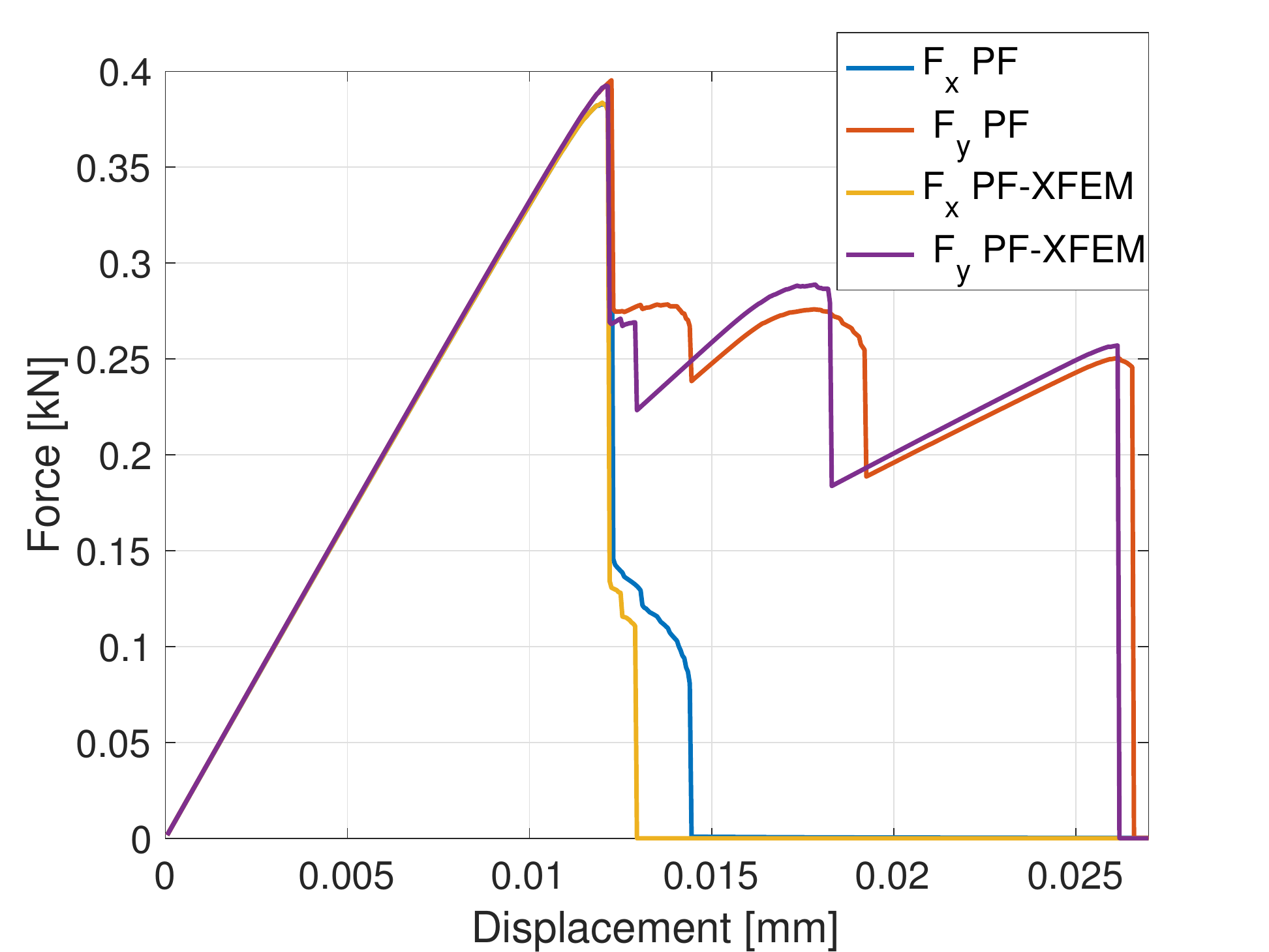}
	\caption{\textit{Multiple cracks test.} Load-displacement curve.}
	\label{fig:multi-loaddisp}       
\end{figure}

\begin{figure}[]
	\centering
	\includegraphics[width=0.55\columnwidth]{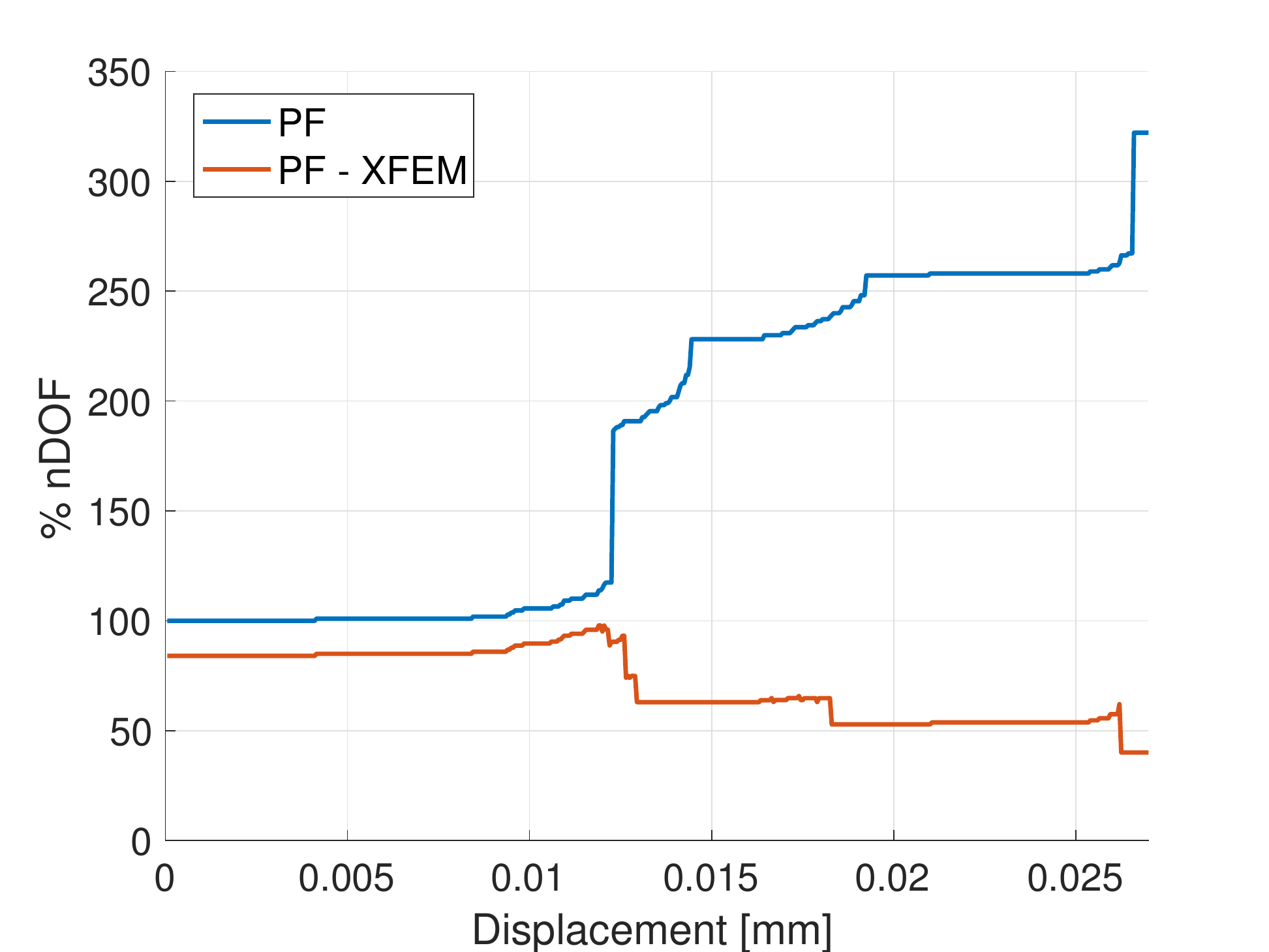}
	\caption{\textit{Multiple cracks test.} Evolution of the number of degrees of freedom.}
	\label{fig:multi-ndofs}       
\end{figure}

Now, we compare the results with the ones obtained by a PF approach. The final partition of the piece for both approaches is shown in Figure \ref{fig:multi-comparison}. The damage field corresponds to the PF simulation, and the final sharp crack for PF-XFEM is plotted in black. The crack patterns are very similar also in this more complex scenario. 
The corresponding load-displacement curves for the horizontal and vertical loads, $F_x$ and $F_y$, in Figure \ref{fig:multi-loaddisp}, again indicate a loss of stiffness for PF in the precracking regime and a steeper descent of the curves for PF-XFEM. The evolution of nDOF is plotted in Figure \ref{fig:multi-ndofs}, showing a substantial decrease in nDOF. For PF-XFEM we observe a reduction of nDOF as crack tips disappear, while in PF the nDOF always increases.


\subsection{Twisting test}\label{sec:twisting}

\begin{figure}[]
	\centering	
	\includegraphics[width=0.5\columnwidth]{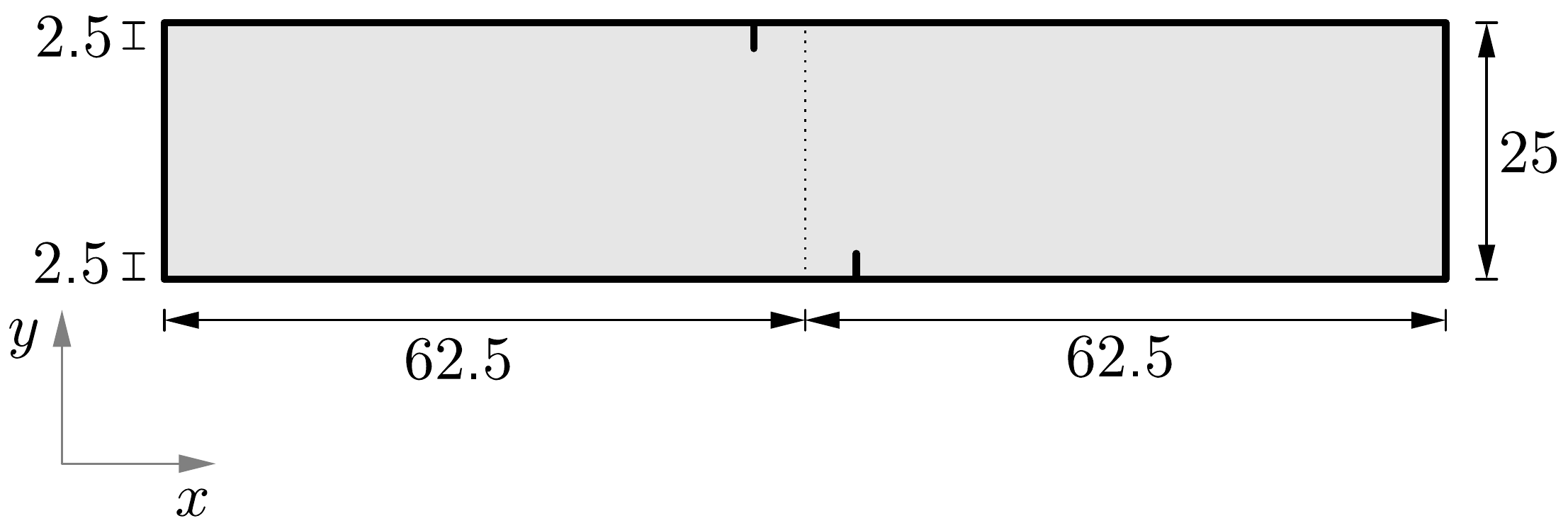}
	
	\vspace{-0.5mm}
	\includegraphics[width=0.5\columnwidth]{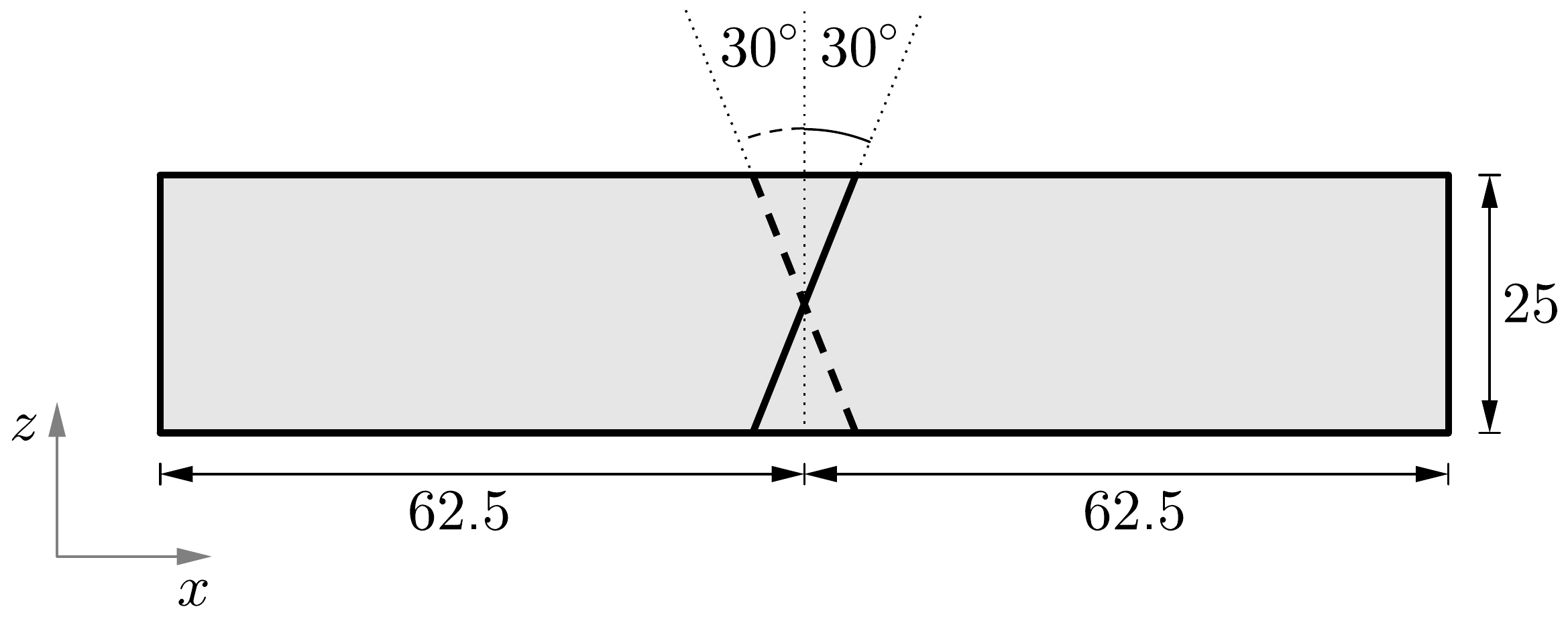}
	\caption{\textit{Twisting test.} Front and top view of the initial geometry of the piece. Lenghts are in mm. }
	\label{fig:twisting-setting}       
\end{figure}

In this example we test the performance of the method in a fully 3D example. This test was also proposed in \cite{MuixiNitsche}. We consider a square section beam in $[0,125]\times[0,25]\times[0,25]$ $\textnormal{mm}^3$, which is clamped on $\{ x = 0 \textnormal{ mm} \}$ and has imposed displacements along $x$ on $\{ x = 125 \textnormal{ mm} \}$.
The beam has two initial notches on $\{ y = 0 \textnormal{ mm} \}$ and $\{ y = 25 \textnormal{ mm} \}$, which are inclined with opposite angles, as shown in Figure \ref{fig:twisting-setting}. 

The parameters are $E = 32$ GPa, $\nu = 0.25$, $G_C = 1.6 \cdot 10^{-4}$ kN/mm and $l = 2$ mm. The load steps take increments $\Delta u_D = 5 \cdot 10^{-4}$ mm. 
We use a uniform hexahedral mesh with element size $h = 3.125$ mm and refinement factor $m = 7$. The distance in the switching criterion is $\delta^* = 2h$. The initial notches are modeled by diffuse cracks and $\Otips$ is the union of elements that contain them.

\begin{figure}[]
	\centering
	\includegraphics[width=0.55\columnwidth]{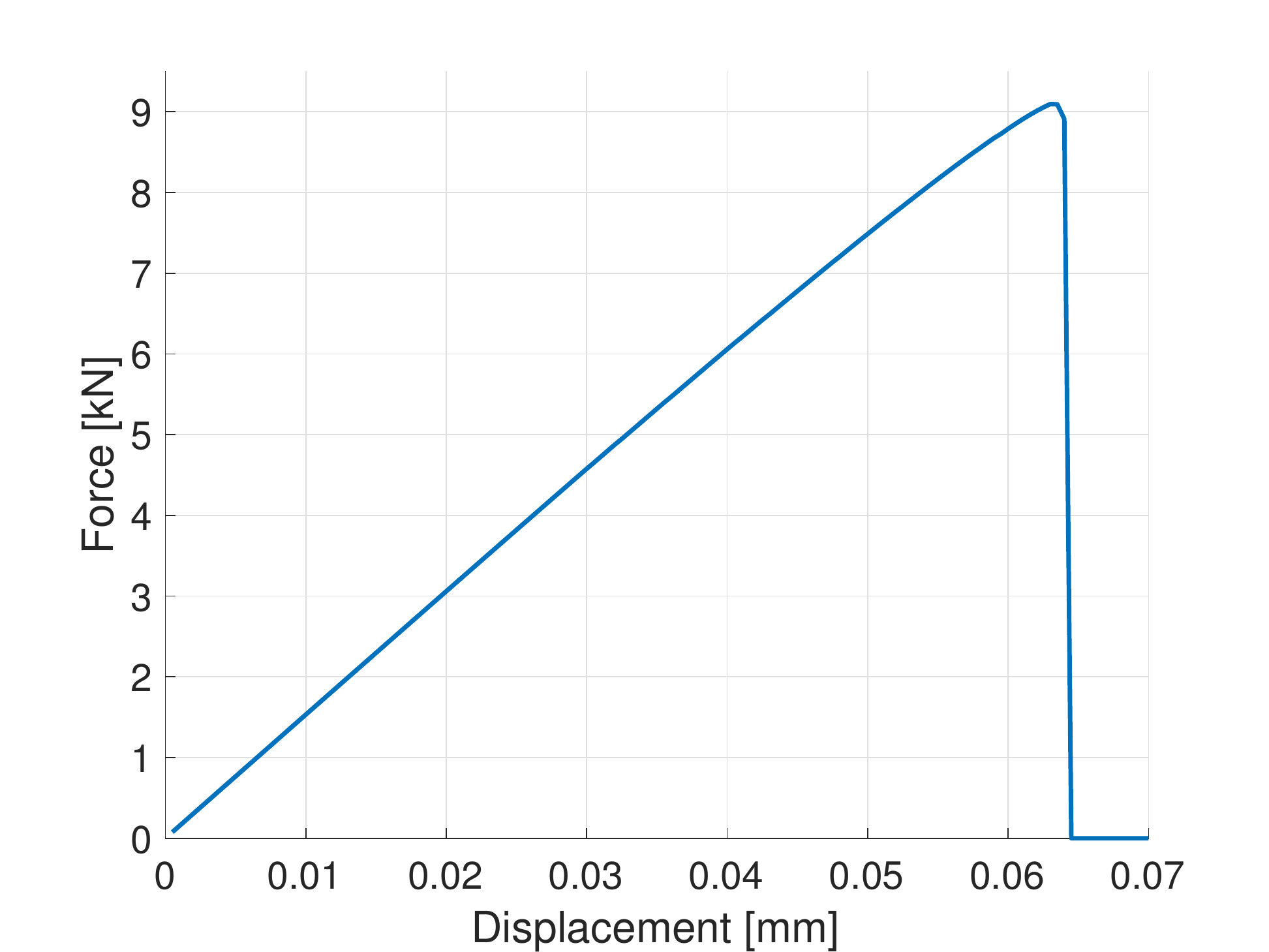}
	\caption{\textit{Twisting test.} Load-displacement curve. The piece completely breaks at $u_D = 0.0645$ mm.}
	\label{fig:twisting-loaddisp}       
\end{figure}

\begin{figure}[]
	\centering
	\begin{subfigure}[b]{\textwidth}
		\centering
		\includegraphics[width=0.58\textwidth]{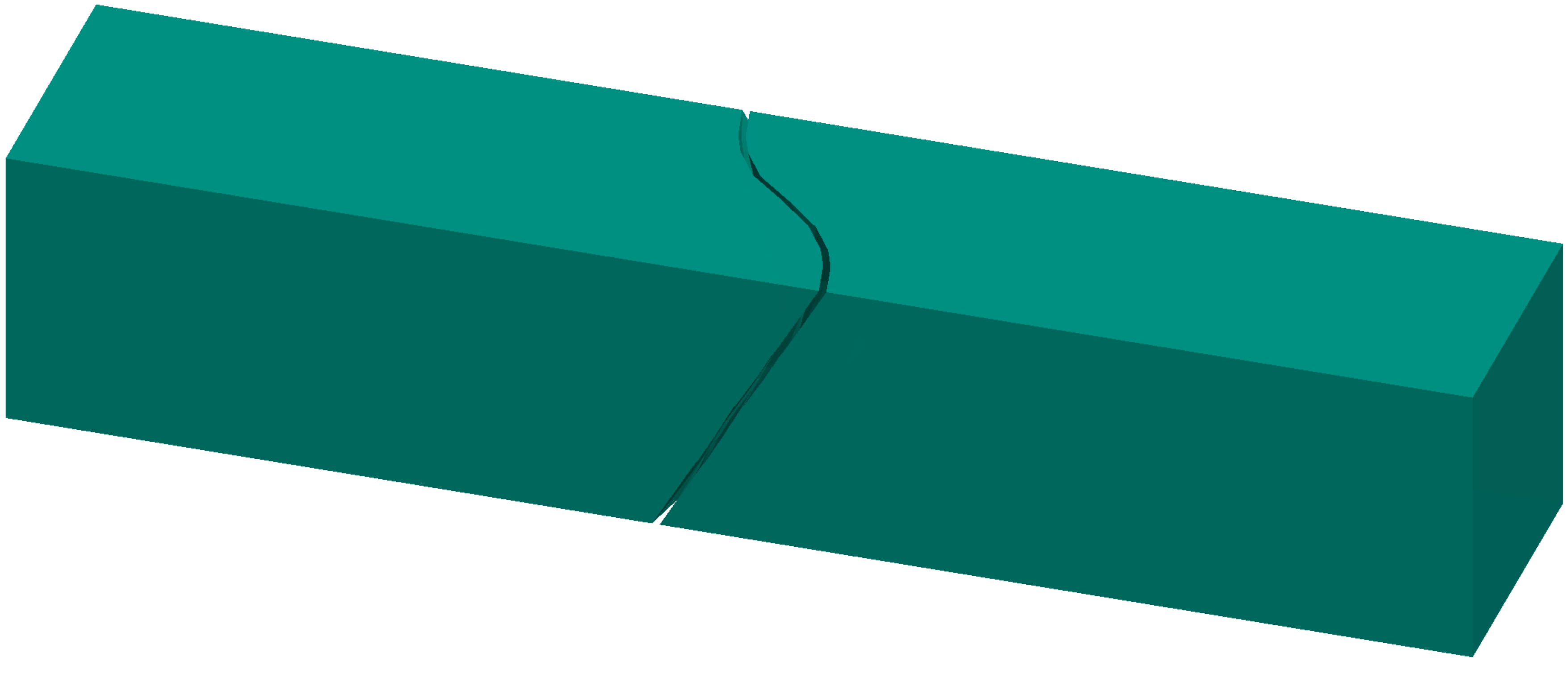}
		\raisebox{2mm}{
			\includegraphics[width=0.08\textwidth]{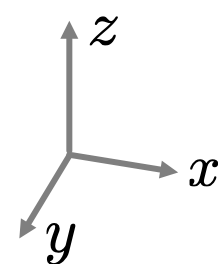}}
	\end{subfigure}
	
	\vspace{5mm}	
	\includegraphics[width=0.65\textwidth]{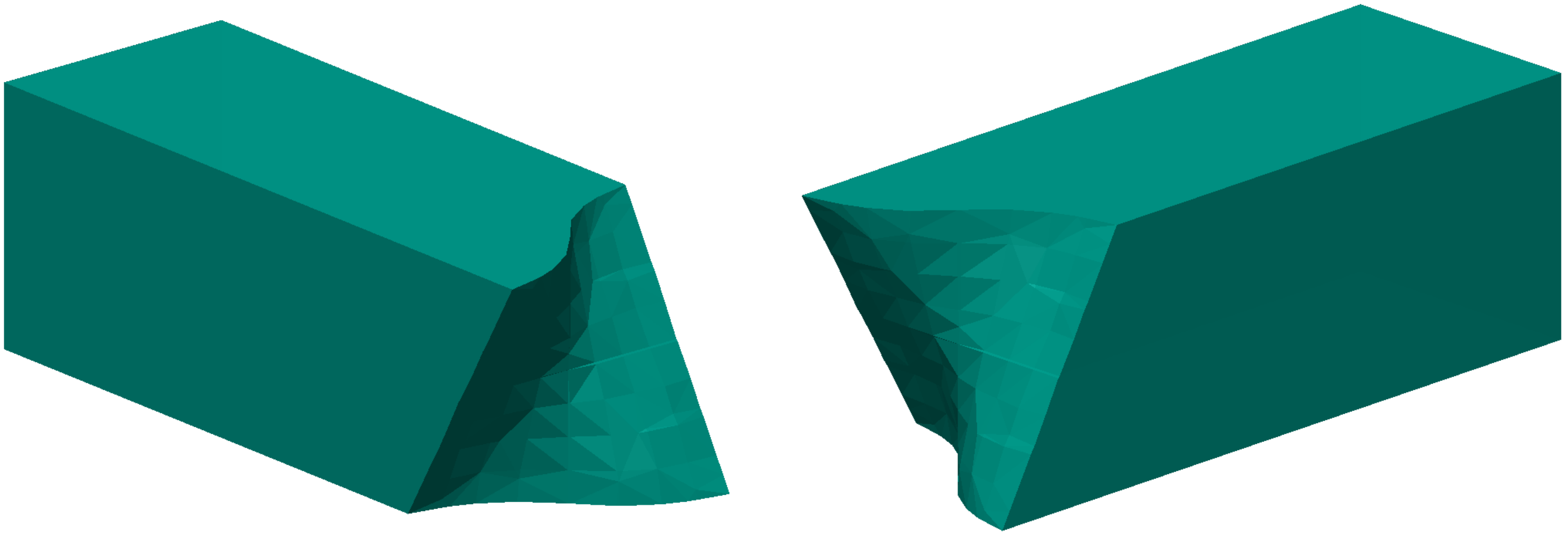}	
	\caption{\textit{Twisting test.} Geometry of the piece at imposed displacement $u_D = 0.0645$ mm. The initial beam breaks into two independent and symmetric halves.}
	\label{fig:twisting-final}       
\end{figure}

In this case, the initial cracks coalesce, completely splitting the beam at a single load step, as can be seen in the load-displacement curve in Figure \ref{fig:twisting-loaddisp}. 
Figure \ref{fig:twisting-final} shows the final geometry  at load step $u_D = 0.0645$ mm. The sharp crack is a twisted surface to match the initial notches, defined by triangular facets. Considering a sharp representation of the crack enables to completely separate the two resulting pieces.

\section{Conclusions} \label{sec:conclusions}

We propose a novel method to simulate fracture which is based on combining a phase-field model in small subdomains around crack tips, $\Otips$, and a discontinuous model in the rest of the domain, $\Oxfem$.  The approach overcomes the limitations of both continuous and discontinuous models.
Propagation is described by the phase-field solution in $\Otips$, while an XFEM approximation explicitly describes the crack opening and enables to use a coarser discretization in almost the whole domain. 
In all examples, a correct definition of the initial partition in $\Otips$ and $\Oxfem$ is crucial to detect crack inception. 

Computationally, the same background mesh is used during the whole simulation. Refined elements are nested in $\Otips$ to capture the phase-field solution and the sharp cracks are introduced via XFEM. The discretization is automatically updated as cracks propagate and remeshing is avoided.
Nitche's method is used to impose continuity of displacements in weak form to mantain a very local refinement with no transitioning elements. 

The robustness of the strategy has been proved in 2D and 3D, including scenarios with branching and coalescence. 
The obtained results are comparable to the ones from a plain phase-field approach, but with an important reduction in the number of degrees of freedom. 
Also, the response of the pieces in the linear elastic regime is more realistic and transmission of forces across cracks is prevented. 
The methodology is an efficient alternative for fracture simulation. 

Further investigation would include enhancing the method by incorporating contact conditions into the XFEM formulation and by exploring other alternatives for the geometrical description of sharp cracks.

\subsection*{Acknowledgements}
This work was supported by the 
Ag\`encia de Gesti\'o d'Ajuts Universitaris i de Recerca training grant FI-DGR 2017, the DAFOH2 project (Ministerio de Ciencia e Innovaci\'on, MTM2013-46313-R) and the Departament d'Innovaci\'o, Universitats i Empresa, Generalitat de Catalunya (2017-SGR-1278).

\bibliographystyle{plain}      
\bibliography{Bibliografia.bib}   

\end{document}